\documentclass[superscriptaddress,aps,preprintnumbers,amsmath,amssymb,sort&compress,nofootinbib,11pt]{revtex4-2}

\usepackage{amsfonts,amsmath,amssymb,ascmac,bm,tensor}
\usepackage{fnpct} 
\usepackage{comment}
\usepackage{ifpdf}
\usepackage{slashed}
\usepackage{color}
\usepackage[mathscr]{eucal}
\usepackage[utf8]{inputenc}
\usepackage{physics}
\usepackage{cancel}

\ifpdf
  \usepackage{graphicx}     
  \usepackage[bookmarksopen,colorlinks=true,linkcolor=bblue,citecolor=bblue,urlcolor=ppink]{hyperref}
\else     
\fi

\definecolor{red}{rgb}{1,0,0}
\definecolor{darkred}{rgb}{0.6,0,0}
\definecolor{darkgreen}{rgb}{0.992447,0.623778,0.034597}
\definecolor{ppink}{rgb}{1,0.4,0.4}
\definecolor{bblue}{rgb}{0.284602,0.317763,0.963947}
\definecolor{purple}{rgb}{0.5 ,0, 0.7}
 
\newcommand{\ee}{\text{e}}
\newcommand{\Pl}{\text{Pl}}
\newcommand{\tmax}{\text{max} }

\newcommand{\CMB}{\text{CMB}}

\makeatletter
\newcommand\footnoteref[1]{\protected@xdef\@thefnmark{\ref{#1}}\@footnotemark}
\makeatother

\allowdisplaybreaks[1]

\begin{document}


\title{
Amplification of Primordial Perturbations from the Rise or Fall of the Inflaton
}

\author{Keisuke Inomata}
\affiliation{Kavli Institute for Cosmological Physics and Enrico Fermi Institute, The University of Chicago, Chicago, IL 60637, USA}
\author{Evan McDonough}
\affiliation{Kavli Institute for Cosmological Physics and Enrico Fermi Institute, The University of Chicago, Chicago, IL 60637, USA}
\affiliation{Department of Physics, University of Winnipeg, Winnipeg, MB R3B 2E9 Canada}
\author{Wayne Hu}
\affiliation{Kavli Institute for Cosmological Physics and Enrico Fermi Institute, The University of Chicago, Chicago, IL 60637, USA}
\affiliation{Department of Astronomy \& Astrophysics, The University of Chicago, Chicago, IL 60637, USA}

\begin{abstract}
\noindent
The next generation of cosmic microwave background, gravitational wave, and large scale structure, experiments will provide an unprecedented opportunity to probe the primordial power spectrum on small scales. An exciting possibility for what lurks on small scales is a sharp rise in the primordial power spectrum: This can lead to the formation of primordial black holes, providing a dark matter candidate or the black holes observed by the LIGO-Virgo collaboration. In this work we develop a mechanism for the amplification of the small-scale primordial power spectrum, in the context of single-field inflation with a step-like feature in the inflaton potential. 
Specifically, we consider both the upward and the downward step in the potential. We also discuss the possibility of the strong coupling between perturbations because the rapid changes of the potential derivatives with the time-dependent field value, caused by the step-like feature, could make the coupling stronger. As a result, we find that the perturbations can remain weakly coupled yet sufficiently enhanced if the step realizes the rapid changes of the potential derivatives in some fraction of an e-fold, $\mathcal O(\mathcal P_{\mathcal R}^{1/2}) \lesssim \Delta N < 1$, where $\mathcal P_\mathcal R$ is the power spectrum of the curvature perturbation at that time.
We also discuss the PBH formation rate from the inflaton trapping at the local minimum, which can occur in the potential with an upward step.
\end{abstract}

\maketitle

\tableofcontents

\section{Introduction}

The paradigm of cosmological inflation \cite{Starobinsky:1980te,Sato:1980yn,Guth:1980zm,Linde:1981mu,Albrecht:1982wi,Linde:1983gd} has been tremendously successful in explaining the observed properties of the universe, from the flatness, homogeneity, and isotropy, to the cosmic microwave background (CMB), and the large scale structure (LSS) of the universe. Single field slow-roll inflation predicts a spectrum of primordial scalar perturbations that is adiabatic, and, on large scales, is nearly-Gaussian and nearly-scale invariant; these predictions have been confirmed by several generations of experiments. Inflation also predicts the existence of primordial gravitational waves, which, as the ``holy grail'' of inflation \cite{Baumann:2008aq}, is a primary target for next generation CMB \cite{Abazajian:2016yjj} and gravitational wave \cite{Bartolo:2016ami} experiments, as well as a small amount of primordial non-Gaussianity, which will be probed by both upcoming CMB  and LSS  experiments \cite{Meerburg:2019qqi}.

The historic successes of inflation, as well as future tests with CMB and LSS, are an exquisite probe of the primordial power spectrum on large scales, corresponding to wavenumbers $k \lesssim 1$\,Mpc$^{-1}$. In contrast, the {\it small-scale} primordial power spectrum remains largely uncharted territory.  There are good physical reasons for this: the diffusion (``Silk'') damping of CMB anisotropies, and the non-linear growth of structure, both serve to mask the power spectrum on small scales, obfuscating the primordial information contained therein.

Despite these challenges, there are exciting prospects for probing the small-scale primordial power spectrum: For example, primordial black holes (PBHs) can be produced by large perturbations on small scales at their horizon entry after the inflation era~\cite{Zeldovich:1967lct,Hawking:1971ei,Carr:1974nx,Carr:1975qj}. PBHs have been studied recently by many authors, both as a dark matter (DM) candidate and as the BHs detected by the LIGO-Virgo collaboration~\cite{Bird:2016dcv,Clesse:2016vqa,Sasaki:2016jop,Kashlinsky:2016sdv,Kashlinsky:2018mnu,Garcia-Bellido:2020pwq} (see Refs.~\cite{Sasaki:2018dmp,Carr:2020gox,Green:2020jor} for recent reviews).  Large primordial perturbations can also produce ultra-compact minihalos (UCMHs), which could emit the gamma rays from their centers depending on the DM properties \cite{Scott:2009tu,Bringmann:2011ut,Nakama:2017qac,Kawasaki:2021yek}. In addition, large perturbations on small scales can be sources of the distortion of CMB spectrum~\cite{Sunyaev:1970eu,Sunyaev:1970er} and the gravitational waves through the non-linear interaction between tensor and scalar perturbations~\cite{tomita1967non,Matarrese:1993zf,Matarrese:1997ay,Ananda:2006af,Baumann:2007zm,Saito:2008jc,Saito:2009jt,Kohri:2018awv,Inomata:2018epa,Adshead:2021hnm} (see Ref.~\cite{Domenech:2021ztg} for recent reviews). The future observation of CMB spectral distortions \cite{Chluba:2019nxa}, e.g., by PIXIE \cite{Kogut:2011xw}, and of a stochastic gravitational wave background, e.g., by SKA \cite{Bull:2018lat}, LISA \cite{LISA:2017pwj}, DECIGO \cite{Kawamura:2020pcg}, or others, will investigate the enhancement of the small-scale perturbations.

These new observational windows, and the past success of the single field inflation paradigm, provide ample motivation to develop the phenomenology of single field inflation on small scales.
In this paper we focus on inflation models that lead to a significant enhancement of primordial perturbations on small scales.
One way of enhancing the perturbations is to introduce a flat region in the inflationary potential~\cite{Ivanov:1994pa,Leach:2000yw,Leach:2001zf,Inoue:2001zt,Tsamis:2003px,Kinney:2005vj,Garcia-Bellido:2017mdw,Ezquiaga:2017fvi,Kannike:2017bxn,Germani:2017bcs,Ballesteros:2017fsr,Hertzberg:2017dkh,Cheng:2018qof,Byrnes:2018txb,Passaglia:2018ixg,Drees:2019xpp,Carrilho:2019oqg,Ng:2021hll}.
During the period of the rolling in the flat region, called the ultra-slow-roll (USR) period, the inflaton is rapidly decelerated due to the Hubble friction, which leads to the perturbation enhancement.
Apart from the flat region in the potential, the perturbation enhancement can be realized by non-canonical kinetic terms~\cite{Romano:2020gtn,Chen:2019zza}, a non-minimal coupling to gravity~\cite{Ezquiaga:2017fvi,Fu:2019ttf,Kawai:2021bye,Kawai:2021edk}, the parametric resonance during the inflation~\cite{Cai:2019bmk}, and the extension to multiple fields~\cite{Inomata:2020xad,Ando:2017veq,Inomata:2017vxo,Palma:2020ejf,Fumagalli:2020adf,Fumagalli:2020nvq}.

In this paper, we focus on canonical single field inflation models with a step-like feature in the potential $V(\phi)$, instead of the flat region. 
This class of models has been extensively studied in the past literature~\cite{Covi:2006ci,Hamann:2007pa,Mortonson:2009qv,Adshead:2011jq,Miranda:2013wxa,Adshead:2014sga,Cai:2015xla,Miranda:2015cea,Kefala:2020xsx,Wolfson:2021zsw} (see also Refs.~\cite{Mishra:2019pzq,Ozsoy:2018flq} for a bump/dip or an oscillatory feature in the potential).
Recently, in Ref.~\cite{Inomata:2021uqj}, we have shown that a downward step in the potential can enhance the power spectrum by $\mathcal O(10^7)$, which is enough for the production of PBHs as DM and the LIGO/Virgo events~\cite{Sasaki:2018dmp}.
Also, the connection between the PBH scenarios and  multiple downward steps has been studied recently in Ref.~\cite{Dalianis:2021iig}.
In terms of the inflationary slow-roll parameters $\epsilon \equiv - \dot{H}/H^2$ and $\eta \equiv {\rm d} \ln \epsilon/{\rm d}N$, with ${\rm d}N\equiv H {\rm d}t$, these models are characterized by a rapid change in $\epsilon$, corresponding to a transient phase of $|\eta| > 1$, whilst maintaining $\epsilon \ll 1 $ at all times.

The physical process at play may be understood from simple energy conservation considerations: When the inflaton encounters a downward step, it experiences a rapid transfer of potential energy into kinetic energy, followed by the dissipation of this kinetic energy via Hubble friction. Conversely, when the inflaton encounters an upward step, there is a rapid conversion of kinetic energy into potential energy. In both cases, across the transition 
an incoming positive frequency mode of the comoving curvature perturbation is rapidly converted to a linear combination of negative and positive frequency modes (with respect to the post-transition background), indicative of particle production (see e.g. Ref.~\cite{Birrell:1982ix}). Evolving the modes through the subsequent cosmological evolution, one finds an enhancement of the primordial power spectrum on the scales close to the horizon at the moment of the transition. 
Note that this particle production makes it possible to realize a power spectrum which is much larger than the prediction in the slow-roll approximation, $\mathcal P_{\mathcal R}\propto 1/\epsilon$.
We will demonstrate this in a concrete inflation model consistent with CMB observations, along with toy models that admit simple analytic solutions.

In this work we also perform a detailed analysis of the perturbativity of the cosmological perturbations. 
Since the coupling constants of higher order operators involving the comoving curvature perturbation are determined by the evolution of the slow-roll parameters, there is a potential strong coupling problem when the slow-roll parameters (e.g., $\epsilon$) undergo a rapid change~\cite{Adshead:2014sga}.  
In particular, the large enhancement of the curvature perturbations makes the coupling stronger.
Once the perturbations are strongly coupled,  linear perturbation theory breaks down.
Since  previous works on the perturbation enhancement with a step feature were based on  linear theory~\cite{Covi:2006ci,Hamann:2007pa,Mortonson:2009qv,Adshead:2011jq,Miranda:2013wxa,Adshead:2014sga,Cai:2015xla,Miranda:2015cea,Kefala:2020xsx,Wolfson:2021zsw,Mishra:2019pzq,Ozsoy:2018flq,Inomata:2021uqj,Dalianis:2021iig}, it is nontrivial whether or not the $\mathcal O(10^7)$ enhancement really occurs with a step-like feature, though we 
provided a qualitative discussion of how to avoid the strong coupling problem in our previous work~\cite{Inomata:2021uqj}. 
One of the main goals of this paper is to show quantitatively the conditions under which strong coupling may be avoided and to provide a concrete potential that can realize the $\mathcal O(10^7)$ enhancement as an example.
Specifically, we demonstrate that a potential with a downward step can amplify perturbations up to levels for the PBH scenarios while leaving the perturbations weakly coupled during the particle production. 
In contrast, in the upward step scenario the perturbations are only marginally weakly coupled during the particle production.
These results rely on our fiducial step form that can be extended to have more than one field-width parameter to describe the step transition.
In an appendix, we show that a tanh-like downward step, which is used in many previous works and has only one parameter for the step field width, cannot avoid the strong coupling issue since the change in $\epsilon$ occurs too quickly, which prevents the enhancement from being studied perturbatively (i.e., using the linearized equations of motion).  These examples illustrate the general conclusion that what is required to avoid strong coupling is 
a condition on the number of efolds $\Delta N$ over which the change in the slow-roll parameters occurs.

In the course of this investigation we make contact with a related but distinct mechanism for the genesis of PBHs, namely through the ``trapping'' of the inflaton in a region of the potential away from the post-inflation vacuum, and the subsequent formation of baby universes \cite{Deng:2017uwc,Garriga:2015fdk}. This naturally arises in the context of an upward step, wherein the incoming kinetic energy of the inflaton would naively (that is, classically) traverse the step, were it not for quantum backreaction. 
Regions of the universe that remain trapped are seen by outside observers, i.e., observers in regions of space in which inflation has ended, as PBHs~\cite{Deng:2017uwc,Garriga:2015fdk,Atal:2019cdz,Atal:2019erb}.

These results serve as a lamp post for the future study of features in the small-scale primordial power spectrum, suggest that care will need to be taken in interpreting future data, and inferences as to the nature of inflation (single field vs. multifield, canonical vs non-canonical, etc.). Finally, these results motivate a detailed comparison of primordial vs. non-primordial (e.g., \cite{Pollack:2014rja,Blinov:2021axd}) mechanisms for the enhancement of perturbations.

The structure of this paper is as follows. In Sec.~\ref{sec:model}, we present a single field inflation model exhibiting a step-like feature, and demonstrate the amplification of perturbations. In Sec.~\ref{sec:up} we present a toy model for an upward step, which exhibits a simple analytic description, and in Sec.~\ref{sec:down} perform a similar analysis for a downward step. 
In Sec.~\ref{sec:nonadiab}, we consider constraints on the model from the strong coupling and the backreaction, utilizing a smoothing of the transition from Sec.~\ref{sec:smoothing} to control their impact.
In Sec.~\ref{sec:trap_inflaton}, we discuss the trapping of the inflaton, and conclude in Sec.~\ref{sec:concl} with a discussion of directions for future work.

\section{Amplification of Perturbations from Step-like Features in the Inflationary Potential}
\label{sec:model}

\subsection{Mechanism}

First, we explain the mechanism for the perturbation enhancement with a potential step.
The evolution of the comoving curvature perturbation is governed by the Mukhanov-Sasaki equation for the curvature perturbation~\cite{Sasaki:1986hm,Mukhanov:1988jd,Miranda:2015cea}, which reads, in Fourier space,
\begin{align}
	\mathcal R_k'' + \left(2 + \eta \right) aH \mathcal R_k' + k^2 \mathcal R_k = 0,
	\label{eq:r_curv_eom_com}
\end{align}
where $a$ is the scale factor and the prime denotes the derivative with respect to the conformal time, $\tau$, defined by ${\rm d}\tau \equiv {\rm d}t/a$ with $\tau=0$ at the end of the inflation era. 
From this equation, the perturbation enhancement can be associated with $\eta$ or equivalently the evolution of $\epsilon$.
To describe the essence, we introduce three $\epsilon$: $\epsilon_i$ as the value just before the step transition, $\epsilon_m$ as the value just after the step transition, and $\epsilon_f$ as the value well after the transition.
Throughout this work, we consider a potential that has almost the same slopes before and after the step, which leads to $\epsilon_f \simeq \epsilon_i$.

We here focus on the case of a sharp step that realizes an almost instantaneous transition from $\epsilon_i$ to $\epsilon_m$, which corresponds to a large $\eta$ during the transition. 
During the large $\eta$ phase, we may approximate Eq.~(\ref{eq:r_curv_eom_com}) by
\begin{equation}
\mathcal{R}_k'' + \frac{\epsilon'}{\epsilon} \mathcal R_k'\simeq 0, 
\label{eq:r_curv_approx}
\end{equation}
up to corrections that scale as $k^2/(\eta aH)^2 $. 
Integrating the equation of motion over the transition, we find a rescaling of ${\cal R}'$,
\begin{align}
\mathcal R_k'(\tau_m) \simeq \frac{\epsilon_i}{\epsilon_m} \mathcal R_k'(\tau_i),
	\label{eq:r_dot_rel}
\end{align}
where $\tau_i$ is the conformal time at $\epsilon=\epsilon_i$ and $\tau_m$ is at $\epsilon = \epsilon_m$. 
This rescaling of ${\cal R}_k '$ implies a mode-mixing across the transition; a hallmark of particle production (see e.g. Ref.~\cite{Birrell:1982ix}). 
In the upward step case, we have $\epsilon_i/\epsilon_m \gg 1$ and therefore a large enhancement of $\mathcal R'$. After the step transition, $\mathcal R$ also gets enhanced by following the enhancement of $\mathcal R'$. 
On the other hand, in the downward step case, we have $\epsilon_i/\epsilon_m \ll 1$ and $\mathcal R'$ gets suppressed with $\mathcal R$ fixed. In this case, the enhancement of curvature perturbation originates from  the following period. In general, the excess kinetic energy due to the downward step naturally makes the following period a USR period, which enhances $\mathcal R$ after the step transition. 
Note that, if the downward step is not sharp and the transition from $\epsilon_i$ to $\epsilon_m$ is adiabatic, $\mathcal R$ gets suppressed during the downward step by $\sqrt{\epsilon_i/\epsilon_m}$ and no perturbation enhancement occurs in the case of $\epsilon_i \simeq \epsilon_f$. 
We shall see that these scaling behaviors are realized in the
upward (Sec.~\ref{sec:up}) and downward step (Sec.~\ref{sec:down})
cases using an analytic approximation for their behavior with the fiducial potential that we introduce next.

\subsection{Fiducial potential}

Here, we introduce our fiducial potential. 
Note again that one of the main goals of this paper is to provide a concrete inflaton potential that can realize the $\mathcal O(10^7)$ enhancement but with enough freedom to explore and avoid the strong coupling between the perturbations.
With this in mind, we determine the fiducial potential, especially the form of the step-like feature.

In this work we focus on a simple class of single field inflation models, namely that in which the inflaton potential can be described by a `bare' potential modulated by a step-like feature.
In this class of models, the inflaton potential can be expressed as,
\begin{align}
	V(\phi) = V_b(\phi) F\left(\phi; \phi_1,\phi_2,	h \right),
	 \label{eq:pot_cmb_to_end_mo}
\end{align}
where $V_b(\phi)$ is the `bare' inflaton potential, and $F\left(\phi; \phi_1,\phi_2,	h \right)$ describes the step-like feature.  For concreteness, we focus on a hilltop inflation model, wherein the potential during inflation can be described locally  as,
\begin{align}
  V_b(\phi) = V_0 \left(1 - \frac{\beta \phi^2/ M_\Pl^2}{1+\phi/\phi_\text{CMB}} \right) ,
  \label{eq:vbase_pot}
\end{align}
where $\phi_\text{CMB}$ is the value of $\phi$ at the moment when the CMB pivot scale ($k = 0.05$\, Mpc$^{-1}$) exits the horizon.  The form of $V_b$ can be expected to differ from the above for $\phi \gg \phi_\text{CMB}$, so as to describe the end inflation, but the precise form is not important for our purposes, as we are interested in the power spectrum on the scales far the horizon scales at the end of the inflation, which is insensitive to the details of the end of inflation.\footnote{
We note that, in the downward step case, the form of $V_b$ should be modified to end inflation at a value of $\phi$ that is much larger than it would be for the same parameters but without the step. This is because the acceleration due to the downward step increases the field distance that the inflation rolls in a fixed 
number of efolds from the CMB scales to the end of inflation.
In Ref.~\cite{Inomata:2021uqj}, we derived the concrete value of $\phi$ at which $V_b$ is modified.
}

The bare potential $V_b$ determines the normalization and the tilt of power spectrum on the CMB scales, irrespective of the feature $F$ as long as it only appears at $\phi >  \phi_{\rm CMB}$. The slow-roll parameters, evaluated at horizon exit of the CMB pivot scale, are given by,
\begin{align}
	\epsilon &\simeq \frac{M_\Pl^2}{2} \left( \frac{1}{V}\frac{\dd V}{\dd \phi} \right)^2 \simeq \frac{9}{32} \frac{\beta^2 \phi_\text{CMB}^2}{M_\Pl^2}, \\
	\eta_V &\equiv M_\Pl^2 \frac{1}{V}\frac{\dd^2 V}{\dd \phi^2} \simeq - \frac{\beta}{4} ,	\\
	\xi^2_V & \equiv M_\Pl^4 \frac{1}{V^2} \frac{\dd V}{\dd \phi} \frac{\dd^3 V}{\dd \phi^3} \simeq
	-\frac{9}{32} \beta^2.
\end{align}
From this we find the tilt of power spectrum at the CMB scale ($n_s$) is determined primarily by the second derivative of the potential, as 
\begin{align}
    n_s -1 \simeq 2 
    \eta_V
    \simeq -\frac{\beta}{2}.
\end{align}
We take $\beta=0.06$ to be consistent with the CMB measurement, $n_s = 0.97$~\cite{Aghanim:2018eyx}.
Then, the third slow-roll parameter becomes $\xi_V^2 = - 1.0\times 10^{-3}$, which is consistent with the Planck results~\cite{Aghanim:2018eyx}.
We relate $V_0$ and $\phi_\text{CMB}$ through the CMB normalization, $V_0/M_\Pl^4 = 24\pi^2 \epsilon(\phi_\text{CMB}) \times 2.1 \times 10^{-9}$~\cite{Aghanim:2018eyx}.

The step feature $F\left(\phi; \phi_1,\phi_2,h \right)$ can describe either an upward or downward step in the potential. We consider a continuous step of the form,
\begin{equation}
	F(\phi;\phi_1,\phi_2,h) =
  1 + h  \left[ S\left( \frac{\phi - \phi_1}{\phi_2-\phi_1} \right) \Theta(\phi-\phi_1) \Theta(\phi_2-\phi) + \Theta(\phi-\phi_2) \right],
\label{eq:f_step}
\end{equation}
where $S(x) \equiv x^2(3- 2x)$  changes from $0$ to $1$ within $0 \leq x \leq 1$.
The parameter $h$ corresponds to the height of the step which is normalized by $V_b$, whereas $\phi_1$ and $\phi_2$ denote the beginning and the end of the step. 
An upward or downward step corresponds to $h>0$ or $h<0$ respectively. 
We shall see in Sec.~\ref{sec:nonadiab} and Appendix~\ref{app:tanh_step}  that this parameterization  provides enough flexibility to avoid strong coupling problems for sharp steps whereas simpler smoothing schemes have problems.

\begin{figure}[h!]
\begin{minipage}{0.49\hsize}
\begin{center}
\includegraphics[width=1\columnwidth]{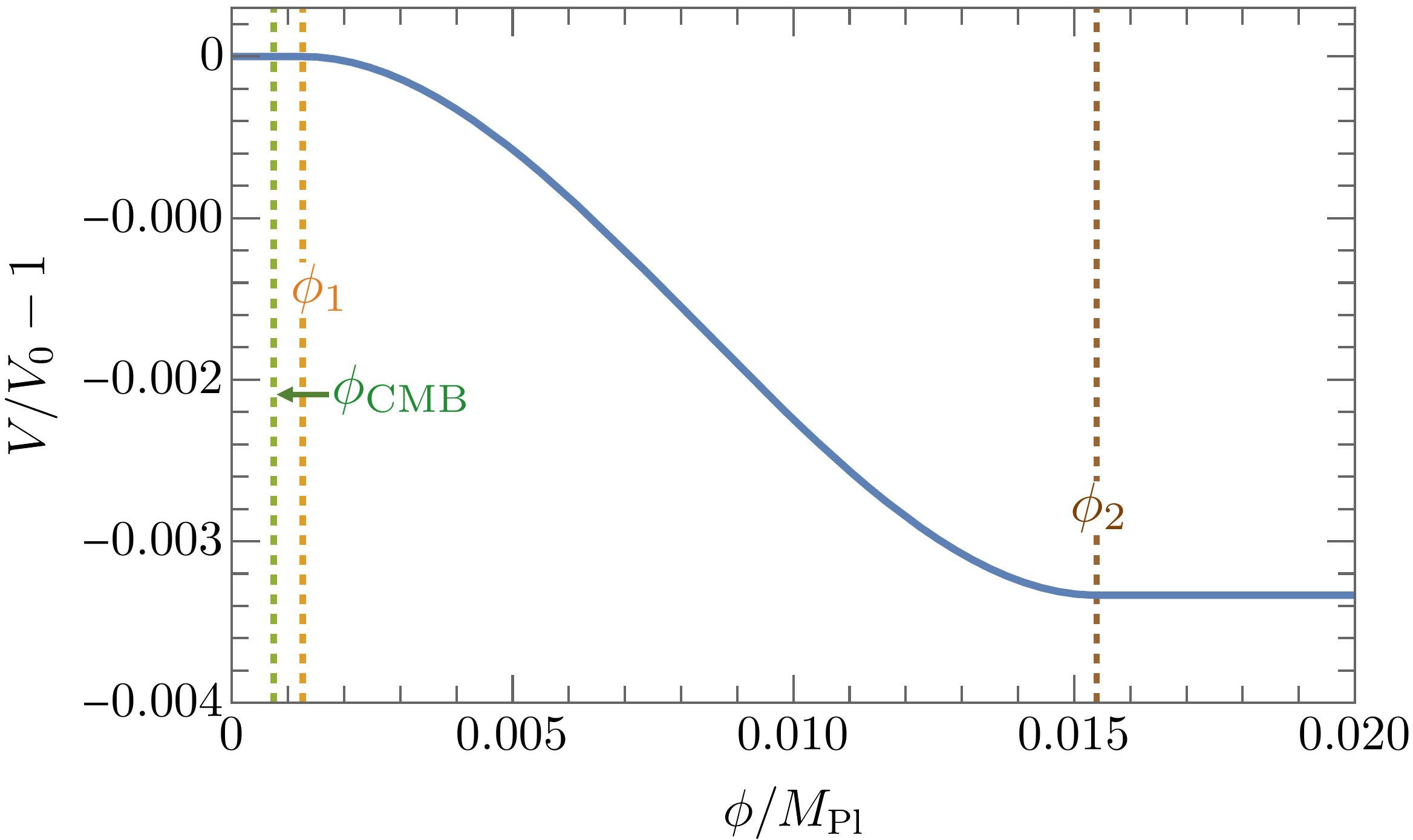}
\end{center}
\end{minipage}
\begin{minipage}{0.49\hsize}
\begin{center}
\includegraphics[width=1\columnwidth]{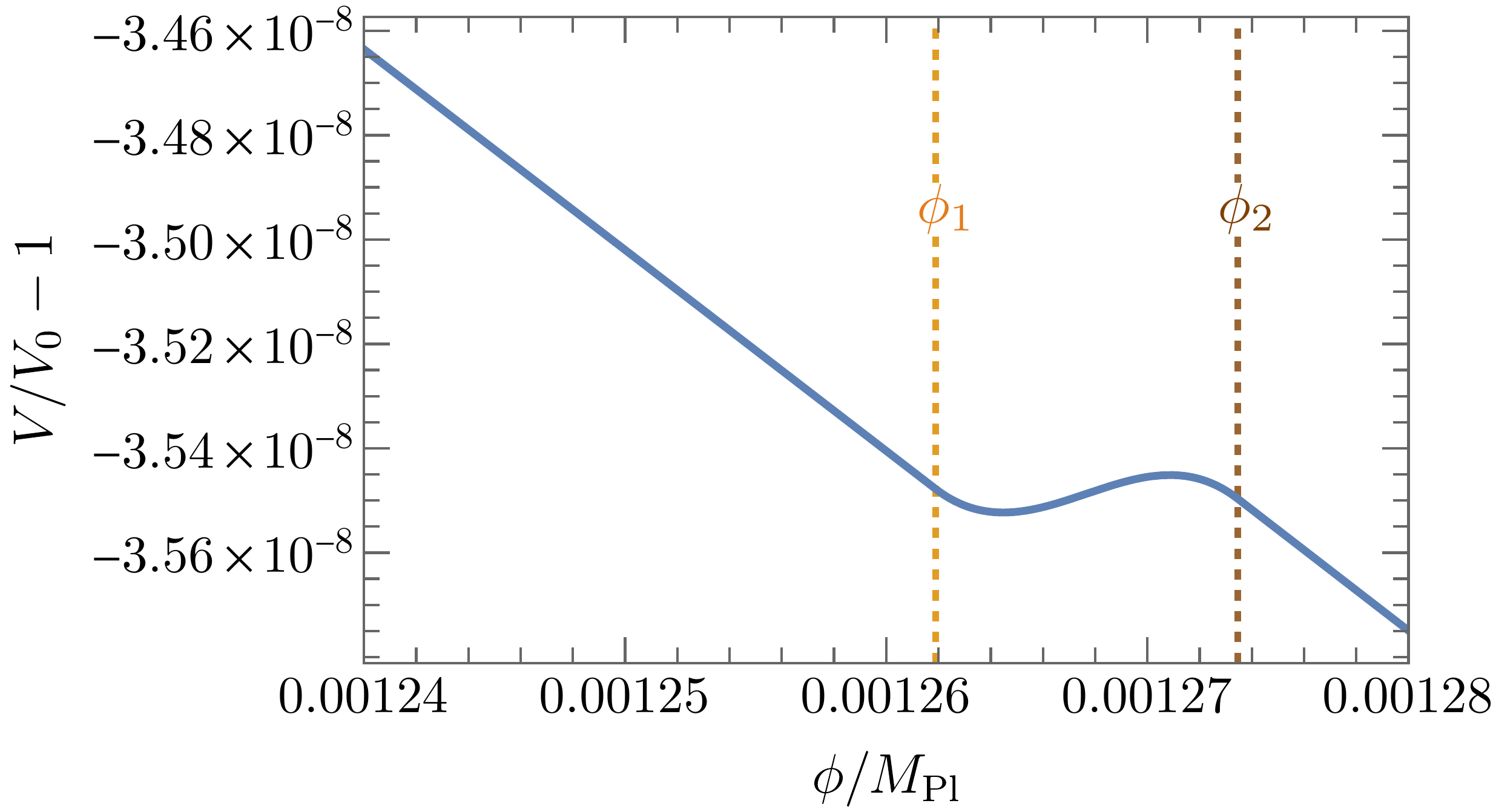}
\end{center}
\end{minipage}
\caption{
The inflaton potentials of Eq.~(\ref{eq:pot_cmb_to_end_mo}) that realizes the large enhancement of perturbations with a downward step (left) and an upward step (right) at $\phi_1 \le \phi \le \phi_2$ highlighted.
The parameters are $n_s = 0.97$, $\epsilon_i=7.43\times 10^{-10}$, and $\epsilon_f = 10^{-9}$ for both steps and $l_d = 0.1 \,(\Delta N_{\rm step}\simeq 0.5)$ and $\epsilon_m = 0.01$ for the downward step and $l_u = 0.3$ and $\epsilon_m = 10^{-4}\,\epsilon_i$ for the upward step.
Note that we take different axis scales in the two figures to highlight the steps at $\phi_1 \le \phi \le \phi_2$ in both cases. 
}
\label{fig:pot_l01}
\end{figure}

This step-like feature can be characterized by four phenomenological parameters, the three $\epsilon$'s ($\epsilon_i, \epsilon_m, \epsilon_f$) and $\Delta N_\text{step}$, the e-folds for the change of $\epsilon_i \rightarrow \epsilon_m$.
In our setup, $\epsilon_i$ is the value at $\phi_1$,  $\epsilon_m$ is the maximum or minimum value associated with the step, and $\epsilon_f$ is given by $\beta^2\phi_\CMB^2/(2M_\Pl^2)$.
In particular, $\epsilon_m$ and $\Delta N_\text{step}$ are controlled by the step height, $h$, and the step width, $\phi_2-\phi_1$, in Eq.~(\ref{eq:f_step}).
Here, we first discuss the relation between $\epsilon_m$ and $h$.
 The energy conservation law can relate $h$ to the $\epsilon$'s as
\begin{equation}
    h \simeq
  \frac{\epsilon_i-\epsilon_m}{3},
  \label{eq:step_height}
\end{equation}
where we have neglected the effect of the Hubble friction as the inflaton traverses the step.
Using this relation, we can use $\epsilon_m$ as an alternative to $h$ in the downward step case.\footnote{Strictly speaking, if we determine the step height through Eq.~(\ref{eq:step_height}), the maximum of $\epsilon$ slightly deviates from $\epsilon_m$ due to the Hubble friction. In this paper, we do not care about this deviation because it hardly changes the final power spectrum in the downward step case. }
On the other hand, in the upward step case, we do not use the relation Eq.~(\ref{eq:step_height}) because $\epsilon_m$ sensitively depends on the real height of the step in the case of $\epsilon_m \ll \epsilon_i$.
Note that the real step height deviates from $h$ because of the tilt of the bare potential.
Although this deviation vanishes as  $\Delta N_\text{step} \rightarrow 0$, it still makes a significant change of $\epsilon_m$ for finite $\Delta N_\text{step}$ in the upward step case.
Moreover in this case, $\epsilon(\phi_2) \ne \epsilon_m$ 
and $\Delta N_\text{step} \ne N_2-N_1$, where $N_a$ is the value at $\phi_a$, since the slope of the bare potential $V_b$ is sufficiently large to alter the position of the local maximum of $V(\phi)$ (see Fig.~\ref{fig:pot_l01}).

Next, we discuss the relation between the e-folds for the transition, $\Delta N_\text{step}$, and the step width, $\phi_2 - \phi_1$. 
For convenience, we define the non-dimensional parameters as $l_d \equiv (\phi_2-\phi_1)/(\sqrt{2 \epsilon_m} M_\Pl)$ and $l_u \equiv (\phi_2-\phi_1)/(\sqrt{2 \epsilon_i} M_\Pl)$.
Throughout this paper, we consider the case of $l_d < \mathcal O(1)$ for the downward step case and $l_u < \mathcal O(1)$ for the upward step case, which lead to $\Delta N_\text{step} < \mathcal O(1)$, because we focus on the case where the inflaton gets significantly accelerated or decelerated within less than $\mathcal O(1)$ e-fold. 
In particular, in the downward step case, we can relate the step width $\phi_2-\phi_1$ (or $l_d$) to $\Delta N_\text{step}$ as 
\begin{align}
  \Delta N_\text{step} \simeq \frac{\ln(\epsilon_m/\epsilon_i)}{\bar \eta} \quad (\text{for downward step}),
  \label{eq:n_step_approx}
\end{align}
where $\bar \eta$ is the value of $\eta$ during the time period in which $\phi$ traverses the step, which as we numerically show, in the downward step case, is well approximated as constant. We may estimate the value of $\bar{\eta}$ for the downward step case by solving the equation of motion for $\phi$ in the neighborhood of $\phi=\phi_1$, where the field has a tachyonic mass $m_\phi^2 = V''(\phi_1) \simeq 3V_0h/(\phi_2- \phi_1)^2$. From this, we can finally express $\bar{\eta}$ as 
\begin{align}
    \bar \eta =  
     -6 - \frac{24hM_\Pl^2}{(\phi_2 - \phi_1)^2} \left( -1 + \sqrt{1- \frac{8 hM_\Pl^2}{ (\phi_2 - \phi_1)^2}} \right)^{-1} \nonumber \\
     = -6 - \frac{4(\epsilon_i - \epsilon_m)}{\epsilon_m l_d^2} \left( -1 + \sqrt{1- \frac{4(\epsilon_i - \epsilon_m)}{3\epsilon_m l_d^2}} \right)^{-1}.
  \label{eq:eta_app_exp_ease}
\end{align}
where we have left the detailed derivation for Appendix~\ref{app:evol_up_down}.\footnote{Meanwhile, in the upward step case, $\eta$ during the rolling up is not constant because the inflaton evolution during that time depends on its initial velocity just before the step. }

\begin{figure}[h!]
\begin{minipage}{0.49\hsize}
\begin{center} 
\includegraphics[width=1\columnwidth]{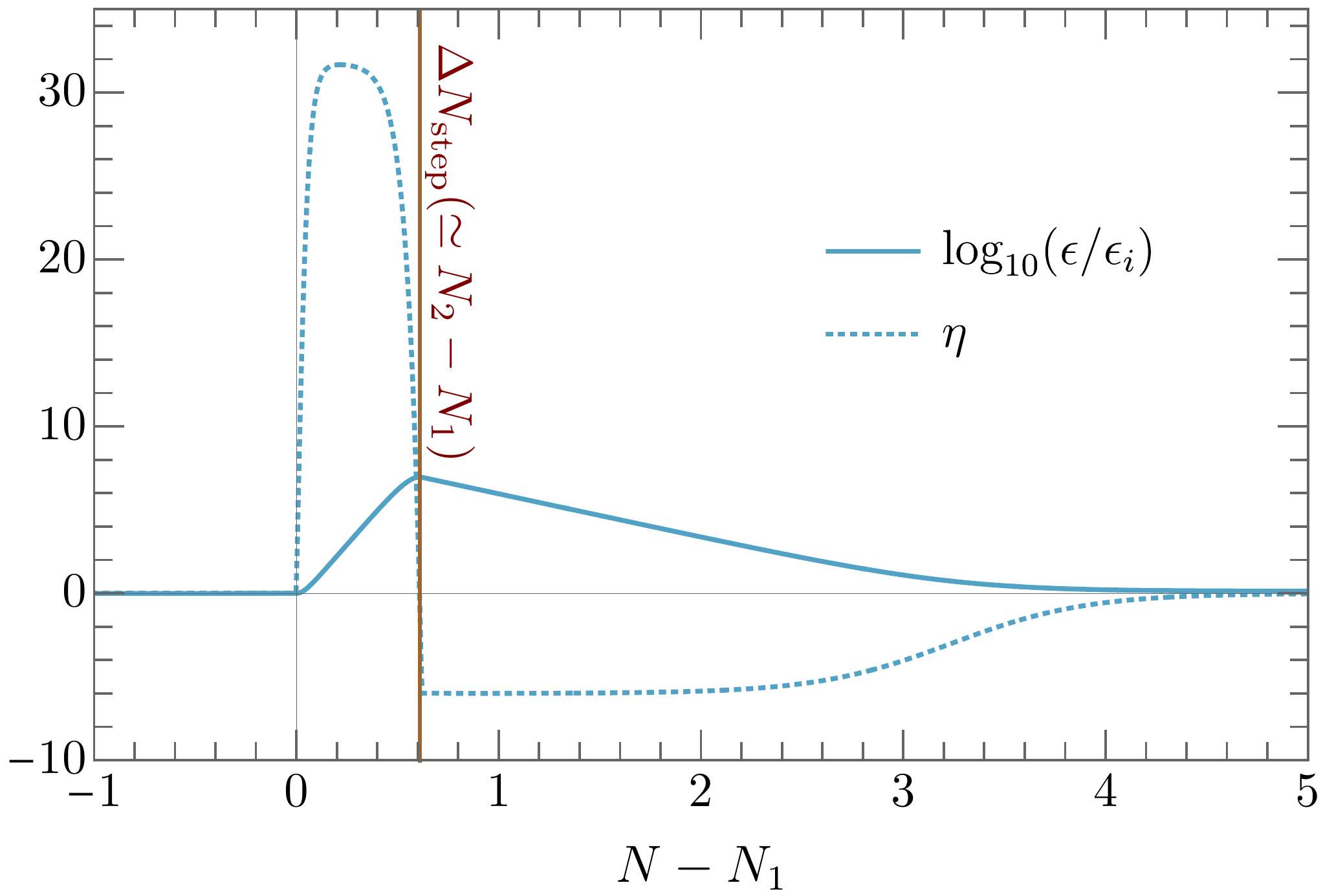}
\end{center}
\end{minipage}
\begin{minipage}{0.49\hsize}
\begin{center}
\includegraphics[width=1\columnwidth]{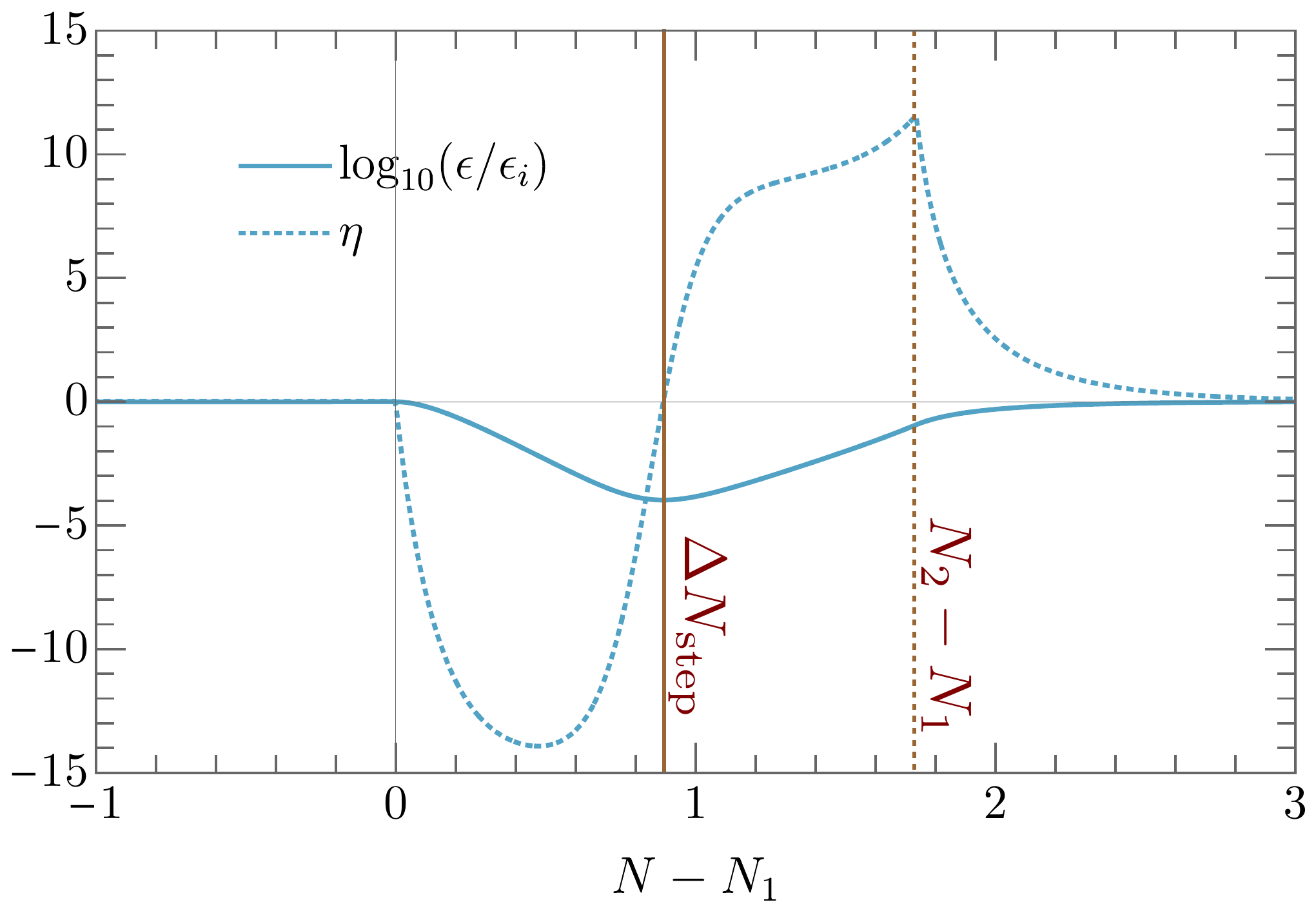} 
\end{center}
\end{minipage}
\caption{
Evolution of $\epsilon$ and $\eta$ in the potential with the downward (left) and the upward step (right), shown in Fig.~\ref{fig:pot_l01}.
The parameters are the same as in Fig.~\ref{fig:pot_l01}.
The vertical brown solid and dotted lines show $\Delta N_\text{step}$ and $N_2 - N_1$, respectively.
Note that we omit the line for $N_2-N_1$ in the left figure because $N_2-N_1 \simeq \Delta N_\text{step}$ in the downward step case.
The approximate form of $\Delta N_\text{step}$, given by Eqs.~(\ref{eq:n_step_approx}) and (\ref{eq:eta_app_exp_ease}), predicts $\Delta N_\text{step} \simeq 0.5$ with $\bar \eta \simeq 32$, which is close to the real value, $\Delta N_\text{step} \simeq 0.6$, in the left figure.
}
\label{fig:eta_ep_evol_up_down}
\end{figure}

As an illustrative example, Fig.~\ref{fig:pot_l01} shows $V(\phi)$, Eq.~\eqref{eq:pot_cmb_to_end_mo}, for a fiducial choice of parameters that exhibit a downward step (left figure) and an upward step (right figure) and are consistent with CMB observations. 
The dynamics of the inflaton evolving in this potential may be understood straightforwardly in terms of the kinetic and potential energy of the inflaton. 
For the downward step case, the numerical evolution of $\epsilon$ and $\eta$, again for the example potential Fig.~\ref{fig:pot_l01} (left), are shown in the left figure of Fig.~\ref{fig:eta_ep_evol_up_down}.
This indicates that  Eq.~(\ref{eq:eta_app_exp_ease}) is a good approximation to the enhancement of $\eta$, and immediately after the step $\eta \simeq -6$, due to the excess kinetic energy (relative to the flatness of the potential) following the downward step. 
On the other hand, for the upward step case, the evolution is shown in the right figure of Fig.~\ref{fig:eta_ep_evol_up_down}.
The inflaton loses kinetic energy as it climbs up the step, leading to a marked reduction in $\epsilon$.
After passing through the top of the potential, the inflaton gets accelerated, which corresponds to the positive $\eta$ period.
The peak of $\eta$ corresponds to the value at $\phi = \phi_2$.
Note that, in the upward step case, the time at $\epsilon=\epsilon_m$ is quite different from the time at $\phi = \phi_2$, unlike in the downward step case.

\begin{figure}[h!]
\centering \includegraphics[width=.6\columnwidth]{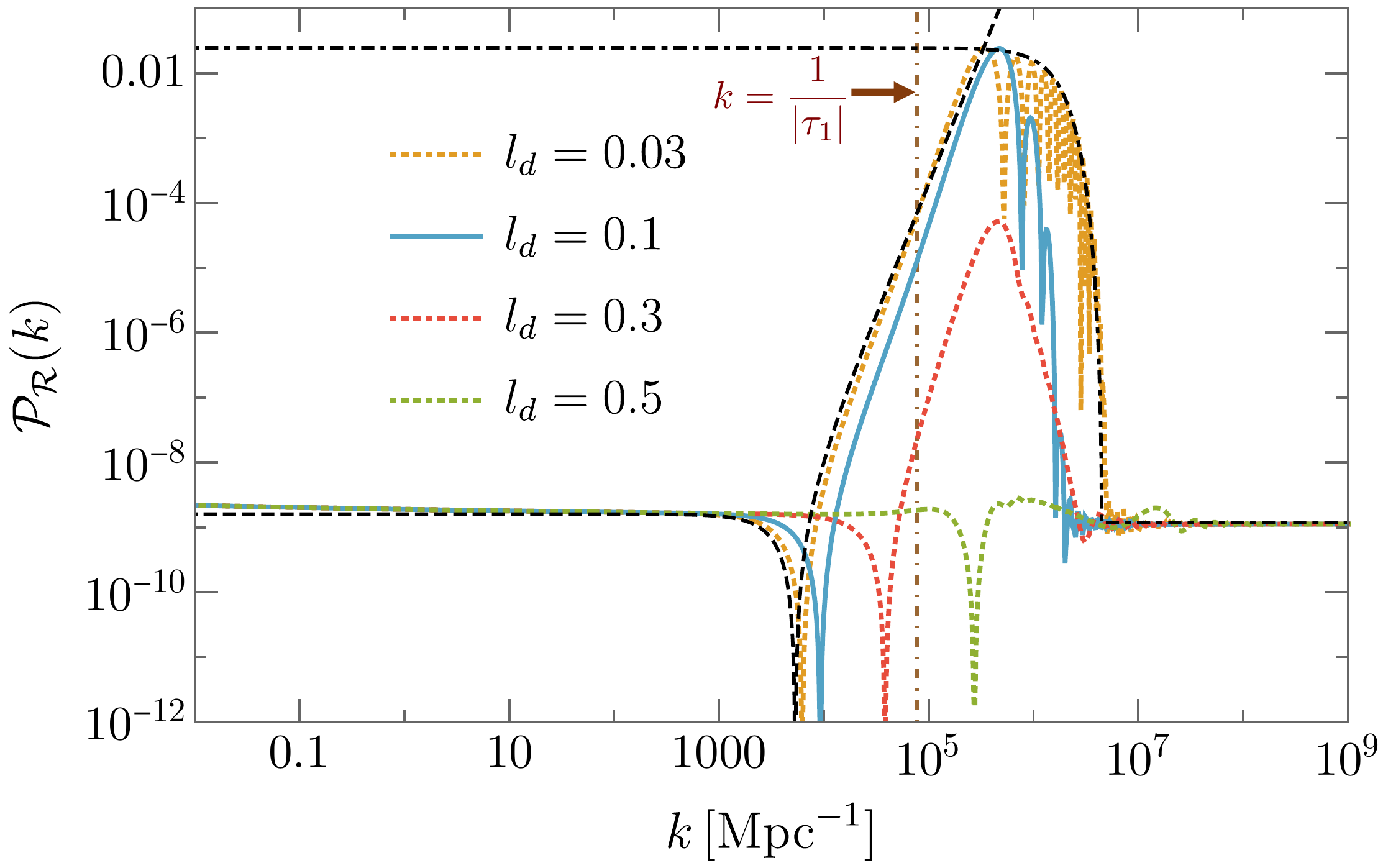}
\caption{ The primordial power spectrum from the potential Eq.~\eqref{eq:pot_cmb_to_end_mo} with a downward step.  
The vertical brown line is the scale corresponding to the horizon scale at $\phi= \phi_1$.
The solid line is the result for the potential shown in the left panel of Fig~\ref{fig:pot_l01}.
The dotted lines are the results for different $l_d$ with the other parameters being the same as in Fig.~\ref{fig:pot_l01}.
The transition e-folds become $\Delta N_\text{step} \simeq 0.2$ in $l_d = 0.03$, $\Delta N_\text{step} \simeq 0.6$ in $l_d \simeq 0.1$, $\Delta N_\text{step} \simeq 2.1$ in $l_d = 0.3$, and $\Delta N_\text{step} \simeq 3.9$ in $l_d = 0.5$.
Overlaid in black lines are analytic approximations, derived in Sec.~\ref{sec:down}, valid on large scales (black dashed), Eq.~(\ref{eq:pr_large_limit_downward}), and small scales (black dot-dashed), Eq.~(\ref{eq:damping}).}
\label{fig:ps_cmb_tra}
\end{figure}

\begin{figure}[h!]
\centering \includegraphics[width=.6\columnwidth]{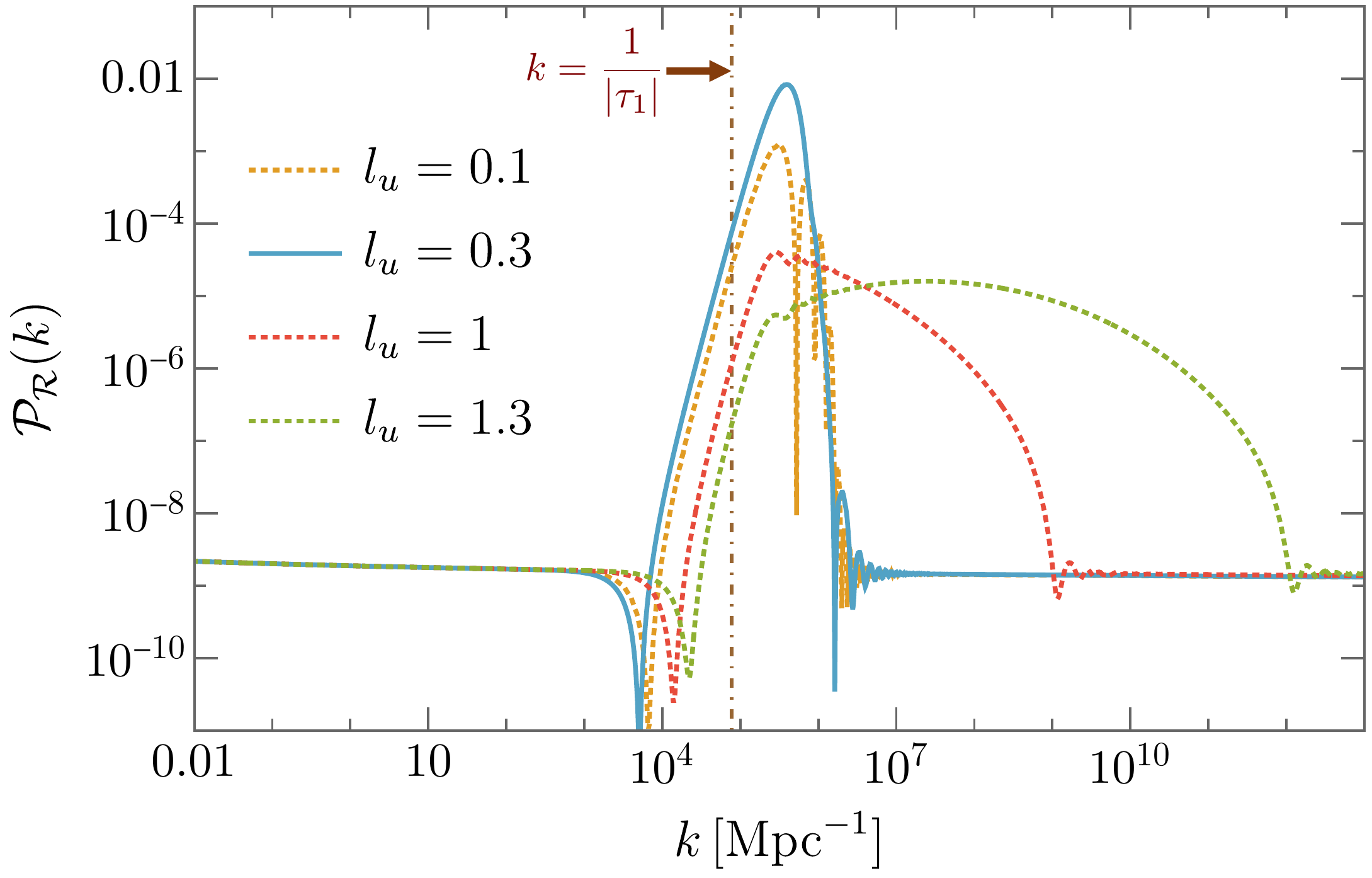}
\caption{ The primordial power spectrum from the potential Eq.~\eqref{eq:pot_cmb_to_end_mo} with an upward step.
The solid line is the result for the potential shown in the right panel of Fig~\ref{fig:pot_l01}.
The dotted lines are the results for different $l_u$ with the other parameters being the same as in Fig.~\ref{fig:pot_l01}.
The transition e-folds become $\Delta N_\text{step} \simeq 0.3$ in $l_u = 0.1$, $\Delta N_\text{step} \simeq 0.9$ in $l_u \simeq 0.3$, $\Delta N_\text{step} \simeq 2.5$ in $l_u = 1$, and $\Delta N_\text{step} \simeq 5.0$ in $l_u = 1.3$.
}
\label{fig:ps_cmb_tra_up}
\end{figure}

We numerically solve Eq.~\eqref{eq:r_curv_eom_com} for the evolution of perturbations during inflation in the potential Eq.~\eqref{eq:pot_cmb_to_end_mo} and compute the power spectrum for the case of a downward step and an upward step. 
Figure~\ref{fig:ps_cmb_tra} shows the power spectra in the potential with a downward step, given in Fig.~\ref{fig:pot_l01}. 
Incidentally, this parameter example realizes the $\mathcal O(10^7)$ enhancement in the power spectrum, required for the PBH scenarios~\cite{Inomata:2017vxo,Sasaki:2018dmp}. The amplitude of the peak of the power spectrum is controlled by $\epsilon_m/\epsilon_f$ (or $h$), while the peak scale is determined by  $\epsilon_i$ (or
$\phi_1$), which then determines the PBH mass. Overlaid on the plot, in dashed black lines, are analytic approximations to the power spectrum, which we derive in Sec.~\ref{sec:down}. 
The spectra for an upward step are shown in Fig.~\ref{fig:ps_cmb_tra_up}, where one can again produce a significant enhancement of perturbations.
We can see that, while the enhancement is of $\mathcal O(\epsilon_i/\epsilon_m)$ in case the step transition occurs over a few e-folds, it can be much larger if the step transition occurs within less than an e-fold.
We note that these results are robust to a smoothing of the potential, e.g., via convolution with a Gaussian, so as to make the step function infinitely differentiable. 
This is demonstrated in Sec.~\ref{sec:nonadiab}.
On top of that, with this smoothed potential, we also discuss in Sec.~\ref{sec:nonadiab} the strong coupling between perturbations, which limits the application of the Mukhanov-Sasaki equation.

In what follows, we will study in detail how the amplification of perturbations exhibited in Figs.~\ref{fig:ps_cmb_tra} and \ref{fig:ps_cmb_tra_up} arises, through toy models that may be solved analytically.

\section{The Rise of the Inflaton (Upward Step)}

\label{sec:up}

We begin with the upward step scenario. 
In this case, the instantaneous deceleration of the inflaton enhances the perturbations more efficiently than gradual deceleration.  On the other hand,
unlike the downward step case, which we discuss in the next section, the upward step case requires a finely-tuned step height to exhibit this enhancement.
This is because almost all of the kinetic energy of the inflaton before the step has to be lost by climbing the step.
The main goal of this section is to show the essence of the perturbation enhancement analytically with a simplified model.

To understand the enhancement of curvature perturbations, we start from the general solution to the Mukhanov-Sasaki equation in a phase of constant $\eta$, given by,
\begin{align}
	\mathcal R_k = C_1 G^{(1)}_\nu(-k \tau) + C_2 G^{(2)}_\nu(-k \tau),
	\label{eq:curv_r_general}
\end{align}
where $\nu = 3/2 + \eta/2$ and $C_1$ and $C_2$ are constant, and  $G^{(1)}_\nu$ and $G^{(2)}_\nu$ are defined with the Hankel functions of the first ($H_\nu^{(1)}$) and the second kind ($H_\nu^{(2)}$) as
\begin{align}
	G^{(j)}_\nu(-k \tau) \equiv  (-k\tau)^\nu H_\nu^{(j)}(-k \tau),
\end{align}
where $j \in (1,2)$.

We here focus on the case where $\epsilon$ changes from $\epsilon_i$ to $\epsilon_f$  with constant $\eta$, whilst satisfying $\epsilon \ll 1$ at all times. Specifically, we parameterize $\eta$ as 
\begin{align}
	\eta = \eta_c \, \Theta(\tau-\tau_1) \Theta(\tau_2 -\tau),
	\label{eq:eta_c_def}	
\end{align}
where $\eta_c (< 0)$ is a constant. 
Then, $\epsilon$ can be expressed as
\begin{align}
	\epsilon = \begin{cases}
	\epsilon_i  & (\tau < \tau_1) \\
	\epsilon_i  \left(\frac{a(\tau_1)}{a(\tau)} \right)^{-\eta_c} = \epsilon_i \left(\frac{\tau}{\tau_1} \right)^{-\eta_c} & (\tau_1 \leq \tau \leq \tau_2 )\\
	\epsilon_f  \left(= \epsilon_i  \left(\frac{\tau_2}{\tau_1} \right)^{-\eta_c} \right) & (\tau_2 < \tau (<0))
	\end{cases},
\end{align}
where we have used the relation  $a\simeq-1/H\tau$. 
Note that this toy model approximates the special case where $\epsilon_m \simeq \epsilon_f$ in the full upward step model of Sec.~\ref{sec:model}. Specifically, this parameterization does not take into account the acceleration phase with positive $\eta$ after the step climbing. However the perturbation enhancement originates from the rapid deceleration of the inflaton and we can estimate the order of the enhancement from this situation with one period of transition.

Imposing the continuity of $\mathcal R$ and $\mathcal R'$ at $\tau_1$ and $\tau_2$, we get 
\begin{align}
\mathcal R_k = 
\begin{cases}
	D_1 G^{(1)}_{3/2}(-k \tau)  +  D_2G^{(2)}_{3/2}(-k \tau)  & (\tau < \tau_1)\\	
	E_1 G^{(1)}_{\nu_c}(-k \tau)  + E_2 G^{(2)}_{\nu_c}(-k \tau)  & (\tau_1 \leq \tau \leq \tau_2)\\
	F_1 G^{(1)}_{3/2}(-k \tau)  + F_2 G^{(2)}_{3/2}(-k \tau) &(\tau_2 < \tau) \\
	\end{cases},
\end{align}
where $\nu_c$ is given by
\begin{align}
	\nu_c = \frac{3}{2} + \frac{\eta_c}{2}.
\end{align}
The coefficients $D_1$ and $D_2$ are determined by the Bunch-Davies initial conditions as\footnote{Note that the solution in $\tau < \tau_1$ can be approximated as $\mathcal R_k \simeq \frac{H}{\sqrt{4k^3 \epsilon_i}} (i + (-k \tau)) \ee^{-i k\tau}$.}
\begin{align}
	 D_1 = - \sqrt{\frac{\pi}{2}} \frac{H}{\sqrt{4k^3 \epsilon_i} M_\Pl} \,\, , \,\, D_2 =0,
   \label{eq:d_coeff}
\end{align}
while the other coefficients are determined by the continuity, and are given in full generality as,
\begin{align}
E_j = D_1 \frac{W[ G^{(1)}_{3/2}(-k \tau_1),{G^{(p_j)}_{\nu_c}}(-k \tau_1)]}
{W[{G^{(j)}_{\nu_c}}(-k \tau_1), G^{(p_j)}_{\nu_c}(-k \tau_1)]},
 \label{eq:e_coeff} 
\end{align}
\begin{align}
F_j = \frac{W[E_1 G^{(1)}_{\nu_c}(- k\tau_2) + E_2 G^{(2)}_{\nu_c}(- k\tau_2), G^{(p_j)}_{3/2}(- k\tau_2)]}
{W[{G^{(j)}_{3/2}}(-k \tau_2), G^{(p_j)}_{3/2}(-k \tau_2)]},
\label{eq:f1_coneff}
\end{align}
where $p_1 =2$, $p_2=1$, and $W$ is the Wronskian, $W[f,g] \equiv fg' - gf'$.
The power spectrum in the late-time limit ($-k \tau \ll 1$) is given by 
\begin{align}
	\mathcal P_\mathcal R =& \frac{k^3}{2\pi^2} \left|\mathcal R_k (|\tau| \ll 1/k) \right|^2 \nonumber \\
	\simeq& \frac{k^3}{2\pi^2} \left( \frac{2}{\pi} \right) |F_1 - F_2|^2, 
	\label{eq:p_r_analytical_2}
\end{align}
where the normalization factor $2/\pi$ comes from the relation $G^{(1)}_{3/2}(x \rightarrow 0) = -i \sqrt{ 2/\pi}$ and $G^{(2)}_{3/2}(x \rightarrow 0) = i \sqrt{ 2/\pi}$.

In the limit of an instantaneous transition, Eq.~(\ref{eq:r_curv_eom_com}) can be approximated as Eq.~\eqref{eq:r_curv_approx}, and we find, analogous to Eq.~\eqref{eq:r_dot_rel},
\begin{align}
	 \mathcal R_k'(\tau_2) \simeq \frac{\epsilon_i}{\epsilon_f} \mathcal R_k'(\tau_1).
	\label{eq:r_dot_rel_2}
\end{align}
The change of $\mathcal R'$ given by  Eq.~(\ref{eq:r_dot_rel_2}) is the origin of the power spectrum enhancement in the upward step case, which can be understood in terms of the subsequent evolution of ${\cal R}$: Although $\mathcal R$ does not change at the instantaneous transition, it grows after the transition following the change of $\mathcal R'$. 
Note that the coefficients $F_1$ and $F_2$ are determined at $\tau_2$.
This means that the particle production completes at $\tau_2$ in our setup.\footnote{
Throughout this paper, we use the word ``particle production'' as the mixing of the positive and the negative frequency modes.
Strictly speaking, the particle picture of the enhanced perturbations can be applicable only when the field evolves adiabatically~\cite{Chung:2018ayg}.
In this sense, our usage of the ``particle production'' is colloquial and refers to the change in the particle occupancy after the non-adiabatic transition.
}
In other words, the change of $\mathcal R$ after $\tau_2$ is not due to the particle production.
If the perturbations are strongly coupled in $\tau_1 < \tau < \tau_2$, the particle production cannot be described by the linearized Mukhanov-Sasaki equation and could possibly be shut off. For simplicity, we assume the strong coupling does not occur in the rest of this section and return to this issue in Sec.~\ref{sec:nonadiab}.
From Eq.~(\ref{eq:r_dot_rel_2}) and the continuity of $\mathcal R$ at the transition, we get the solution in the limit of   $|\tau_2 - \tau_1| \rightarrow 0$,
\begin{align}
	\mathcal R_k(\tau > \tau_2) = F_{1,\text{lim}} G^{(1)}_{3/2}(-k\tau) + F_{2,\text{lim}} G^{(2)}_{3/2}(-k\tau),
\end{align}
where
\begin{align}
	\label{eq:f1_coeff_lim}
	F_{1,\text{lim}} &=  - \frac{\sqrt{\pi} H}{\sqrt{8k^3 \epsilon_i} M_\Pl} \left[ \frac{i(\epsilon_f-\epsilon_i)+(-k \tau_1) (\epsilon_f+\epsilon_i)}{(-2k \tau_1) \epsilon_f} \right], \\
	\label{eq:f2_coeff_lim}	
	F_{2,\text{lim}} &=- \frac{\sqrt{\pi}H}{\sqrt{8k^3 \epsilon_i} M_\Pl}\left(\frac{\epsilon_f- \epsilon_i}{\epsilon_f} \right) \left( \frac{i+ (-k \tau_1)}{-2k\tau_1} \right) \ee^{-2ik\tau_1}.
\end{align}
Note $\tau_1 = \tau_2$ in this limit.
Then, from Eq.~(\ref{eq:p_r_analytical_2}), the power spectrum is given by
\begin{align}
	\mathcal P_\mathcal R 
	=& \frac{k^3}{2\pi^2} \left( \frac{2}{\pi} \right) \left(|F_{1,\text{lim}}|^2 -2 \text{Re}[ F_{1,\text{lim}}^* F_{2,\text{lim}} ] + |F_{2,\text{lim}}|^2 \right).
	\label{eq:p_r_analytical_lim}
\end{align}
For the perturbations on superhorizon scales at the transition ($-k\tau_1 \ll 1$), the power spectrum becomes 
\begin{align}
	\label{eq:p_approx_super}
	\mathcal P_{\mathcal R} \simeq& \frac{H^2}{8 \pi^2 \epsilon_i M_\Pl^2 } \left[ 1 + \left( \frac{2}{3} - \frac{2 \epsilon_i}{3 \epsilon_f} \right) (-k \tau_1)^2 
	+ \left( -\frac{13}{45} + \frac{ \epsilon_i^2}{9 \epsilon_f^2} + \frac{8 \epsilon_i}{45 \epsilon_f} \right) (-k\tau_1)^4 \right] + \mathcal O((-k\tau_1)^6).
\end{align}
In particular, the $\mathcal O((-k\tau_1)^4)$ term describes the growth of the power spectrum in the upward step case. 
It suffices to stop at this order given the matching conditions for the two
solutions  at the transition (see Ref.~\cite{Byrnes:2018txb} and an exceptional case in Ref.~\cite{Carrilho:2019oqg}, not relevant here, where those solutions allow a steeper  $k^5 (\text{log}\,k)^2$ growth).

On the other hand, for the perturbations on subhorizon scales at the transition ($-k\tau_1 \gg 1$), the power spectrum can be approximated as 
\begin{align}
	\label{eq:p_r_ana_small_limit}
	\mathcal P_{\mathcal R} \simeq& \frac{H^2}{8 \pi^2 \epsilon_i M_\Pl^2 } \left( \frac{ \epsilon_i^2 +  \epsilon_f^2}{2 \epsilon_f^2} + \frac{\epsilon_i^2 - \epsilon_f^2}{2 \epsilon_f^2} \cos(-2 k \tau_1) \right).
\end{align}
The physical origin of the oscillation of the power spectrum on the scales of $-k\tau_1 \gg 1$ can be seen in Eq.~(\ref{eq:r_dot_rel_2}).
The curvature perturbation on $-k\tau_1 \gg 1$ oscillates before the transition and its time derivative also does.
Since the effect of the step transition scales $\mathcal R'$ at the transition, which determines particle production through the two amplitudes on the other side,  the power spectrum in $k|\tau_1| \gg 1$ exhibits an oscillatory  feature.

\begin{figure}[h!]
\begin{center}
\includegraphics[width=0.6\columnwidth]{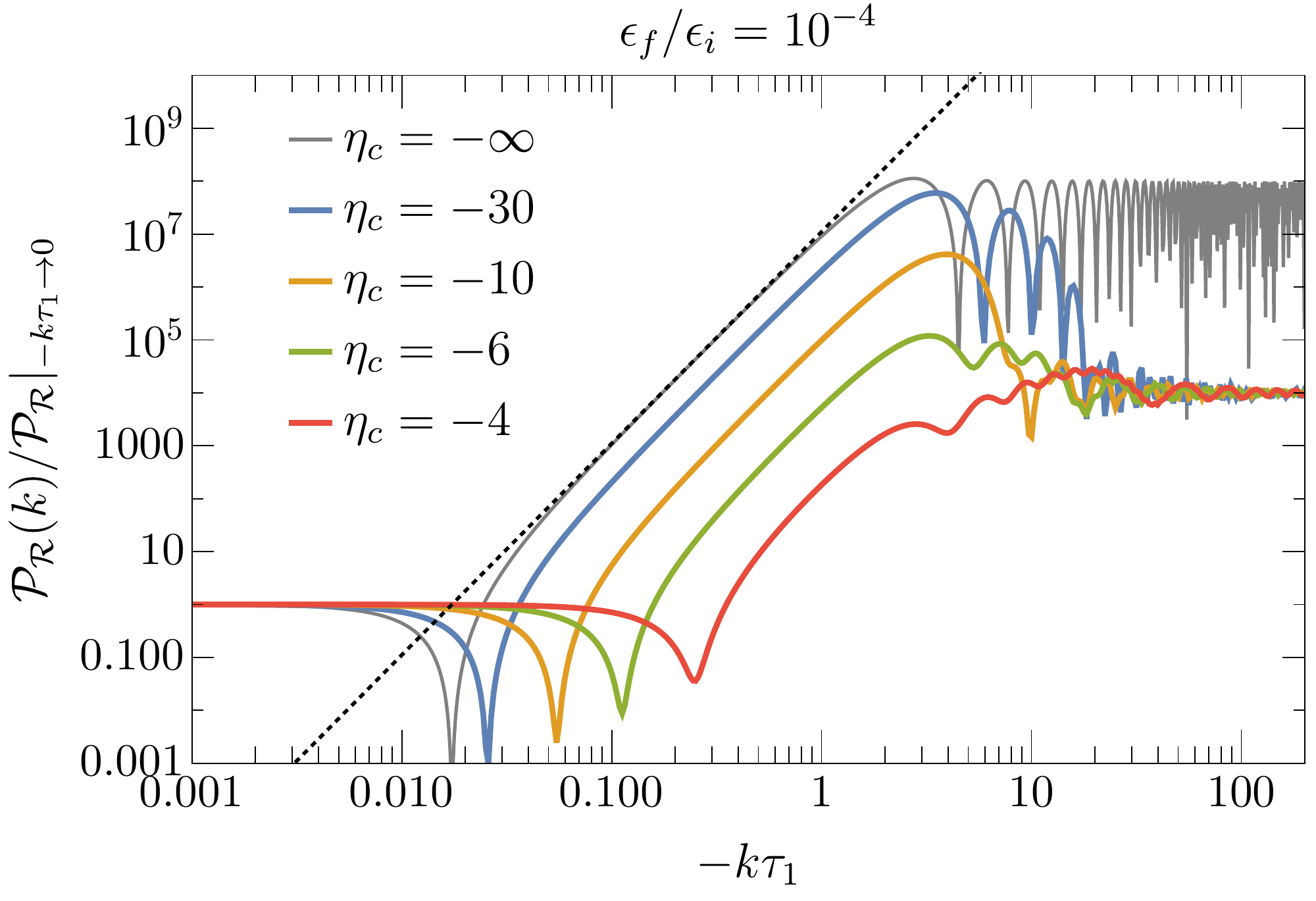}  
\end{center}
\caption{ 
	The power spectra with different $\eta_c$ in our toy upward step model, where the evolution of $\eta$ is given by Eq.~(\ref{eq:eta_c_def}).
  The power spectra are normalized by the value in the limit of large scale, $-k\tau_1 \rightarrow 0$.
  Note that, in terms of the parameterization in Sec.~\ref{sec:model}, this toy model has $\epsilon_m = \epsilon_f$, in contrast to the model in Sec.~\ref{sec:model}, where $\epsilon_f \simeq \epsilon_i$.
In the case of the instantaneous transition limit ($\eta_c = -\infty$), we use the simple forms of $F_{1,\text{lim}}$ and $F_{2,\text{lim}}$, given by Eqs.~(\ref{eq:f1_coeff_lim}) and (\ref{eq:f2_coeff_lim}).
  We take $\epsilon_f/\epsilon_i = 10^{-4}$ for all lines.
  For comparison, we also plot the approximation form, $\epsilon_i^2/(9 \epsilon_f^2) (-k\tau_1)^4$, from Eq.~(\ref{eq:p_approx_super}) with a black dotted line.
}
\label{fig:ps_one_period}
\end{figure}

The  solution Eq.~(\ref{eq:p_r_analytical_2}) to this toy model, i.e.\ the background specified by Eq.~\eqref{eq:eta_c_def}, is shown in Fig.~\ref{fig:ps_one_period}. 
From this one may appreciate that a large negative $\eta_c$ leads to a large enhancement of the perturbations, peaked around $-k\tau_1 \simeq 3$.  The peak of the power spectra can be much larger than the value in the slow-roll approximation with $\epsilon_f$, given by $H^2/(8\pi^2 \epsilon_f M_\Pl^2)$.  At the same time, we can also see that there is an asymptotic value of the peak height for a large $\eta_c$. 
The black dotted line shows that the dominant term $\epsilon_i^2/(9 \epsilon_f^2) (-k\tau_1)^4$ in the approximate form, Eq.~(\ref{eq:p_approx_super}), fits the growth of the power spectrum well.
Notice that the maximal enhancement is $\mathcal O(\epsilon_i^2/\epsilon_f^2)$, which is much larger than the  $\mathcal O(\epsilon_i/\epsilon_f)$, expected from the slow-roll approximation.
On the very small scales, the results in Fig.~\ref{fig:ps_one_period} asymptote to the value in the slow-roll approximation, $H^2/(8\pi^2 \epsilon_f M_\Pl^2)$, except for $\eta_c = -\infty$. 
This can be understood from the fact that the particle production occurs only if the timescale of the step transition is much smaller than the timescale of the perturbation oscillation, $1/k$.
This is why the case of $\eta_c = -\infty$ does not asymptote to the slow-roll value in a finite $k$.

Finally, the toy model in this section corresponds to the case of $\epsilon_m = \epsilon_f$ in the model in Sec.~\ref{sec:model}.
The main difference is that only the model in Sec.~\ref{sec:model} has the inflaton acceleration phase after the step climbing, which increases $\epsilon$ such that $\epsilon_m \rightarrow \epsilon_f (\simeq \epsilon_i)$.
This acceleration phase decreases the power spectrum before the perturbation exits the horizon, in proportion to $1/\epsilon$.
As $\Delta N_\text{step} \rightarrow 0$, $\epsilon$ grows more quickly and leads to a larger value of $\epsilon$ at horizon exit of the peak mode.
This is why the peak of the power spectrum for $l_u = 0.1$ is smaller than that for $l_u = 0.3$ in Fig.~\ref{fig:ps_cmb_tra_up}.
Because of this acceleration phase before the horizon exit, the enhancement in the model in Sec.~\ref{sec:model} (Fig.~\ref{fig:ps_cmb_tra_up}) is a bit smaller than $\mathcal O(\epsilon_i^2/\epsilon_m^2)$ (still larger than $\mathcal O(\epsilon_i/\epsilon_m)$) and the small-scale power spectrum approaches the same value as the large-scale one. 
Since the effects of this acceleration phase depend on the shape of the upward step, we do not go into details of this aspect in this paper.
Still, we can conclude that the origin of the enhancement larger than $\mathcal O(\epsilon_i/\epsilon_m)$ in Fig.~\ref{fig:ps_cmb_tra_up} mainly comes from the enhancement of $\mathcal R'$ given by Eq.~(\ref{eq:r_dot_rel_2}).

\section{The Fall of the Inflaton  (Downward Step)}
\label{sec:down}

Now we consider a downward step. In this case, the inflaton gains the kinetic energy from the downward step and its velocity becomes much larger than implied by the slow-roll approximation. Because of this, the transition in $\epsilon$ is followed by a second period approximated as a USR period, with $\eta=-6$. Specifically, we can parametrize $\eta$ as 
\begin{align}
	\eta = \eta_c \Theta(\tau-\tau_1) \Theta(\tau_2 -\tau) -6 \Theta(\tau -\tau_2).
	\label{eq:eta_two_periods_th_exp}
\end{align}
The slow-roll parameter $\epsilon$ changes from $\epsilon_i$ to $\epsilon_m$ with $\eta=\eta_c$ in $\tau_1 < \tau < \tau_2$ and, after that, continues decreases with $\eta = -6$. 
Similarly to the previous section, once $\epsilon_i/\epsilon_m$ is fixed, $\tau_2/\tau_1$ and $\tau/\tau_2$ are given by 
\begin{align}
  \tau_2/\tau_1 =& (\epsilon_i/\epsilon_m)^{1/\eta_{c}},  \\
  \tau/\tau_2 =& (\epsilon_m/\epsilon(\tau))^{-1/6} \quad (\tau > \tau_2),
\end{align}
where $\epsilon(\tau)$ is the value at $\tau$.
Then, the solution of the curvature perturbation is given by,
\begin{align}
\label{eq:r_sol_two_periods}
\mathcal R = 
\begin{cases}
	D_1 G^{(1)}_{3/2}(-k \tau) & (\tau < \tau_1)\\	
	E_1 G^{(1)}_{\nu_{c}}(-k \tau)  + E_2 G^{(2)}_{\nu_{c}}(-k \tau)  & (\tau_1 \leq \tau \leq \tau_2)\\
	H_1 G^{(1)}_{-3/2}(-k \tau)  + H_2 G^{(2)}_{-3/2}(-k \tau)  & (\tau_2 \leq \tau )
	\end{cases}.
\end{align}
Note that we do not put the end of the USR period in this setup. 
This is because the USR phase induced by the downward step ends adiabatically and the final amplitude of the superhorizon curvature perturbation is expected to be that at $\epsilon(\tau) = \epsilon_f$ in the above setup.\footnote{If we end the USR period with a step function in Eq.~(\ref{eq:eta_two_periods_th_exp}), there is an additional period of particle production that further amplifies curvature perturbations,  which is not the case for the true downward step scenario.}
From the continuity of $\mathcal R$ and $\mathcal R'$, the coefficients, $D_1$, $E_1$ and $E_2$ are given by Eqs.~(\ref{eq:d_coeff}) and (\ref{eq:e_coeff}), and $H_1$ and $H_2$ are given by 
\begin{align}
    H_j &= \frac{W[E_1 G^{(1)}_{\nu_{c}}(- k\tau_2) + E_2 G^{(2)}_{\nu_{c}}(- k\tau_2), G^{(p_j)}_{-3/2}( -k \tau_2)]}{W[G^{(j)}_{-3/2}(-k \tau_2),G^{(p_j)}_{-3/2}(-k \tau_2)]},
\end{align}
where we again define $W[f,g] = fg'-gf'$.

\begin{figure} 
\centering \includegraphics[width=0.6\columnwidth]{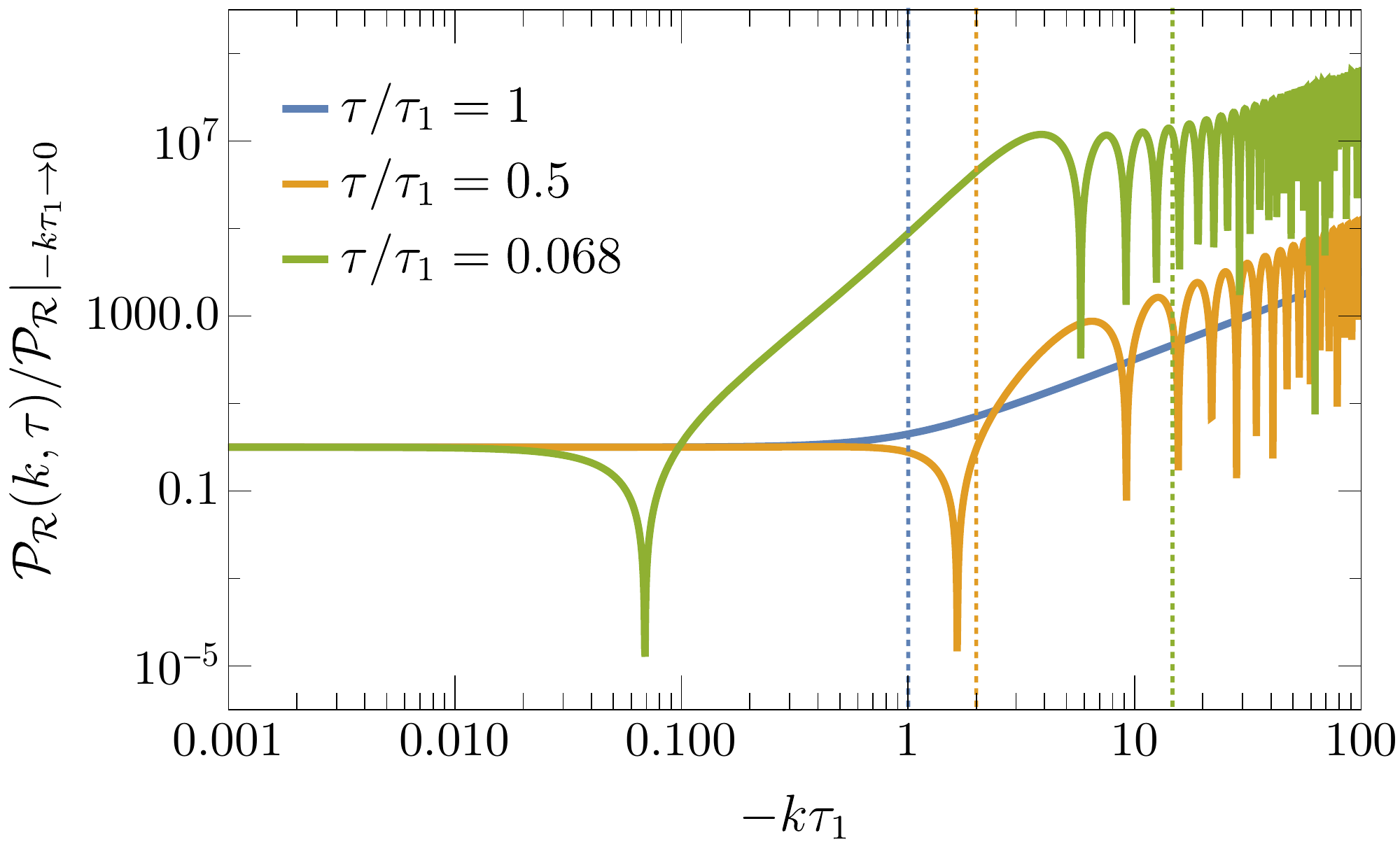}
\caption{ The evolution of the power spectra in the case with an instantaneous downward step and the subsequent USR period.
The power spectra are normalized by the value in the limit of large scale, $-k\tau_1 \rightarrow 0$.
The concrete expression of the power spectrum is given by Eq.~(\ref{eq:p_r_f_lim}).
We take $\epsilon_m/\epsilon_i = 10^7$ for all the lines.
Note that $\tau/\tau_1 = 0.068$ corresponds to the time when $\epsilon = \epsilon_i$.
The vertical lines are the horizon scales at $\tau/\tau_1 = 1, 0.5$ and $0.068$ from left to right.
}
\label{fig:ps_th_exp}
\end{figure}

Similarly to the previous section, we here take the limit of $\eta_c = \infty$ to understand the properties of the enhancement. 
Then, $H_1$ and $H_2$ become
\begin{align}
	H_{1,\text{lim}} =& - \frac{H\sqrt{\pi}}{4 \sqrt{2\epsilon_i k^3} M_\Pl} \frac{ 3\epsilon_m + (2\epsilon_m + \epsilon_i) (-k\tau_1)^2 + i(\epsilon_m + \epsilon_i) (-k\tau_1)^3 }{\epsilon_m}, \\
	H_{2,\text{lim}} =& \frac{H\sqrt{\pi}}{4 \sqrt{2\epsilon_i k^3} M_\Pl} \ee^{-2ik\tau_1} \frac{(1-i(-k\tau_1)) (3 \epsilon_m - 3 i \epsilon_m(-k\tau_1) + (\epsilon_i - \epsilon_m)(-k\tau_1)^2 ) }{\epsilon_m},
\end{align}
where note again $\tau_2 = \tau_1$ in this limit.
The time-dependent power spectrum is given by 
\begin{align}
	\mathcal P_\mathcal R(k,\tau) 
	=& \frac{k^3}{2\pi^2} |\mathcal R_k(\tau)|^2 \nonumber \\
	=& \frac{k^3}{2\pi^2}  \left| H_{1,\text{lim}} G^{(1)}_{-3/2}(-k \tau)  + H_{2,\text{lim}} G^{(2)}_{-3/2}(-k \tau) \right|^2.
	\label{eq:p_r_f_lim}
\end{align}
Figure~\ref{fig:ps_th_exp} shows the power spectrum with different $\tau$.
From the figure, we can see that, while the particle production completes at $\tau_2$, the curvature perturbations grow in time during the USR phase. 
Also, we can see that the enhanced power spectrum oscillates with constant amplitude between $k=-1/\tau_1$ and $k=-1/\tau$, i.e.\ between the horizon scale at the transition and the (smaller) horizon scale at a given later time.

In the large-scale limit $k\ll -1/\tau_1$, the power spectrum can be approximated as 
\begin{equation}
	\mathcal P_{\mathcal R}(k,\tau) \simeq \frac{H^2}{8\pi^2 \epsilon_i M_\Pl^2} \left[1 - 2 c_2(\tau) (-k\tau_1)^2 + c_4(\tau) (-k\tau_1)^4 \right],
  \label{eq:pr_large_limit_downward}  
\end{equation}	
where we have kept terms up to $\mathcal O((-k\tau_1)^4)$ and
\begin{align}
c_2(\tau)  ={}&   \frac{1}{10} \left(\frac{\epsilon_m}{\epsilon(\tau)} \right)^{-1/3} + \frac{1}{15} \left[ -10 + \left(\frac{\epsilon_m}{\epsilon(\tau)} \right)^{1/2}  \right]  +   \frac{1}{3} \left( \frac{\epsilon_i}{\epsilon_m} \right) \left[ -1 + \left(\frac{\epsilon_m}{\epsilon(\tau)} \right)^{1/2} \right],
\nonumber\\
 \simeq{}&
\frac{1}{15} \left(\frac{\epsilon_m}{\epsilon(\tau)}\right)^{1/2}
\\
c_4(\tau)  = {}&
 \frac{1}{1575} \left\{ 
	7\frac{\epsilon_m}{\epsilon(\tau)} \left( 1 + 5\frac{\epsilon_i}{\epsilon_m} \right)^2 - 3 \frac{\epsilon(\tau)}{\epsilon_m} \left[ \left( \frac{\epsilon_m}{\epsilon(\tau)} \right)^{1/3} \left( -9 + 140 \left( \frac{\epsilon_m}{\epsilon(\tau)} \right)^{1/3} \right) \right. \right. \nonumber \\
  & \left. + 70 \frac{\epsilon_i}{\epsilon_m} \left( \frac{\epsilon_m}{\epsilon(\tau)} \right)^{2/3}\right] 
	+ \left( 700 - 84 \left( \frac{\epsilon_m}{\epsilon(\tau)} \right)^{1/6} - 230 \left( \frac{\epsilon_m}{\epsilon(\tau)} \right)^{1/2} \right)   \nonumber \\
  & \left.
    - 140 \frac{\epsilon_i}{\epsilon_m} \left[ -5 + 3 \left( \frac{\epsilon_m}{\epsilon(\tau)} \right)^{1/6} + \left( \frac{\epsilon_m}{\epsilon(\tau)} \right)^{1/2} \right]
	- 175 \left( \frac{\epsilon_i}{\epsilon_m} \right)^2 \left[ -1 + 2 \left( \frac{\epsilon_m}{\epsilon(\tau)} \right)^{1/2} \right] \right\}
	 \nonumber \\
 \simeq{}&
 \frac{1}{225} \frac{\epsilon_m}{\epsilon(\tau)} \simeq c_2^2,
\end{align}
where the final approximation assumes
$\epsilon(\tau)/\epsilon_m \ll 1$, which is appropriate well after the step transition.
Notice that the power spectrum becomes close to zero when $c_2 (-k\tau_1)^2 \simeq 1$, around which a near zero crossing (or rotation by $\pi$ in the complex plane) of $\mathcal R$ occurs (see \cite{Passaglia:2018ixg} for a discussion of the highly suppressed out-of-phase residual).
As in the upward step case, the
matching conditions for the two solutions at the transition
determine that this quartic order in the expansion suffices.

On the other hand, the perturbations which were inside the horizon at the transition become the following expression after their horizon exit (see also \cite{Miranda:2015cea}):
\begin{align}
	\mathcal P_{\mathcal R}(k,\tau) &\simeq \frac{H^2}{8\pi^2 \epsilon_i M_\Pl^2} \frac{\epsilon_m}{\epsilon(\tau)} \frac{1 - \cos(-2k\tau_1) + \left(\frac{\epsilon_i}{\epsilon_m}\right)^2 (1+\cos(-2k\tau_1)) }{2} \quad (1/|\tau_1| \ll k \ll 1/|\tau| ) \nonumber \\
	&\simeq \frac{H^2}{8\pi^2 \epsilon_i M_\Pl^2} \frac{\epsilon_m}{\epsilon(\tau)} \frac{1 - \cos(-2k\tau_1) }{2} \quad (1/|\tau_1| \ll k \ll 1/|\tau| ) ,
	\label{eq:pr_small_limit_downward}
\end{align}
where we have assumed $\epsilon_i/\epsilon_m \ll 1$ in the final line. From this expression, we may appreciate that the enhancement of power spectrum is of $\mathcal O(\epsilon_m/\epsilon(\tau))$ and its oscillation frequency is $|\tau_1|/\pi$ in $k$.
After $\epsilon$ adiabatically reaches $\epsilon_f$ and becomes constant, the final power spectrum for $k \gg -1/\tau_1$ after the horizon exit of each mode becomes
\begin{align}
    \mathcal P_{\mathcal R}(k) &\simeq \frac{H^2}{8\pi^2 \epsilon_i M_\Pl^2} \frac{\epsilon_m}{\epsilon_f} \frac{1 - \cos(-2k\tau_1) }{2}.
  \label{eq:pr_small_limit_downward_2}
\end{align}
Note that this expression is valid even for the perturbations on subhorizon scales at $\epsilon = \epsilon_f$.
The result in the case of $\tau/\tau_1 = 0.068$ in Fig.~\ref{fig:ps_th_exp} shows that, while the superhorizon power spectrum around $3 \lesssim -k\tau_1 \lesssim 10$ is almost constant in $k$ up to the oscillation, the subhorizon power spectrum grows in $k$.
However, this subhorizon growth in $k$ is just due to the subhorizon evolution of the perturbations, which is not related to the enhancement mechanism itself. 
The subhorizon perturbations decrease until their horizon exit and freeze after then, which results in the final power spectrum given by Eq.~(\ref{eq:pr_small_limit_downward_2}) even for such perturbations.
Here, we note that the large perturbation enhancement in a downward step with $\epsilon_i \simeq \epsilon_f$ requires a very small $\epsilon_i (< 10^{-7})$ because $\epsilon_m$ cannot be much larger than unity. This restricts the form of the base potential to ones where the power spectrum tilt on the CMB scales is mainly determined by $\eta$, though this restriction is easily avoided if we allow $\epsilon_f \ll \epsilon_i$.

While the result Eq.~(\ref{eq:pr_small_limit_downward_2}) applies in the limit of $\eta_c \rightarrow \infty$, the perturbations in the case of a finite $\eta_c$ do not get enhanced by the downward step transition on very small scales since the modes oscillate during the transition. 
Because of this, the power spectrum on small scales effectively gets damped with a finite $\eta_c$.
We may understand this damping of high-$k$ modes analytically, as follows. The perturbation enhancement can be interpreted as the tachyonic instability due to the negative mass of the inflaton in the downward step. In the conformal Newtonian gauge, the equation of motion for the inflaton fluctuation can be expressed as~\cite{Weinberg:2008zzc} 
\begin{align}
	\delta \phi_k'' + 2 \mathcal H \delta \phi_k' + a^2 \frac{\partial^2 V}{\partial \phi^2} \delta \phi_k + k^2 \delta \phi_k = -2a^2 \Phi_k \frac{\partial V}{\partial \phi} + 4 \phi' \Phi_k',
	\label{eq:eom_inflaton_pertb}
\end{align}
where we take the following notation of the metric perturbations:
\begin{align}
	\dd s^2 = -a^2 (1 + 2 \Phi) \dd \tau^2 + a^2 (1 - 2 \Phi) \delta_{ij} \dd x^i \dd x^j.
\end{align}
Here, we neglect the contributions from the metric perturbations because they are suppressed by $\epsilon$ during the inflation era~\cite{Dodelson:1282338}. Then, we can rewrite Eq.~(\ref{eq:eom_inflaton_pertb}) as
\begin{align}
	\delta \phi_k'' + 2 \mathcal H \delta \phi_k' + a^2 m^2 \delta \phi_k + k^2 \delta \phi_k \simeq 0,
\end{align}
where $m^2$ is the effective mass of the inflaton, given by $m^2 = \partial^2 V/\partial \phi^2$.
Since the timescale of the downward step transition is shorter than the Hubble timescale at that time, we also neglect the Hubble friction term and just focus on the exponential growth due to the negative mass of the inflaton during the step transition.
Then, we can approximate that the inflaton perturbation grows $\delta \phi_k \propto \ee^{\int \dd \tau \sqrt{-m^2a^2 - k^2}}$.
Here, we also assume constant $m^2$ during the downward step transition, which corresponds to the constant $\eta$ as we will see below.
Then, we define the damping factor for the power spectrum as
\begin{align}
	D(k) \equiv \ee^{\lambda_\text{eff}(k) (\tau_2-\tau_1)} \left( \frac{\epsilon_m}{\epsilon_i} \right)^{-1/2},
	\label{eq:damping_fac}
\end{align}
where we have normalized $D(k)$ as $D(k) \rightarrow 1$ in $\eta \rightarrow \infty$, as we will see below, and $\lambda_\text{eff}$ is the effective frequency of the inflaton fluctuation during the downward step transition, defined as 
\begin{align}
	\lambda_\text{eff}(k) \equiv \sqrt{-k^2 - m^2 a^2}.
\label{eq:eff_mass_1}	
\end{align}
The mass $m$ may be related to the slow-roll parameter $\eta$ via the background equations of motion (see Appendix~\ref{app:evol_up_down} for details), as
\begin{align}
	m^2 \simeq -\frac{H^2}{4} \left( 6 \eta + \eta^2 \right).
  \label{eq:eta_m_relation}
\end{align}
With this relation, we can rewrite Eq.~(\ref{eq:eff_mass_1}) as 
\begin{align}
	\lambda_\text{eff}(k) \simeq \sqrt{-k^2 +\frac{1}{4\tau^2_1} \left( 6 \eta + \eta^2 \right)} ,
\end{align}
where we have neglected the evolution of $a$ during the downward step transition.
We may now rewrite the damping factor as 
\begin{align}
	D(k,\tau_1, \eta) \simeq&  \exp\left[ \sqrt{-k^2 +\frac{1}{4\tau^2_1} \left( 6 \eta + \eta^2 \right)} (-\tau_1) \frac{1}{\eta} (-\log(\epsilon_i/\epsilon_m)) \right] \left( \frac{\epsilon_m}{\epsilon_i} \right)^{-1/2} \nonumber \\
	=& \left( \frac{\epsilon_m}{\epsilon_i} \right)^{ \sqrt{-\frac{(-k\tau_1)^2}{\eta^2} + \frac{3}{2\eta} + \frac{1}{4}} - \frac{1}{2}},
	\label{eq:damping}
\end{align}
where we have used the relation $\tau_2-\tau_1 \simeq \tau_1 \log(\epsilon_i/\epsilon_m)/\eta$, valid in the limit of large $\eta$.
We can see $|D(k,\tau_1, \eta)|^2 \rightarrow 1$ for $\eta \rightarrow \infty$. 
On the other hand, in the small-scale limit ($k \gg |\eta/\tau_1|$), we have $|D(k,\tau_1, \eta)|^2 \rightarrow \epsilon_i/\epsilon_m$.
Multiplying the approximate form in Eq.~(\ref{eq:pr_small_limit_downward_2}) by $|D(k,\tau_1, \eta)|^2$, we can take into account the damping of the power spectrum. 
Indeed, comparing to Fig.~\ref{fig:ps_cmb_tra}, we may appreciate that the damping factor fits the damping of the power spectrum very well.

\section{Bounds on Non-Adiabatic Evolution}
\label{sec:nonadiab}

In the previous sections, we have discussed the perturbation enhancement with the linear perturbation theory. 
However, once the perturbations are strongly coupled, the perturbation theory can no longer give reliable results.
In particular, the sharp feature of the potential generally causes stronger coupling between perturbations.
In this sense, we still need to be careful about whether or not the $\mathcal O(10^7)$ enhancement can be really realized by a step-like feature.
The goal of this section is to show that our fiducial potential can avoid the strong coupling problem with an appropriate modification and is a successful example for the $\mathcal O(10^7)$ enhancement.

\subsection{Smoothing Out of the Step Transition}
\label{sec:smoothing}
So far, we have seen that the large enhancement can be realized by the concrete potential, Eqs.~(\ref{eq:pot_cmb_to_end_mo})-(\ref{eq:f_step}), which includes jumps in $V''$ at $\phi_1$ and $\phi_2$.
However, in realistic situations, the second and higher derivatives of the potential are expected to transit smoothly there. 
Moreover jumps in $V''$ would automatically lead to strong coupling problems as we discuss in the following subsections.
In this subsection, we first show  that with a smoothing out of the higher derivatives of the potential at $\phi_1$ and $\phi_2$ the enhanced power spectrum remains largely unchanged as long as the transition  occurs within much less than one e-fold.

\begin{figure} 
\centering \includegraphics[width=0.6\columnwidth]{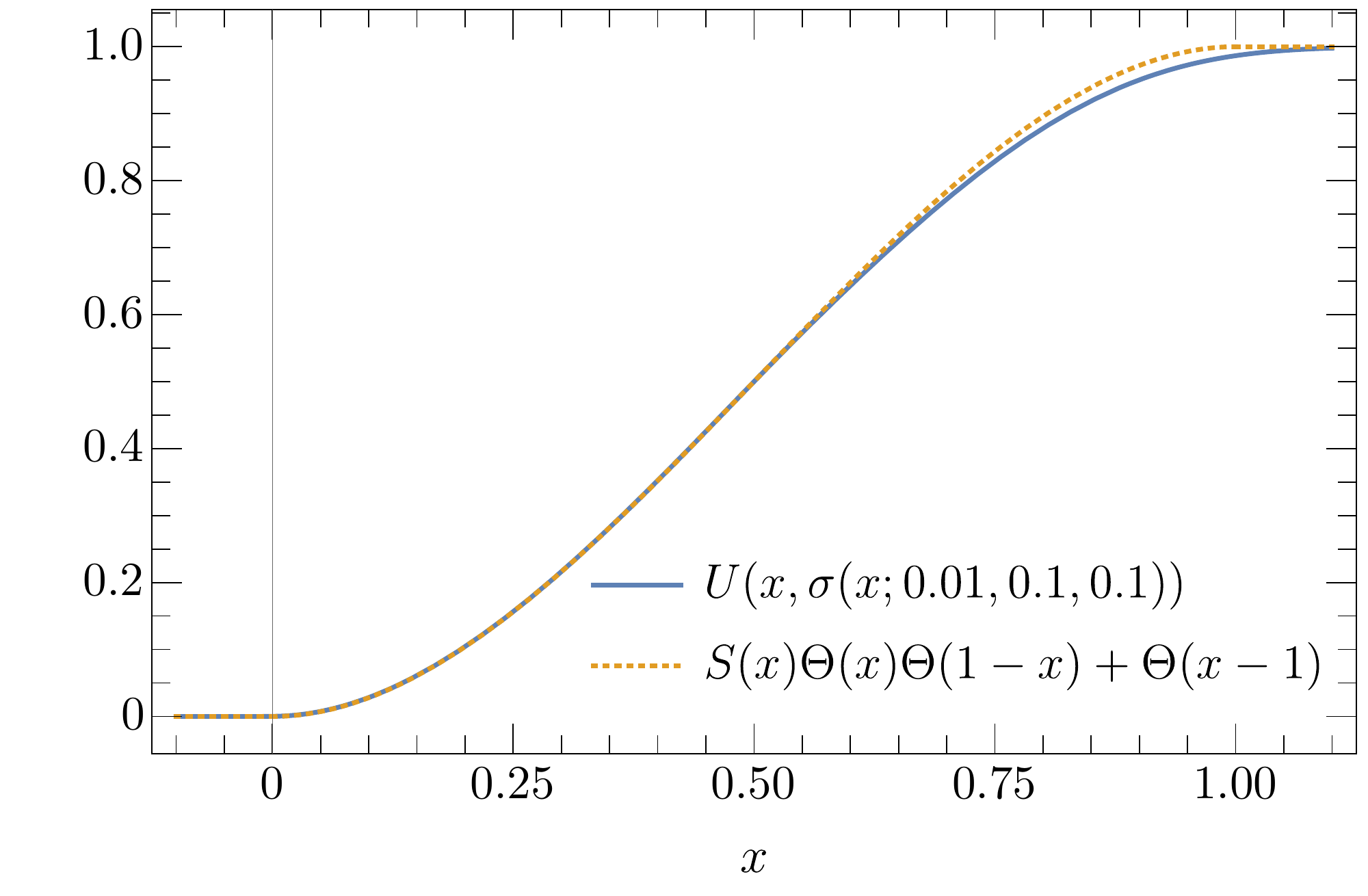} 
\caption{ Comparison between the smoothed function $U(x;\sigma_1,\sigma_2)$ and the function with jumps in $V''$.
We take $\sigma_1=0.01$ and $\sigma_2= \sigma _m =0.1$ for $U(x,\sigma(x))$.
}
\label{fig:comp_t_ease}
\end{figure}

Specifically we consider a smoothed version of the original potential which is infinitely differentiable:
\begin{align}
  V(\phi) =& V_0 \left(1-\frac{\beta \phi^2/ M_\Pl^2}{1+\phi/\phi_\text{CMB}}  \right) G(\phi; \phi_1, \phi_2, h, \Delta \phi_1, \Delta \phi_2, \Delta \phi_m) + V_\text{end}(\phi; \phi_\text{end}),
   \label{eq:pot_cmb_to_end_mo_2}
\end{align}
where $G$ is defined by
\begin{align}
  G(\phi; \phi_1, \phi_2, h , \Delta \phi_1, \Delta \phi_2, \Delta \phi_m) \equiv 1- h U\left( \frac{\phi-\phi_1}{\phi_2-\phi_1}; \sigma\left( \frac{\phi-\phi_1}{\phi_2-\phi_1}; \frac{\Delta \phi_1}{\phi_2-\phi_1}, \frac{\Delta \phi_2}{\phi_2-\phi_1}, \frac{\Delta \phi_m}{\phi_2-\phi_1} \right) \right).
\end{align}
The function $U$ is a smoothed version of the step in
Eq.~(\ref{eq:f_step})
defined by 
\begin{align}
  U(x; \sigma) \equiv& \int^{\infty}_{-\infty} \dd y \frac{1}{\sqrt{2\pi} \sigma} \ee^{-\frac{(x-y)^2}{2\sigma^2}} \left[ S(y) \Theta(y) \Theta(1-y) + \Theta(y-1) \right],
\end{align}
and the smoothing width 
$\sigma(x;\sigma_1,\sigma_2,\sigma_m)$ is defined by
\begin{align}
  \sigma(x;\sigma_1,\sigma_2,\sigma_m) \equiv \sigma_1 + \frac{1}{2}(\sigma_2-\sigma_1)\left( 1+\text{tanh}\left( \frac{x-1/2}{\sigma_m} \right) \right).
\end{align}
Figure~\ref{fig:comp_t_ease} shows the function $U$.
The parameters $\Delta \phi_1$ and $\Delta \phi_2$ correspond to the field widths for the smoothing out of the transition at $\phi_1$ and $\phi_2$, respectively.  The field smoothing width is then itself smoothly interpolated between the two values by a step-like function with width $\Delta\phi_m$.  This ensures that all derivatives of $U$ are finite and continuous.
Conversely, if we take the limit of $\Delta \phi_1 \rightarrow 0$ and $\Delta \phi_2 \rightarrow 0$, we reproduce the original potential, given by Eqs.~(\ref{eq:pot_cmb_to_end_mo})-(\ref{eq:f_step}).

\begin{figure}  
\centering \includegraphics[width=0.6\columnwidth]{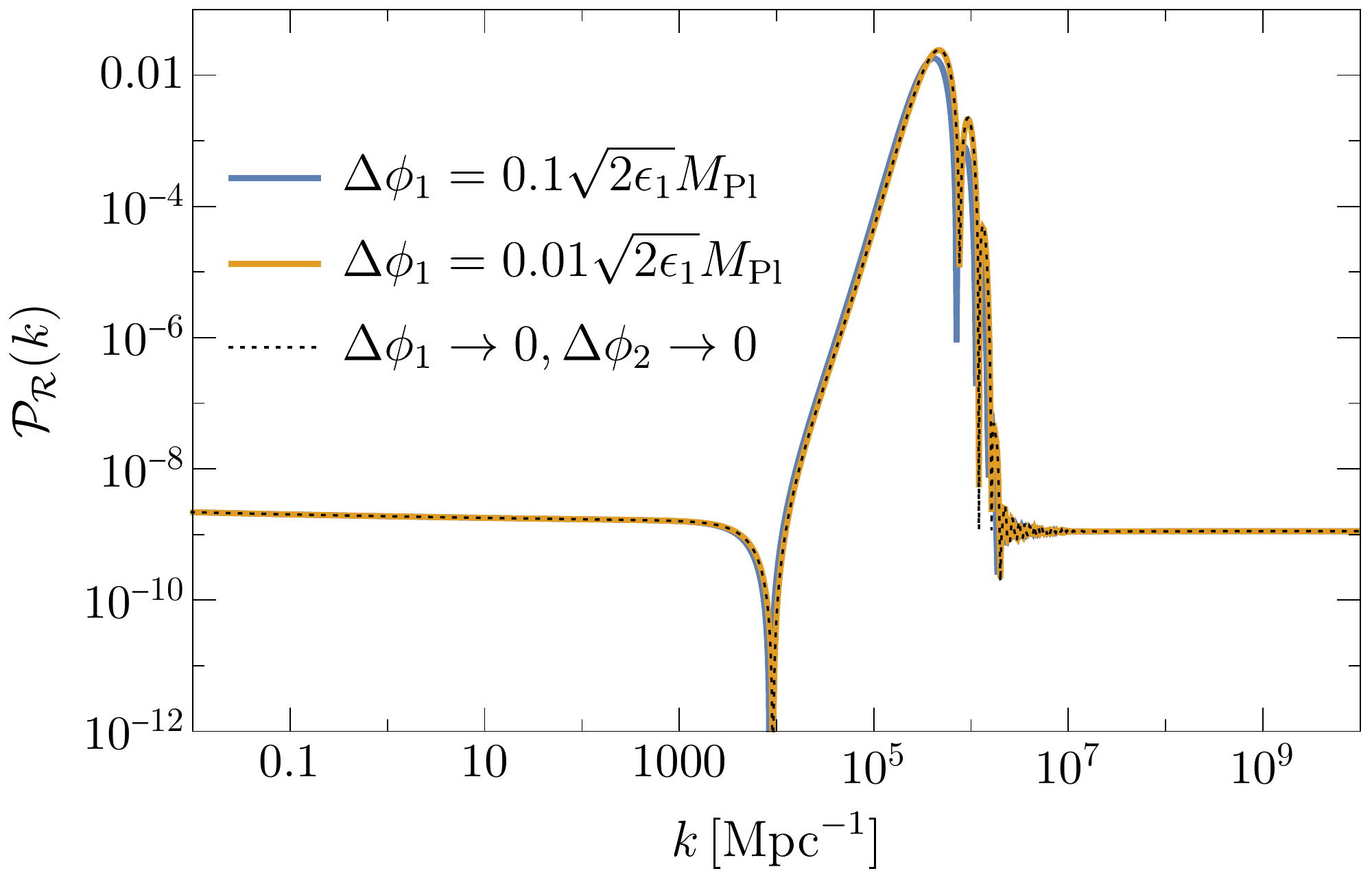}
\caption{ The power spectra for the downward step case with different smoothing scales $\Delta \phi_1$. 
The black dotted line shows the spectrum before the smooth out.
We take $\Delta \phi_2 = \Delta \phi_m = 0.01 \sqrt{2\epsilon_m} M_\Pl$ except for the black dotted line.
The other parameters are the same as in the left panel of Fig.~\ref{fig:pot_l01}.
}
\label{fig:ps_type3_mo}
\end{figure}

\begin{figure}  
\centering \includegraphics[width=0.6\columnwidth]{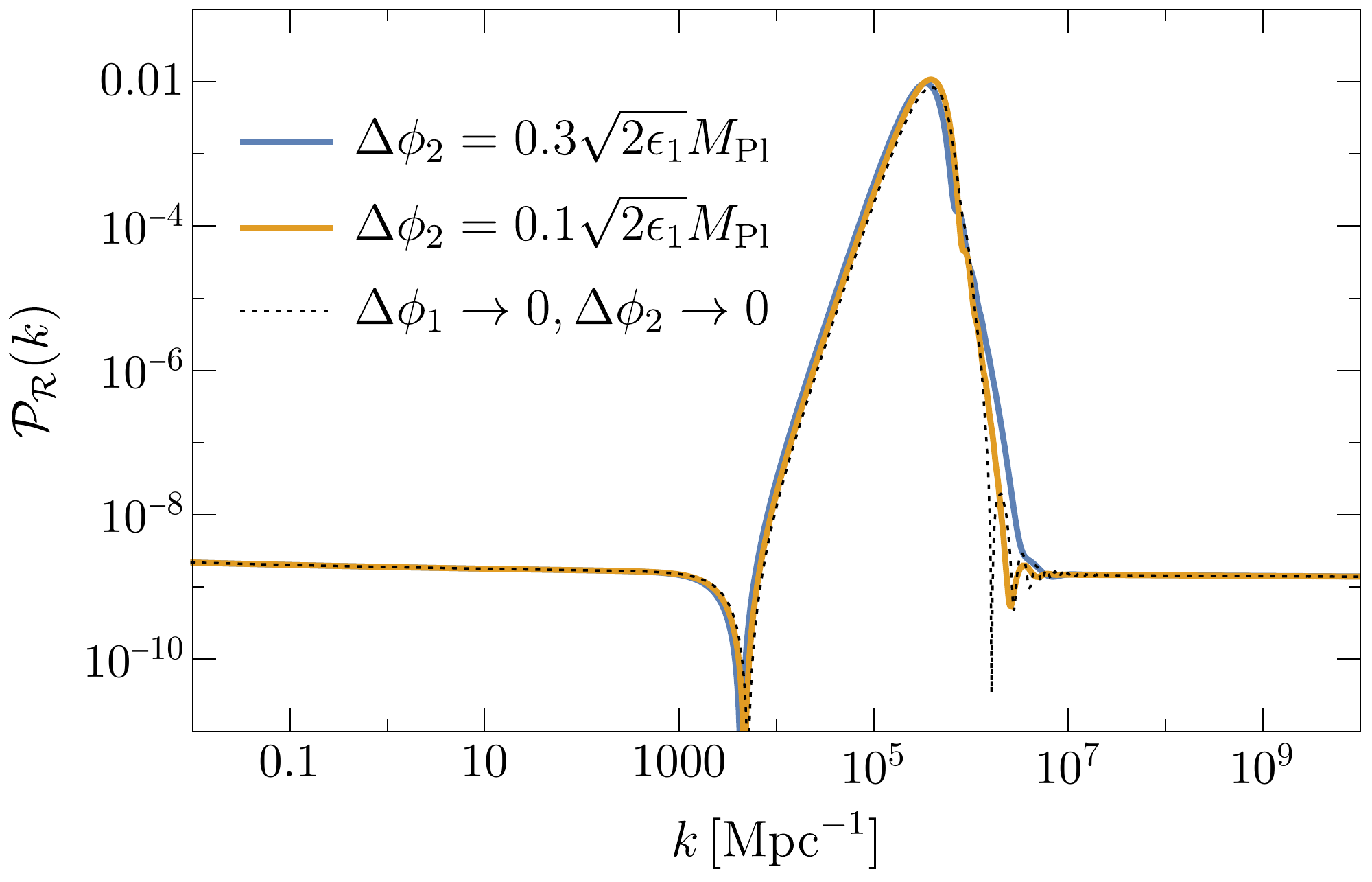}
\caption{ 
The power spectra for the upward step case with different smoothing scales $\Delta \phi_2$.
The black dotted line shows the spectrum before the smoothing out.
We take $\Delta \phi_1 = 0.01 \sqrt{2\epsilon_i}M_\Pl$ and $\Delta \phi_m = \Delta \phi_2$ for all lines.
The other parameters are the same as in the right panel of Fig.~\ref{fig:pot_l01}.
}
\label{fig:ps_smooth_up}
\end{figure}

Figures~\ref{fig:ps_type3_mo} and \ref{fig:ps_smooth_up} show the power spectra for the downward and the upward step cases, respectively.
In the downward step case, the inflaton velocities at $\phi_1$ and $\phi_2$ can be approximated as $\sqrt{2\epsilon_i}M_\Pl$ and $\sqrt{2\epsilon_m}M_\Pl$, respectively.
On the other hand, in the upward step case, both the velocities can be approximated as  $\sqrt{2\epsilon_i}M_\Pl$ because of the acceleration phase after the potential climbing. 
For this reason, we focus on the downward step case where $\Delta \phi_1 \ll \sqrt{2\epsilon_i} M_\Pl$ and $\Delta \phi_2 \ll \sqrt{2\epsilon_m}M_\Pl$, and the upward step case where $\Delta \phi_1, \Delta \phi_2 \ll \sqrt{2\epsilon_i} M_\Pl$.
In Figs.~\ref{fig:ps_type3_mo} and \ref{fig:ps_smooth_up}, we can see that, if the transition in $V''$ still occurs within much less than one e-fold, the smoothing out does not change the power spectrum so much. 

At the same time, the smoothing out with a too small field width could cause the strong coupling of perturbations, which invalidates linear perturbation theory.
In the following sections, we will check whether the strong coupling problem occurs in our setup.

\subsection{Strong Coupling}

We now consider the limits of perturbation theory. We begin with the simple requirement that linearized perturbation theory remains a good approximation to the full dynamics.
In terms of the inflaton, this can be phrased as the requirement that perturbation theory in inflaton fluctuations, $\phi(\bm x,t) = \phi(t) + \delta \phi(\bm x,t)$, is well behaved, such that the equation of motion for $\delta \phi(\bm x,t)$ is well approximated by the perturbative expansion of the Klein-Gordon equation where potential $V(\phi)$ and its derivatives are evaluated at the background field value $\phi(t)$. 
Specifically, this requires
$\delta \phi$ to obey
\begin{align}
  \delta \phi'' + 2 \mathcal H \delta \phi' - \nabla^2 \delta \phi + a^2 \frac{\partial^2 V}{\partial \phi^2} \delta \phi  = -a^2 \sum_{n=3} \frac{1}{(n-1)!} V^{(n)} (\delta \phi)^{n-1} , 
  \label{eq:eom_inflaton_pertb_higher}
\end{align}
where $V^{(n)} \equiv \partial^n V(\phi)/\partial \phi^n$ and we have neglected the metric perturbations again. 
Hence qualitatively, we require a sufficiently smooth potential for the right hand side to be negligible. 
If it is negligible then we may use the (linearized) Mukhanov-Sasaki equation, Eq.~(\ref{eq:r_curv_eom_com}) for the curvature perturbation.

We can make the requirements for perturbativity of the curvature perturbation more precise by working in the effective field theory (EFT) of inflation~\cite{Cheung:2007st}. 
In the EFT of inflation, the curvature perturbations are related to the Goldstone boson ($\pi$) due to the time translation symmetry breaking through $\mathcal R = -H \pi$, which is valid at linear order in perturbations.
Assuming a canonical inflation model and neglecting the subleading terms in $\epsilon$, we get the action for $\pi$ in the EFT of inflation~\cite{Behbahani:2011it}:
\begin{align}\label{eq:Spi_can}
S_{\rm \pi} \simeq \int d^4 x   \sqrt{- g} (-M_\Pl^2) \left[ 3H^2(t+\pi) + \dot H(t+\pi) + \dot{H} (t + \pi) \left( (1+\dot{\pi})^2-\frac{ (\partial_i \pi)^2}{a^2}\right)\right],
\end{align}
where the dots here mean the derivative with respect to the physical time, $t$.
Note that all features of $V(\phi)$ are completely subsumed into $H(t+\pi)$. 
After performing the Taylor expansion and integrating the Lagrangian by parts, we obtain the $n$-th order action in $\pi$:
\begin{align}
  S_n \simeq \int \dd^4 x a^3 M_\Pl^2 \left[ - \frac{1}{(n-2)!} H^{(n-1)} \pi^{n-2} \left( \dot \pi^2 - \frac{(\partial_i \pi)^2}{a^2} \right) + \frac{3}{n!}(2 H H^{(n)} - \partial^n_t (H^2))\pi^n \right],
  \label{eq:s_n_pi}
\end{align}
where $H^{(n)} = \partial_t^{n} H$ and the Hubble parameter is evaluated on the background, i.e.\ at  $t$ (not $t+\pi$).

Therefore in the EFT of inflation, perturbativity is related to the smoothness of $H(t)$ which is equivalent to the smoothness of $V(\phi)$.
To see how $V^{(n)}$ is incorporated in $H^{(n+1)}$, we here write down explicit expressions for the first few derivatives,
\begin{align}
  H^{(2)} 
  & = -6 H \dot H + \frac{1}{M_\Pl^2} \dot \phi V^{(1)}, \label{eq:h_2_ana}\\
  H^{(3)} 
  &= -6 (\dot H^2  + H H^{(2)}) + \frac{1}{M_\Pl^2} \left( -3 H \dot \phi V^{(1)} - \left({V^{(1)}}\right)^2 + \dot \phi^2 V^{(2)} \right), \label{eq:h_3_ana}\\
  H^{(4)} 
  & =  -6(3 \dot H H^{(2)} + H H^{(3)}) + \frac{1}{M_\Pl^2} \left( -3 \dot H \dot \phi V^{(1)} -3 H \ddot \phi V^{(1)} -9 H \dot \phi^2 V^{(2)} - 4 V^{(1)} V^{(2)} \dot \phi + \dot \phi^3 V^{(3)} \right). \label{eq:h_4_ana}
\end{align}
Note that $V^{(n)}$ is incorporated in $H^{(n+1)}$ in general.

The terms in Eq.~(\ref{eq:s_n_pi}) that are explicitly proportional to $\pi^n$ are suppressed by powers of $\epsilon$ relative to the $n$-th order terms arising from the series expansion of $H(t+\pi)$. The $n$-th order Lagrangian for $\pi$ can be approximated as,
\begin{align}
  \mathcal L_n(\pi) \simeq  -\frac{M_\Pl^2}{(n-2)!} H^{(n-1)} \pi^{n-2} \left( \dot \pi^2 - \frac{(\partial_i \pi)^2}{a^2} \right).
\end{align}
As a simple diagnostic for strong coupling of $\pi$ (and by extension ${\cal R}$), we consider the relative size of ${\cal L}_n(\pi)$ and ${\cal L}_2(\pi)$.
We impose the following condition for weak (non-strong) couplings: 
\begin{align}
  \label{eq:ln_o_l2}
  &\left| \frac{\mathcal L_{n}}{\mathcal L_2} \right| < 1 \quad (\text{for all } n(>2)) \\
  \Rightarrow \quad&
  \left| \frac{H^{(n-1)}}{(n-2)! H^{n-2} \dot H} \mathcal R^{n-2}(\tau)\right| <  1 \quad (\text{for all } n(>2)).
  \label{eq:no_sc_cons_eft}
\end{align}
We note that this condition is a naive estimation, and might be modified depending on the precise definition of strong coupling. One alternative way to proceed would be to make use of standard In-Out formalism tools, such as the optical theorem  (see, e.g., Appendix E in Ref.~\cite{Baumann:2011su}), or its recent In-In formulation \cite{Goodhew:2020hob}. Our situation differs from the analysis performed in those works in that the field evolves non-adiabatically through the region of interest, and the mode functions differ substantially from the naive expectation in de Sitter space, meaning that any analysis must ultimately be performed numerically. In light of this, we will use the simple diagnostic above, and perform numerical analysis.

The typical amplitude of $\mathcal R^{(n-2)}$ can be approximated as its standard deviation, given by $\expval{\mathcal R^{2(n-2)}}^{1/2}$.
Assuming the Gaussian power spectrum, we can rewrite this as 
\begin{align}
\expval{\mathcal R^{2(n-2)}}^{1/2} =& \sqrt{(2n-5)!!} \expval{\mathcal R^2}^{(n-2)/2}  \nonumber \\
\simeq&  \sqrt{(2n-5)!!} \, \mathcal P_{\mathcal R}^{(n-2)/2}(k_*,\tau),
\end{align} 
where $n!!$ is the double factorial ($n!! = n (n-2) (n-4) \cdots 1$). 
The $k_*$ depends on the situation. 
Before the step transition, $k_*$ is the smallest scale that we want to calculate with the Mukhanov-Sasaki equations because the power spectrum on subhorizon scales is a monotonically increasing function in $k$ at that time. 
We denote this smallest scale (or the largest value of $k$) by $k_\tmax$.
Since we are interested in the power spectrum enhancement due to the step transition, this $k_\tmax$ should be larger than the largest $k$ on which the particle production occurs due to the non-adiabatic evolution. 
On the other hand, after the step transition, the situation becomes different.
The power spectrum on subhorizon scales is no longer a monotonically increasing function in $k$. 
Instead, it has a peak at $k_\text{peak}$, which is associated with the particle production.
In this case, the $k_*$  of interest should be $k_\text{peak}$ when $\mathcal P_\mathcal R(k_\text{peak}) > \mathcal P_\mathcal R(k_\tmax)$.

For convenience, we define the following quantities:
\begin{align}
  A_n(k,\tau) \equiv \frac{H^{(n-1)}}{(n-2)! H^{n-2} \dot H} \sqrt{(2n-5)!!} \, \mathcal P_{\mathcal R}^{(n-2)/2}(k,\tau).
  \label{eq:a_n_def}
\end{align}
With this, we can rewrite the non-strong coupling condition, Eq.~(\ref{eq:ln_o_l2}), as $|A_n(k_\text{peak},\tau)|, |A_n(k_\tmax,\tau)| < 1$ for all $n>2$.
Although we determine $k_\text{peak}$ as the peak scale of the power spectrum on superhorizon scales at late time, this can be different from the real peak scale especially in the early stage of the step transition, where the curvature perturbations are not significantly enhanced yet.
At the same time, due to the insufficient enhancement, we can also expect that the $\mathcal P_\mathcal R (k_\text{peak})$ is not so different from the power spectrum at the real peak scale at that time.
Because of this, our characterization of  $k_\text{peak}$ suffices  in the following.  
From Figs.~\ref{fig:ps_cmb_tra} and \ref{fig:ps_cmb_tra_up}, we can see that $k_\tmax = 10 k_\text{peak}$ is larger than the largest $k$ of the enhancement in the case of $l_d = 0.1$ (downward step) and $l_u = 0.3$ (upward step).
In practice therefore, we check the strong coupling condition for $k_\text{peak}$ and $k_\tmax = 10 k_\text{peak}$.

\begin{figure}[h!]
\begin{minipage}{0.49\hsize}
\begin{center}
\includegraphics[width=1\columnwidth]{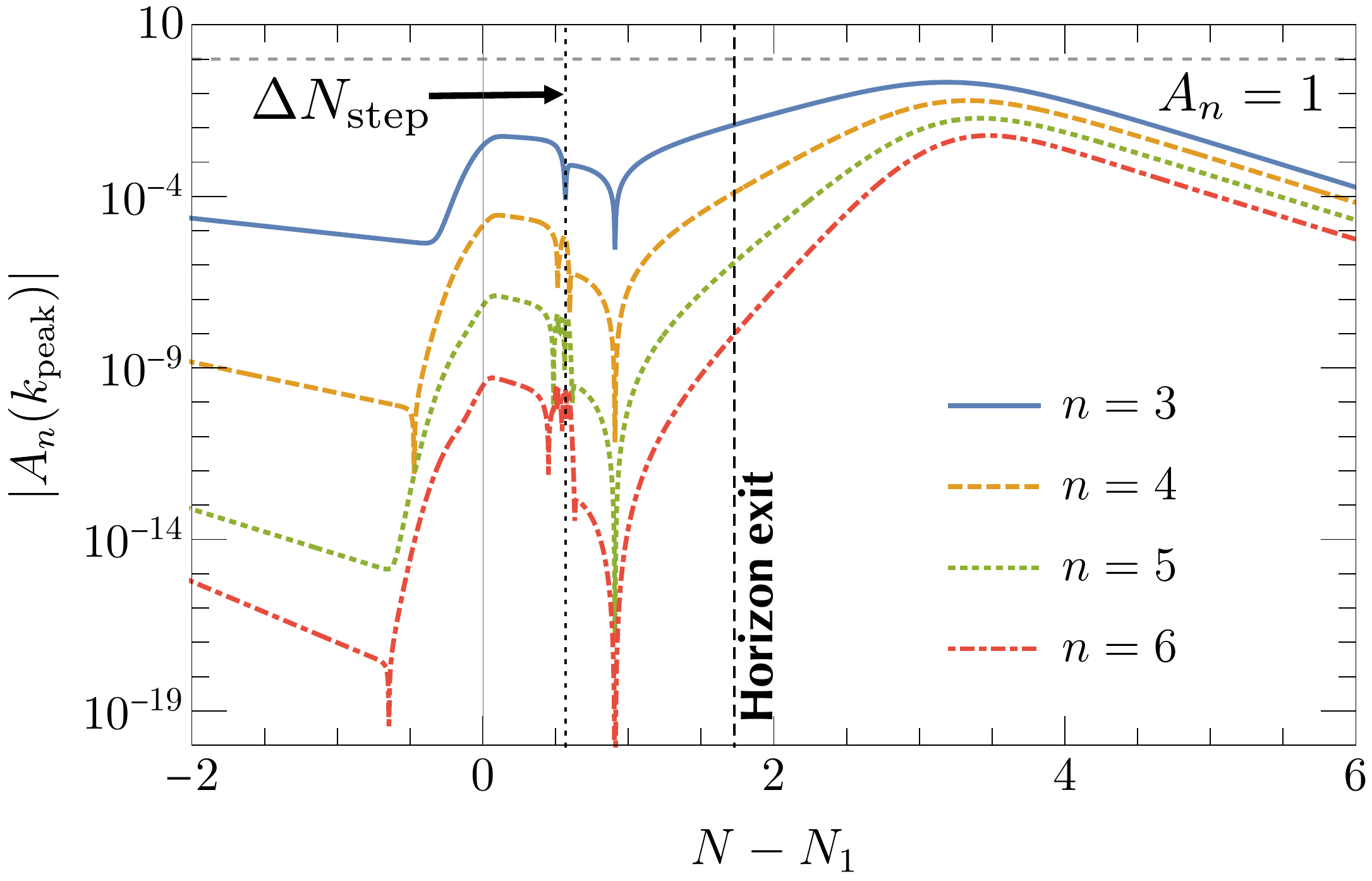}
\end{center}
\end{minipage}
\begin{minipage}{0.49\hsize}
\begin{center}
\includegraphics[width=1\columnwidth]{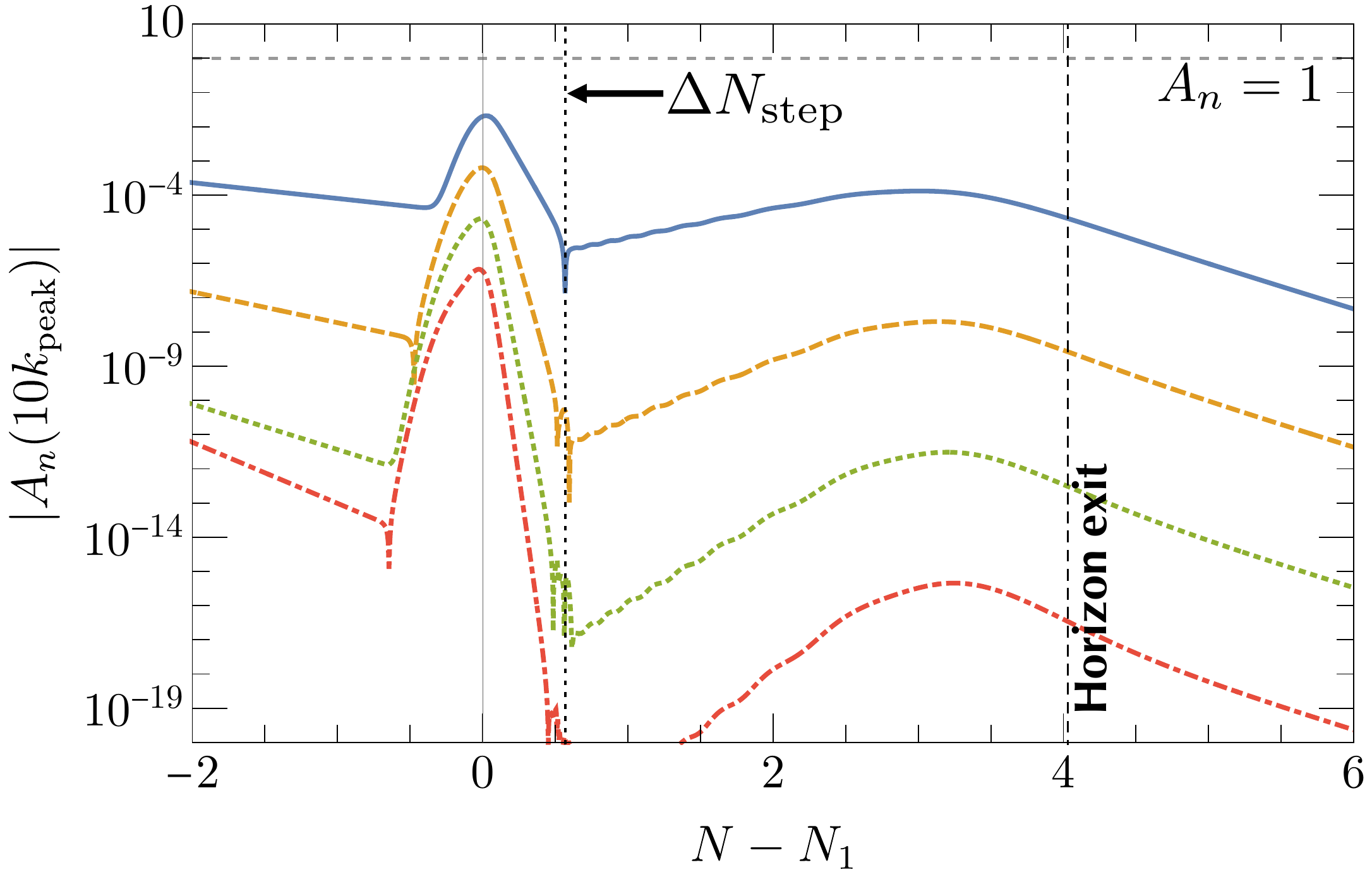}
\end{center}
\end{minipage}
\caption{
The evolution of the strong coupling parameter  $A_n(k_\text{peak})$ (left) and $A_n(10k_\text{peak})$ (right) in the downward step case.
We take the same parameters as the blue line in Fig.~\ref{fig:ps_type3_mo} ($\Delta \phi_1 = 0.1\sqrt{2\epsilon_i}M_\Pl, \Delta \phi_2 = \Delta \phi_m = 0.01 \sqrt{2\epsilon_m} M_\Pl$).
The horizontal black dashed line is $A_n = 1$, the nominal threshold of strong coupling.
The vertical dotted lines are at $\Delta N_\text{step} (\simeq N_2-N_1)$.
The vertical dashed lines are at the e-folds at the horizon exit of the scales of 
$k_\text{peak}$ or $10 k_\text{peak}$. Notice $|A_n|\ll 1$ inside the horizon and $|A_n| <1$ always for both modes.}
\label{fig:lnol2_down}
\end{figure}

\begin{figure}[h!]
\begin{minipage}{0.49\hsize}
\begin{center}
\includegraphics[width=1\columnwidth]{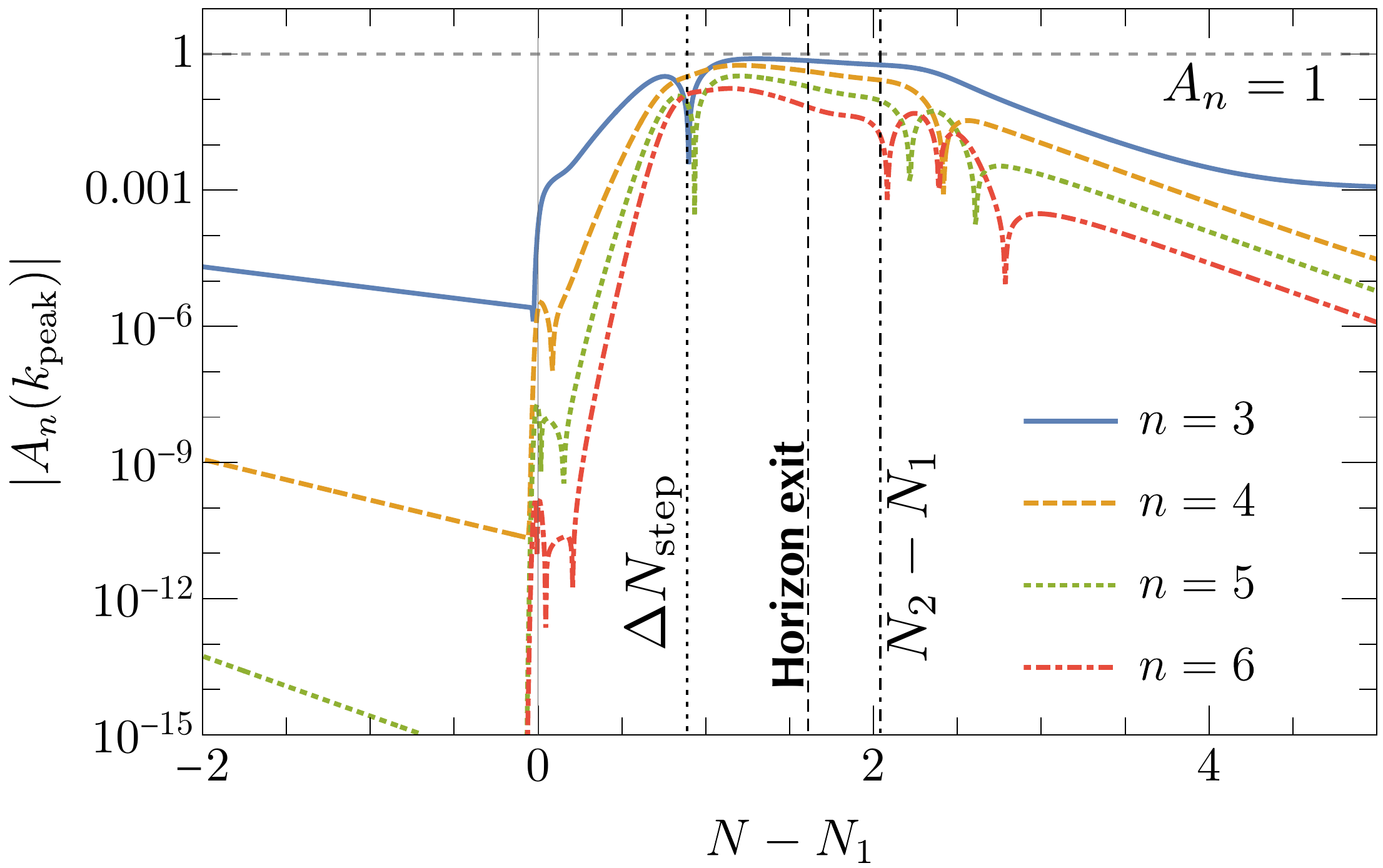}
\end{center}
\end{minipage}
\begin{minipage}{0.49\hsize}
\begin{center}
\includegraphics[width=1\columnwidth]{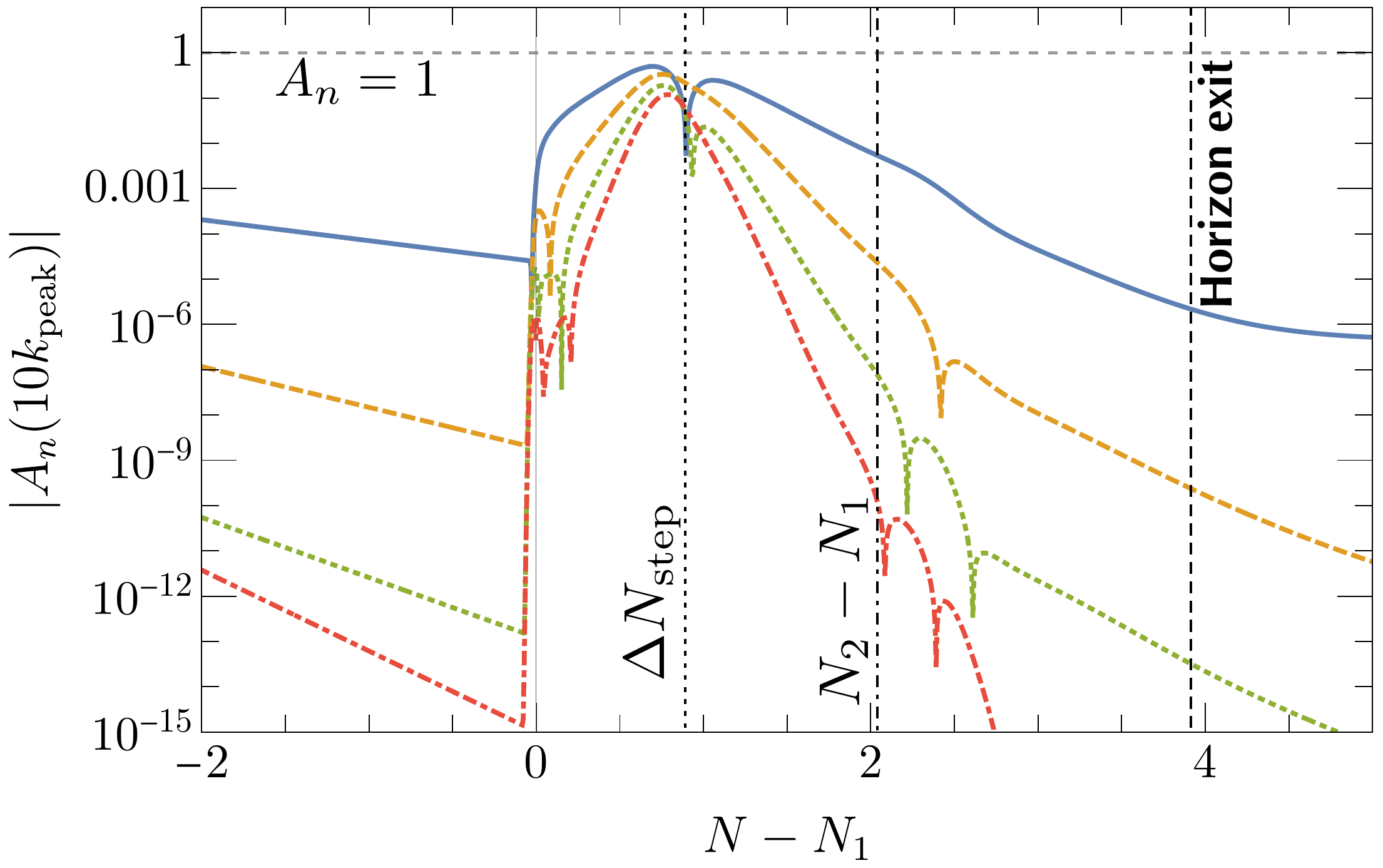}
\end{center}
\end{minipage}
\caption{
The evolution of $A_n(k_\text{peak})$ (left) and $A_n(10k_\text{peak})$ (right) in the upward step case.
The other parameters are the same as the orange line in Fig.~\ref{fig:ps_smooth_up} ($\Delta \phi_1 = 0.01 \sqrt{2\epsilon_i}M_\Pl, \Delta \phi_2 = \Delta \phi_m = 0.1 \sqrt{2\epsilon_i}M_\Pl$).
}
\label{fig:lnol2_up_1em4}
\end{figure}

Figures~\ref{fig:lnol2_down} and \ref{fig:lnol2_up_1em4} show the evolution of $|A_n(k_\text{peak})|$ and $|A_n(10 k_\text{peak})|$ for $3 \leq n \leq 6$.
Figure~\ref{fig:lnol2_down} shows the results in the downward step case, where the left figure is for $A_n(k_\text{peak})$ and the right one for $A_n(10 k_\text{peak})$.
For the result of $A_n(k_\text{peak})$, the peak around $N-N_1 \sim 0$ is due to relatively large values of $H^{(n)}$ associated with the rolling down of the inflaton but only achieves an amplitude of  $|A_n| \lesssim 10^{-2}$.
After the rolling down, the universe enters the USR period, during which the curvature perturbations grow. 
The origin of the other peak around $N-N_1 \sim 3$ comes from this growth of the curvature perturbations when the mode of $k_\text{peak}$ is superhorizon.
After the USR period, the inflaton gets in the slow-roll attractor and $H^{(n)}$ becomes small, which decreases $|A_n(k_\text{peak})|$ in time.
The peak value of $|A_3|$ in the USR phase is $\sim 0.2$ and therefore we can expect that the linear perturbation theory describes the step transition very well and can marginally describe the subsequent USR evolution of the perturbations.  Furthermore the $k \sim 10k_\text{peak}$ mode is even less strongly coupled. 
Since $\mathcal P_\mathcal R(k_\text{peak}) > \mathcal P_\mathcal R(10 k_\text{peak})$ is satisfied once the curvature perturbations get enhanced, $|A_n(10 k_\text{peak})|$ is larger than $|A_n(k_\text{peak})|$ only before the enhancement. 
From these figures, we can see that, although $|A_n(10 k_\text{peak})|$ can be larger than $|A_n(k_\text{peak})|$ before the perturbation enhancement, the maximum value of $|A_n(10 k_\text{peak})|$ (around $N - N_1 \sim 0$) is smaller than that of $|A_n(k_\text{peak})|$ (around $N - N_1 \sim 3$).

Here, we make some remarks.
First, we should keep in mind that, even if $|A_n|$ becomes closer to unity during the USR period with other parameters, the strong coupling of perturbations does not prevent the non-adiabatic particle production because it completes before the USR period, which amplifies the curvature perturbations mainly on superhorizon scales where gravitational interactions rather than scalar field interactions dominate.
In that case, although the precise calculation of the power spectrum requires more careful methods, such as the stochastic formalism~\cite{Pattison:2017mbe,Biagetti:2018pjj,Ezquiaga:2018gbw,Ezquiaga:2019ftu,Figueroa:2020jkf,Pattison:2021oen} and the Hartree factorization~\cite{Cheng:2021lif}, the perturbation enhancement definitely occurs because the particle production itself can be described with the linear perturbation theory.
Second, strictly speaking, we need to check the infinite number of the higher order contributions, $A_{n}$ even for $n > 6$, but, in our model, the smoothing out is done with the tanh function (see Sec.~\ref{sec:smoothing}) and therefore we can expect the similar behavior even for $n>6$.
We will come back to this issue later in this section.
Third, as mentioned above, the strong coupling condition, given by Eq.~(\ref{eq:no_sc_cons_eft}), could have some uncertainties at the quantitative level. However, given $|A_n| \lesssim \mathcal O(0.01)$ at the particle production, we can expect the conclusion that the particle production occurs would not change even if the strong coupling scales are defined more precisely.

Figure~\ref{fig:lnol2_up_1em4} shows the results in the upward step case.
Contrary to the downward step case, the largest value of $|A_n(k_\text{peak})|$ near $N-N_1 \sim 1$ is due to the step transition itself, namely from the decrease of $\dot H$ (or $\epsilon$).
The dip around $N-N_1 \sim 1$ comes from the sign change of $\eta$.
 On the other hand, similar to the downward step case, the maximum $|A_n(10k_\text{peak})|$ is comparable to or smaller than $|A_n(k_\text{peak})|$.  Since the peak of all the $|A_n|$ are slightly smaller than unity, the strong coupling of the perturbations is marginally avoided.

However, we should keep in mind that our strong coupling bound is a rough estimate and that  it is only marginally satisfied at the main particle production event itself for the fiducial upward step, unlike that of the downward step.   Therefore there is still the possibility that strong coupling shuts down particle production even in our fiducial case for the upward step.
To properly calculate the perturbations that are strongly coupled, we need to perform the lattice simulation (see Ref.~\cite{Caravano:2021pgc} for the lattice simulation during the inflation).
Since the main cause of the large $|A_n|$ is the decrease of the inflaton velocity, we can realize a weaker coupling (a smaller $|A_n|$) with a larger $\epsilon_m/\epsilon_i$, though the perturbation enhancement is correspondingly reduced.

\begin{figure}[h!]
\centering \includegraphics[width=.6\columnwidth]{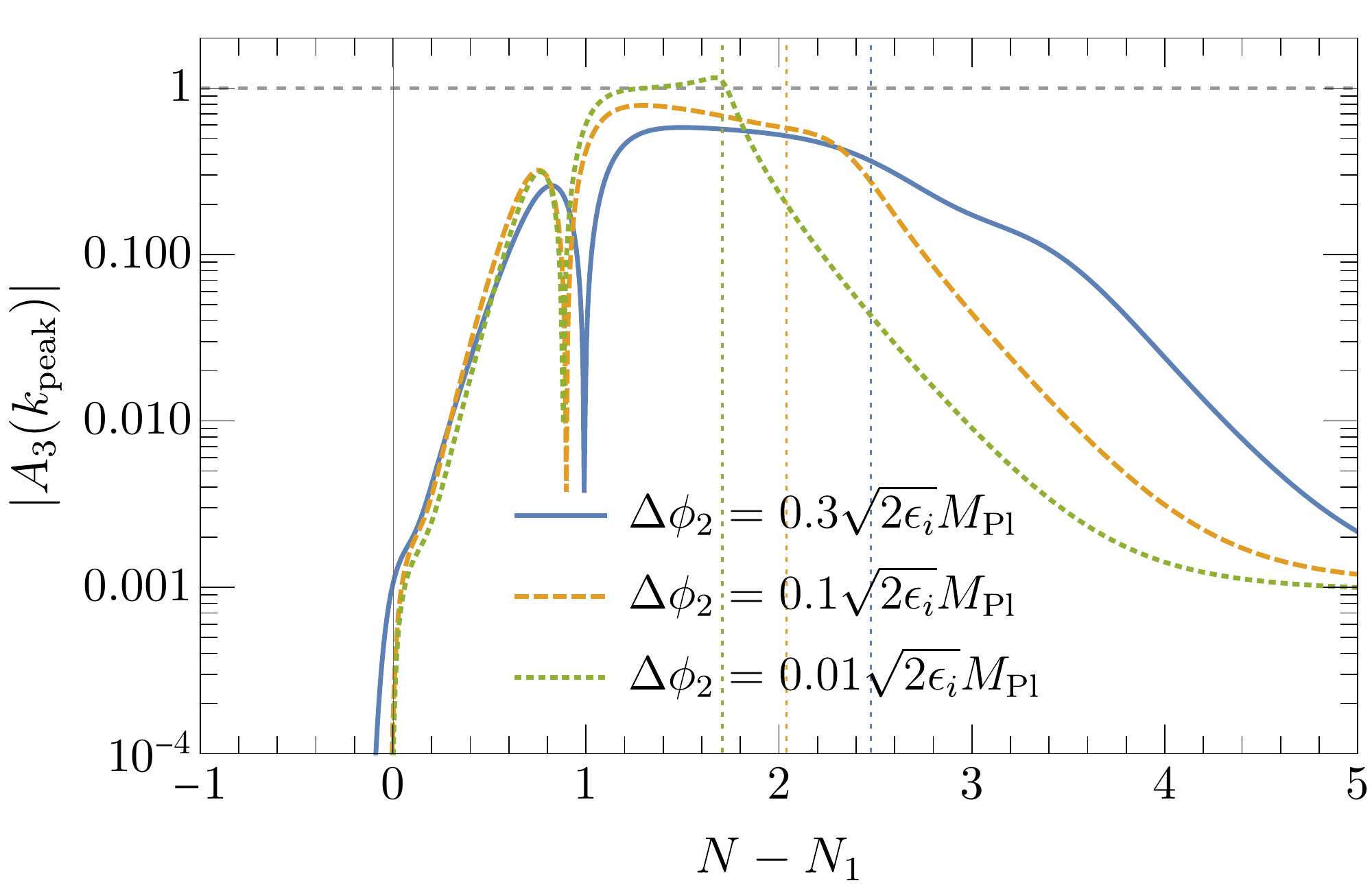}
\caption{ The evolution of $A_3(k_\text{peak})$ in the upward step case with different $\Delta \phi_2$.
We take $\Delta \phi_m = \Delta \phi_2$ for all plots.
For the other parameters, we take the same parameters as in Fig.~\ref{fig:lnol2_up_1em4} (or the orange line in Fig.~\ref{fig:ps_smooth_up}).
The vertical dotted lines represent $N_2-N_1$ for $\Delta \phi_2/(\sqrt{2\epsilon_i}M_\Pl) = 0.01, 0.1$ and $0.3$ from left to right.
}
\label{fig:a3_delta2}
\end{figure}

Smoothing the upward step also does not qualitatively change the strong coupling bound without correspondingly reducing the power spectrum.
Figure~\ref{fig:a3_delta2} shows $|A_3|$ with different $\Delta \phi_2$, which we showed in Fig.~\ref{fig:ps_smooth_up} 
to have little impact on the power spectrum.
From this figure, we can see that a larger smoothing width correspondingly only slightly decreases $|A_3|$ near $N-N_1 \sim 1$. 
This is because the large $|A_3|$ (and other $|A_n|$) there mainly comes from the decrease of the inflaton velocity and the following perturbation growth, which are not directly related to the smoothing at $\phi_2$.
In other words, the largest value of $|A_n|$ is mainly determined by the sharpness of the step, characterized by $\phi_2-\phi_1$ and $h$.
There are three periods for the particle production around $N \simeq N_1, N \simeq N_1+\Delta N_\text{step}$, and $N \simeq N_2$, where $\eta$ changes rapidly and the mixing of the positive and the negative modes occurs (see Fig.~\ref{fig:eta_ep_evol_up_down}).
The main contribution to the perturbation enhancement comes from the particle production around $N_1 + \Delta N_\text{step}$, which can be understood with the discussion in Sec.~\ref{sec:up}.
On the other hand, the particle production at $N_2$ does not change the peak height of the power spectrum because the peak scale exits the horizon before $N=N_2$.
This is why Fig.~\ref{fig:ps_smooth_up} shows that the peak height of the power spectrum does not depend on $\Delta \phi_2$.
The effect of the particle production at $N_2$ can only be seen in the oscillatory feature around the cutoff scale of the perturbation enhancement, shown in Fig.~\ref{fig:ps_smooth_up}.

Here, we discuss the higher order contribution in $\pi$ using $A_n$ with $n>3$, which one might think may be larger than $|A_3|$ especially at $\phi_1$ and $\phi_2$ because the $A_n$ depends on the higher derivatives of the potential.
This scaling is regulated by the smoothing field width ($\Delta \phi_1$, $\Delta \phi_2$). 
From Eqs.~(\ref{eq:h_2_ana})-(\ref{eq:h_4_ana}), we can see that $H^{(n)}$ includes the term $\dot \phi^{n-1}V^{(n-1)}/M_\Pl^2$. 
Since we perform the smoothing out with a steplike function where the smoothing scale transitions from $\Delta\phi_1$ to $\Delta \phi_2$ 
in Sec.~\ref{sec:smoothing}, we can approximate the order of $V^{(n)}$ as $V^{(n)}(\phi_{1,2}) \simeq V_b(\phi_{1,2})\,  \partial^n G(\phi)/\partial \phi^n|_{\phi=\phi_{1,2}}  \simeq \mathcal O(h V_{b} /((\phi_2-\phi_1)^2 \Delta \phi_{1,2}^{n-2}))$.

From this, we can get a rough necessary condition for the convergence of 
\begin{align}
A_n = \mathcal O\left( \frac{H^{(n-1)}}{H^{n-2} \dot H} \mathcal P_{\mathcal R}^{(n-2)/2} \right)
\end{align}
in the limit of $n \rightarrow \infty$
as 
\begin{align}
&\frac{A_{n+1}}{A_n} = \mathcal O \left( \frac{H^{(n)}}{H H^{(n-1)}} \mathcal P_{\mathcal R}^{1/2}\right) < 1 \nonumber \\
\Rightarrow \quad 
   &\Delta \phi_{1,2} > \frac{\dot \phi_{1,2}}{H} \mathcal P_{\mathcal R}^{1/2} \nonumber \\
\Rightarrow \quad & 
  \Delta \phi_{1,2} > \sqrt{2 \epsilon(\phi_{1,2})}M_\Pl \mathcal P_{\mathcal R}^{1/2},
\end{align}
where we have neglected the prefactor dependent on $n$ for simplicity.
If this condition is not satisfied, the perturbation could be strongly coupled at the higher order in $\pi$.
In the following, let us focus on whether the peak-scale perturbations are strongly coupled at $\phi_1$ or $\phi_2$.
In the downward step case, the curvature perturbations do not grow before $\phi=\phi_2$ and the peak scale is smaller than the horizon scales in $\tau_1 < \tau < \tau_2$ by a factor $\mathcal O(1)$.
From this, we can roughly estimate $\mathcal P_\mathcal R (k_\text{peak})^{1/2} \sim \mathcal O(10^{-4})$ before $\phi=\phi_2$.
Then, we can derive the rough condition for the step width for our downward step case:
\begin{align}
  \Delta \phi_{1,2} > \mathcal O(10^{-4}) \sqrt{2 \epsilon(\phi_{1,2})}M_\Pl \quad (\text{for downward step}).
\end{align}
This condition is satisfied in our fiducial examples for the downward step case.
On the other hand, for the upward step case, the situation is a bit complicated because the curvature perturbations grow by $\phi=\phi_2$.
Then, we obtain
\begin{align}
  \begin{split}
  \Delta \phi_{1} &> \mathcal O(10^{-4}) \sqrt{2 \epsilon(\phi_{1})}M_\Pl \\
  \Delta \phi_{2} &> \mathcal P_{\mathcal R}^{1/2}(k_\text{peak}) \sqrt{2 \epsilon(\phi_{2})}M_\Pl \\  
  \end{split}
  \quad (\text{for upward step}).
\end{align}
At least in Fig.~\ref{fig:lnol2_up_1em4}, the peak scale $k_\text{peak}$ is (almost) on superhorizon at $\phi=\phi_2$.
Because of this, we can substitute $\mathcal P_{\mathcal R}^{1/2}(k_\text{peak}) \simeq 10^{-1}$ from Fig.~\ref{fig:ps_cmb_tra_up}.
Also, from Fig.~\ref{fig:eta_ep_evol_up_down}, we can see $\epsilon(\phi_2)$ is between $\epsilon_i$ and $\epsilon_m$. 
Then, we finally find that our fiducial case in Fig.~\ref{fig:lnol2_up_1em4} marginally satisfies the above condition.

Finally, we should keep in mind that the above results depend on the shape of the potential step. 
As another example, we take a step form of $\tanh (x)$ and discuss the strong coupling issues in Appendix~\ref{app:tanh_step}.
Remarkably, in the downward step with the tanh step, the perturbations are more strongly coupled due to a large $\eta$.
This is mainly because the tanh step case has only one field-width parameter for the smoothness of all potential derivatives at all field values during the transition and the full ($\mathcal O(\epsilon_m/\epsilon_i)$) enhancement requires the field width to be much less than $\sqrt{2\epsilon_i} M_\Pl$.  This is the primary reason we have taken a more flexible step transition where the various critical field regions carry their own smoothness scale.
See the Appendix for detail.

\subsection{Backreaction}

Distinct but complementary to constraints from perturbativity of the action is the backreaction of the enhanced perturbations on the background evolution.  As a simple consistency check, we may impose that the energy density of the enhanced perturbations, which ostensibly is transferred from the kinetic energy of the homogeneous inflaton field, is less than the kinetic energy of the inflaton.

The total energy density in the enhanced perturbations, i.e.~from particle production, is given by~\cite{Adshead:2014sga}
\begin{align}
	\rho_p &\simeq   \frac{1}{2} \expval{\dot{\delta \phi}^2 + \frac{(\partial_i \delta \phi)^2}{a^2}} - \rho_\text{vac} \nonumber \\ 
  &=  \int \dd \ln k \left[ \frac{1}{2} \left( \mathcal P_{\dot \delta\phi}(k) + \frac{k^2}{a^2} \mathcal P_{\delta \phi}(k)\right) - \tilde \rho_\text{vac}(k) \right] \nonumber \\
  &\simeq \tilde \rho_{p}(k_\text{peak}) -  \tilde \rho_\text{vac}(k_\text{peak}),
	\label{eq:rho_p_formula}
\end{align}
where $\tilde \rho_{p} \equiv \frac{1}{2}(\mathcal P_{\dot \delta\phi} + k^2/a^2 \mathcal P_{\delta \phi})$, $k_\text{peak}$ is the peak scale of $\tilde \rho_p$, and the $\rho_\text{vac}$ is the contribution from the vacuum fluctuations, whose power spectrum is given by $\tilde \rho_\text{vac} \simeq H^4 (k\tau)^4/(2\pi)^2$.
Meanwhile, the kinetic energy of the inflaton is given by 
\begin{align}
	\rho_\text{kin} &\equiv \frac{1}{2} \dot \phi^2  = \epsilon H^2 M_\Pl^2.
\end{align}
To avoid the backreaction to the background evolution, the energy density of the enhanced perturbation must be smaller than the kinetic energy of inflaton, $\rho_p < \rho_\text{kin}$~\cite{Adshead:2014sga}.

\begin{figure}  
\centering \includegraphics[width=0.6\columnwidth]{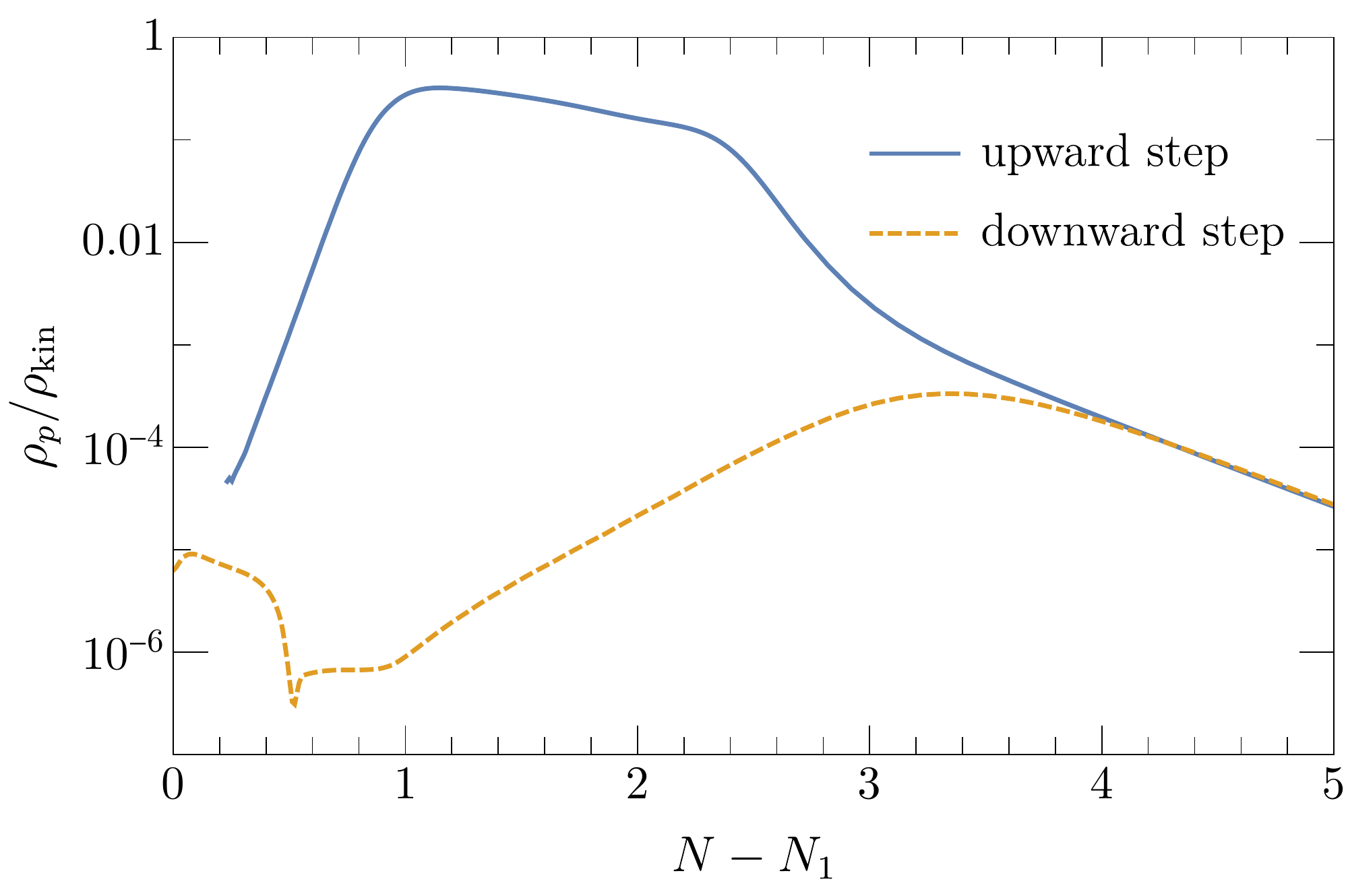}
\caption{ The time dependence of the energy density of the enhanced perturbations normalized by the kinetic energy of the inflaton at each time.
The parameters are the same as in Fig.~\ref{fig:lnol2_down} for the downward step case and Fig.~\ref{fig:lnol2_up_1em4} for the upward step case.
}
\label{fig:e_cons_up_down}
\end{figure}

Figure~\ref{fig:e_cons_up_down} shows the evolution of $\rho_p/\rho_\text{kin}$ in the case of the upward step and the downward step case.
In the calculation, we take the flat gauge, in which $\mathcal R = \delta \phi/(\sqrt{2\epsilon}M_\Pl)$.
The no plot for the upward step case in $N-N_1 \lesssim 0.4$ implies that there is no peak of $\tilde \rho_p$ in $k$ then. 
That is, the perturbation enhancement has not occurred yet.
The peaks around $N-N_1 \sim 1$ for the upward step case roughly correspond to the minimum of the inflaton kinetic energy.
On the other hand, the peak around $N-N_1 \sim 3$ for the downward step case corresponds to the end of the USR period, that is, the end of the growth of the curvature perturbations.
In Fig.~\ref{fig:e_cons_up_down}, all of the lines are below unity, which indicates that the backreaction of the enhanced perturbation on the background evolution is small on average in our fiducial parameter sets. 
On the other hand, as we shall see next, since for the upward step the kinetic energy in the background is so small at the top of the step, there can be fluctuating regions that fail to climb the step.

\section{Trapping the Inflaton in an Upward Step: PBH's as Baby Universes}
\label{sec:trap_inflaton}

To complete this work we consider a phenomenon that is unique to an upward step: the possibility that the inflaton may be unable to climb the step, and instead become {\it trapped} and unable to roll down the potential. 
In particular, while in the previous sections we have discussed the inflaton dynamics neglecting the backreaction of the quantum fluctuations, the backreaction can actually cause the interesting phenomenon.
If the inflaton decelerates due to the quantum backreaction just before the step, the inflaton could fail to climb the potential and be trapped at the local minimum, even in a potential that is fine-tuned to allow the classical inflaton (without the backreaction) to successfully climb the step. 
Quantum backreaction opens the possibility that a fraction of the universe might fail to climb the step, while the rest of the universe evolves in accordance with classical intuition. In this case, observers in the post-inflationary universe see the trapped regions as PBHs \cite{Garriga:2015fdk,Deng:2017uwc}.

According to the detailed analysis based on the Kramers-Moyal equation in Ref.~\cite{Biagetti:2018pjj}, the distribution of the quantum-backreaction-induced fluctuations of the normalized inflaton velocity asymptotes to a Gaussian distribution with variance given by\footnote{The variance asymptotes to this value after the universe experiences the inflation era for large e-folds ($N\gg 1$)~\cite{Biagetti:2018pjj}. In this paper, to discuss the connection to the PBH scenarios, we mainly focus on the case where the universe experiences the inflation era for $N\gg 1$ before the transition and therefore we can safely use the asymptotic form of the distribution.}
\begin{align}
	\expval{(\Delta \Pi)^2} = \frac{3H^2}{8 \pi^2},
\end{align}
where $\Pi$ is defined as 
\begin{align}
	\Pi \equiv \frac{\dd \phi}{\dd N} \simeq \frac{\dot \phi}{H}.
\end{align}
Since the Hubble parameter is almost constant, we can approximate $\Delta \Pi \simeq \Delta\dot \phi/H$.
Then, we can regard the distribution of the inflaton velocity fluctuations as the Gaussian distribution with the variance given by 
\begin{align}
	\expval{(\Delta \dot \phi)^2} = \frac{3H^4}{8 \pi^2}.
\end{align}
From this result, we can also see that the variance of the kinetic energy of the inflaton is given by
\begin{align}
 \expval{(\Delta K)^2} &\equiv \expval{\left(\frac{1}{2} \left( \dot {\bar\phi} + \Delta \dot \phi \right)^2 - \frac{1}{2} \dot{\bar \phi}^2 \right)^2} \nonumber \\
 	&\simeq \dot{\bar \phi}^2  \expval{(\Delta \dot \phi )^2} \nonumber \\
	&= \frac{3 H^4} {4 \pi^2} \bar K, 	
\label{eq:vari_kinetic}	
\end{align}
where $\dot {\bar\phi}$ indicates the background value, which does not include the quantum backreaction, and the background kinetic energy is defined as $\bar K \equiv \frac{1}{2}\dot{\bar \phi}^2 (=\epsilon_i H^2 M_\Pl^2)$.

Hereafter, for simplicity, we consider the case where the transition from $\epsilon_i$ to $\epsilon_m$ occurs rapidly enough that we can neglect the effect of the Hubble friction while the inflaton climbs the upward step.
In this case, if we neglect the quantum backreaction, the kinetic energy becomes $\epsilon_m H^2 M_\Pl^2$ soon after the transition.
In other words, the inflaton just before the step has an excess of kinetic energy of $\epsilon_m H^2 M_\Pl^2$ in addition to that required to climb the step at the classical level without the backreaction.
From this, we can express the condition of the fluctuations for the inflaton trapping (the failure to climb the step) as
\begin{align}
	-\Delta K > \epsilon_m H^2 M_\Pl^2.
\end{align}
For convenience, we define the normalized fluctuation of the kinetic energy as $E_K \equiv \Delta K/\bar K$ and rewrite the condition as 
\begin{align}
	E_K < - \frac{\epsilon_m}{\epsilon_i}.
\end{align}
From Eq.~(\ref{eq:vari_kinetic}), we can derive the variance of $E_K$ as
\begin{align}
	 \sigma_{E_K}^2 \equiv \expval{E_K^2} &= \frac{3H^2}{4\pi^2 \epsilon_i M_\Pl^2 } \nonumber \\
	 &= 6 \mathcal P_{\mathcal R}(k \ll -1/\tau_1),
\end{align}
where we have used the relation $\mathcal P_\mathcal R \simeq H^2/(8\pi^2 \epsilon_i M_\Pl^2)$, valid on the scales much larger than the horizon scales at the step transition.
Then, we can express the probability of the inflaton trapping as
\begin{align}
	\beta_\text{trap} &=  \int^{-\epsilon_m/\epsilon_i}_{-\infty} \dd E_K \frac{1}{\sqrt{2\pi} \sigma_{E_K}} \exp\left( - \frac{E_K^2}{2\sigma_{E_K}^2} \right) \nonumber \\
	&\simeq \frac{1}{\sqrt{2\pi}} \frac{\sigma_{E_K}}{\epsilon_m/\epsilon_i} \exp\left( -\frac{(\epsilon_m/\epsilon_i)^2}{2 \sigma_{E_K}^2} \right),
	\label{eq:beta_trap}
\end{align}
where the approximate equality is valid only when $\sigma_{E_K}/(\epsilon_m/\epsilon_i) \ll 1$.

Here, let us explain the fate of the inflaton-trapping region.
Because of  causality, the initial size of the inflaton-trapping region is expected to be comparable to the horizon at the transition.
Also, these regions are surrounded by the other ordinary regions, where the inflaton succeeds in climbing the potential step.
This case is similar to the consequences of bubble nucleation through the tunneling to the true vacuum during the inflation, which is discussed in Refs.~\cite{Garriga:2015fdk,Deng:2017uwc}.
The inflaton-trapping region in our case corresponds to the bubble produced through the tunneling in the references. 
Since the inflation lasts for $\mathcal O(10)$ e-folds after the transition, the inflaton-trapping bubble is above the horizon
when the other ordinary regions end inflating.
This situation corresponds to the ``supercritical'' case in Refs.~\cite{Garriga:2015fdk,Deng:2017uwc}, whose consequences are as follows.
The bubbles are surrounded by radiation or dust after the inflation of the other ordinary regions.
At that time, while the universes inside the supercritical bubbles exponentially expand due to the positive energy of their (false) vacuum, the surrounding radiation or dust region does not expand exponentially. 
This twisted situation leads to the two universes connected to each other by a wormhole.
After a while, the observer outside the bubble sees the bubble size comparable to the horizon size for the observer and the wormhole closes around that time. 
Then, the bubble region is finally seen as a BH for the observer outside the bubble, though the observer inside the bubble still feels the universe expanding exponentially.
One might worry about the lifetime of the inflaton-trapping bubble because a too-short lifetime does not allow the supercritical bubble.
In Appendix~\ref{app:bounce_action}, we calculate the bounce action for the tunneling rate from the local minimum to the true vacuum and show that the tunneling rate is small enough for the supercritical bubble at least in our fiducial setup.

In short, the inflaton-trapping bubble finally becomes a PBH similarly to the BH production by the supercritical bubble. 
This possibility is already pointed out in Refs.~\cite{Atal:2019cdz,Atal:2019erb}.
It was shown in Refs.~\cite{Garriga:2015fdk,Deng:2017uwc} that the mass of the PBH in the supercritical case is comparable to the total mass in the horizon of the other ordinary regions at the time when the bubble re-enters the horizon.
The PBH mass is then of the same order as that produced by the large density perturbations at the peak of the spectrum.

While the inflaton trapping can be a seed of a PBH, the enhanced density perturbations can also be the seed.
When very large density perturbations enter the horizon, they can collapse to PBHs because of their own gravitational attraction force.
This PBH formation mechanism is the most common in the studies of the inflation models for the PBH scenarios.
Here, we explain the PBH production rate by the enhanced density perturbations and compare it with the PBH production rate by the inflaton trapping.
In the following, for simplicity, we assume that the curvature perturbations follow the Gaussian distribution.
Based on the Press-Schechter formalism for the large density perturbations, the PBH production rate with mass of $M$ is given by~\cite{Green:2004wb}\footnote{We neglect the overall factor 2, given in Ref.~\cite{Green:2004wb}, because its necessity is not clear.}
\begin{align}
	\beta_p(M) &= \int_{\delta_\text{c}} \dd \delta \frac{1}{\sqrt{2 \pi} \sigma(M)} \exp\left( -\frac{\delta^2}{2  \sigma^2(M)} \right) \nonumber \\
	& \simeq \frac{1}{\sqrt{2\pi}} \frac{\sigma(M)}{\delta_\text{c}} \exp\left( -\frac{\delta_\text{c}^2}{2 \sigma^2(M)} \right),
	\label{eq:beta_def}
\end{align}
where $\delta_\text{c}$ is the threshold of the density contrast for PBH production.
$\sigma^2$ is the variance of the density contrast given by 
\begin{align}
	\sigma^2(R) = \int^\infty_0 \frac{\dd q}{q} \tilde W^2(q;R) T^2(qR) \frac{16}{81} (qR)^4 \mathcal P_{\mathcal R}(q),
\end{align}
where $R$ is the smoothing scale related to the PBH mass, $\tilde W$ is a window function in Fourier space, and $T(x)$ is the transfer function of the gravitational potential, which is given by
\begin{align}
	T(x) = 3 \frac{\sin(x/\sqrt{3}) - (x/\sqrt{3}) \cos(x/\sqrt{3})}{(x/\sqrt{3})^3}.
\end{align}
As a fiducial example of the window function, we take the real-space top-hat form:\footnote{See Refs.~\cite{Ando:2018qdb,Young:2019osy,Tokeshi:2020tjq} for the discussion on the choice of the window function.}
\begin{align}
	\tilde W(q; R) = 3 \frac{\sin(qR) - (qR) \cos(qR)}{(qR)^3}.
\end{align}
In this case, we have $\sigma^2(M) \simeq \mathcal P_{\mathcal R}(k(M))$ for a scale-invariant spectrum~\cite{Ando:2017veq,Ando:2018qdb}, where $k(M)$ denotes the scale of perturbation that produces a PBH with $M$.
In addition, the non-linear relation between the curvature perturbation and the density perturbation modifies Eq.~(\ref{eq:beta_def})~\cite{Kawasaki:2019mbl,DeLuca:2019qsy,Young:2019yug} and the modification can be roughly taken into account by just using $\sigma^2(M) \simeq \mathcal P_{\mathcal R}(k(M))/1.4^2$ without changing the form of Eq.~(\ref{eq:beta_def})~\cite{Kawasaki:2019mbl}.
For simplicity, we use this relation and roughly estimate the production rate of the peak mass as 
\begin{align}
	\beta_p(M_\text{peak})
	& \simeq \frac{1}{\sqrt{2\pi}} \frac{\mathcal P_{\mathcal R}^{1/2}(k_\text{peak})}{1.4 \delta_\text{c}} \exp\left( -\frac{(1.4 \delta_\text{c})^2}{2 \mathcal P_{\mathcal R}(k_\text{peak})} \right).
	\label{eq:beta_def_peak}
\end{align}
Substituting $\delta_\text{c} = 0.51$ as a fiducial value for a PBH production in a radiation era~\cite{Young:2019osy}, we rewrite this equation as 
\begin{align}
	\beta_p(M_\text{peak})
	& \simeq \frac{(2.0 \mathcal P_{\mathcal R}(k_\text{peak}))^{1/2}}{\sqrt{2\pi}} \exp\left( -\frac{1}{3.9 \mathcal P_{\mathcal R}(k_\text{peak})} \right).
	\label{eq:beta_def_peak_2}
\end{align}
Comparing this equation with the inflaton-trapping probability given by Eq.~(\ref{eq:beta_trap}) and taking into account the relation $\mathcal P_{\mathcal R}(k_\text{peak}) < (\epsilon_i/\epsilon_m)^2 \mathcal P_\mathcal R(k \ll -1/\tau_1)$ (see Eq.~(\ref{eq:p_r_ana_small_limit})), we see $\beta_\text{trap} > \beta_p$, which means that the PBHs from the inflaton trapping region could possibly be produced more than the ones from the enhanced perturbations.
However, we should keep in mind the following issues.

First, we need to be careful about the validity of the Gaussian distribution assumption in the above calculation of $\beta_p$.
Generally speaking, the non-Gaussianity could significantly change the PBH abundance because the large perturbations that produce PBHs are sensitive to the deviation from the Gaussian distribution~\cite{Byrnes:2012yx,Young:2013oia}.
In Refs.~\cite{Atal:2019cdz,Atal:2019erb}, the authors discuss the probability of the inflaton trapping and the non-Gaussianity caused by the potential bump approximated as $\propto -(\phi - \phi_c)^2$ with $\phi_c$ being the position of the bump.
On the other hand, we consider the potential step, which is different from the bump with the quadratic potential.
In particular, the rapid change of the second derivative of the potential induces the particle production, which leads to the large perturbation enhancement in our model.
Because of this, the analysis on the non-Gaussianity in our case requires another approach.
We leave this issue for future work.\footnote{The non-Gaussianity associated with the downward step is discussed in Ref.~\cite{Davies:2021loj}.}

Second, the mass spectra of the two types of PBHs are expected to be different.
The mass spectrum of the PBHs from the large perturbations is determined by the profile of the power spectrum of curvature perturbations.
On the other hand, the PBH mass spectrum from the inflaton trapping would depend on the probability distribution of the shape of the inflaton trapping bubbles. 
Generally speaking, the bubbles could be different from the spherical shape, which is expected not to prevent the PBH formation, but to change the PBH masses even if the bubbles are produced at the same time.
We leave a more detailed analysis of the mass spectrum associated with the inflaton trapping to future work.

\section{Conclusion}
\label{sec:concl}

The small-scale primordial power spectrum presents an opportunity for next-generation experiments to dramatically increase our knowledge of the very early universe. With this in mind, we have developed a scenario of inflationary cosmology, wherein a step in the inflaton potential, either downward or upward, is imprinted in the primordial power spectrum by a sharp rise followed by oscillations on small scales. The model is that of single-field canonical inflation, and does not necessitate additional fields or non-canonical kinetic terms.
Compared with the previous works~\cite{Covi:2006ci,Hamann:2007pa,Mortonson:2009qv,Adshead:2011jq,Miranda:2013wxa,Adshead:2014sga,Cai:2015xla,Miranda:2015cea,Kefala:2020xsx} that discuss the power spectrum features associated with the potential step, the new point of this paper is that we have mainly considered a relatively large change in the kinetic energy of the inflaton due to the step,  which can lead to a large ($\lesssim \mathcal O(10^7)$) enhancement in the power spectrum while avoiding strong coupling issues.
Note that the linear perturbation theory cannot give reliable results for strongly coupled perturbations.
In this work, to show that the $\mathcal O(10^7)$ enhancement can be realized by a step-like feature with the linear theory, we take the fiducial form of the potential that can avoid the strong coupling.
More generally,
the balance between enhancement and strong coupling requires the step transition to occur within a finite fraction of an e-fold, $\mathcal O(\mathcal P_{\mathcal R}^{1/2}(k_\text{peak},\tau)) \lesssim \Delta N < 1$, where $\tau$ here means the time when the salient aspects for particle production occur.
Because of the acceleration or deceleration of the inflaton during the transition this general e-fold criterion places a more complicated requirement on smoothness in field space, which our model satisfies whereas more naive approaches like a $\tanh$ step cannot.
These results give a general conclusion that a step-like feature can really realize the $\mathcal O(10^7)$ enhancement, required for the PBH scenarios.

This large enhancement is motivated by the PBH scenarios for dark matter, or BHs observed by the LIGO-Virgo collaboration from the collapse of rare fluctuations during the radiation or matter dominated epoch.
We have also discussed the probability of the inflaton trapping at the local minimum that appears in the upward step case. 
This inflaton trapping occurs due to the quantum backreaction to the inflaton velocity and ultimately also leads to the production of a PBH.
As a result, we have found that the PBHs from the inflaton trapping could be produced more than the PBHs from the collapse of the large density perturbations. At the same time, we should keep in mind that this result is based on our assumption that the density perturbations follow the Gaussian distribution, which would be modified in models with a non-slow-roll period. We leave the study of the non-Gaussianity associated with the upward step transition to future work.

 In addition to providing the black holes observed by the LIGO-Virgo collaboration, or the black holes of PBH dark matter, the perturbation enhancement in this scenario brings with it complementary observables, in the form of induced gravitational waves~\cite{tomita1967non,Matarrese:1993zf,Matarrese:1997ay,Ananda:2006af,Baumann:2007zm,Saito:2008jc,Saito:2009jt,Kohri:2018awv,Inomata:2018epa,Adshead:2021hnm,Domenech:2021ztg} and CMB spectral distortions~\cite{Fixsen:1996nj,Chluba:2012we,Kohri:2014lza}. 
 Apart from them, the downward step variant is falsifiable through detection of B-mode polarization of CMB: the requisite growth in the slow-roll parameter $\epsilon$ implies a small initial value, which in turn implies a tensor-to-scalar ratio $r=16\epsilon$ well below the sensitivity of next generation experiments. A comprehensive sensitivity forecast of future observational probes of the model is left to future work.

The model also serves as a stepping stone for future studies of particle production during inflation. 
For example, a spectator field that experiences such a step can be expected to undergo a similar non-adiabatic evolution and particle production. 
In particular, for the spectator field identified as an axion, the authors in Ref.~\cite{Cheng:2018yyr,Ozsoy:2020ccy,Ozsoy:2020kat} showed that a copious production of gauge fields occurs and the produced gauge fields induce large density perturbations and gravitational waves (see also Refs.~\cite{Anber:2009ua,Barnaby:2011qe,Adshead:2015pva}). 
The spectator field can also be a curvaton~\cite{Enqvist:2001zp,Lyth:2001nq,Moroi:2001ct} and the enhancement of power spectrum could occur even in this case.
Besides, the particle production associated with the step-like feature may be interesting as a genesis mechanism for particle dark matter that is ``completely dark'' \cite{Kolb:2020fwh}.
 Finally, it will also be interesting to explore the connection of the single field model presented here to multi-field models, with the single field model realized as one possible trajectory in multidimensional field space. Multi-field models generically exhibit isocurvature perturbations, which may also be enhanced, and in certain models (e.g., \cite{Palma:2020ejf,McDonough:2020gmn,Cai:2021yvq}) can modify the tensor-to-scalar ratio. We leave this and other interesting topics to future work.

\acknowledgments

The authors thank Alexander Vilenkin for helpful comments.
The authors were supported by the Kavli Institute for Cosmological Physics at the University of Chicago through an endowment from the Kavli Foundation and its founder Fred Kavli.
E.M. and W.H. were supported by U.S. Dept. of Energy contract DE-FG02-13ER41958.  W.H. was additionally supported by  the Simons Foundation.

\appendix

\section{Evolution of $\eta$ in Downward Step}
\label{app:evol_up_down}

In this appendix, we analytically discuss the evolution of $\eta$ in the potential with a downward step, given by Eq.~(\ref{eq:pot_cmb_to_end_mo}) with $h<0$.
During the downward step, the inflaton has a large tachyonic mass. Here, we relate $m$ to $\eta$ through the equation of motion of the inflaton, given by 
\begin{align}
  \ddot \phi + 3 H \dot \phi+ m^2 \phi = 0.
\end{align}
Then, we derive the solution
\begin{align}
  \phi \simeq C_1 \ee^{s_{+}(t-t_1)} + C_2 \ee^{s_{-}(t-t_1)},
\end{align}
where $s_\pm$ is defined by 
\begin{align}
  s_\pm \equiv \frac{3 H}{2} \left( -1 \pm \sqrt{1-\frac{4}{9} \frac{m^2}{H^2}} \right).
\end{align}
In the case of a large tachyonic mass, the first term with $s_+$ quickly dominates the evolution of $\phi$.
Then, we can safely approximate the evolution of the inflaton during the downward step transition as 
\begin{align}
  \phi \simeq C_1 \exp\left( \frac{3 H}{2} \left( -1 + \sqrt{1-\frac{4}{9} \frac{m^2}{H^2}} \right) (t-t_1) \right).
\end{align}
Note that this approximation leads to the constant $\eta$ during the downward step transition, unlike during the upward step one.
With this equation, we can approximate $\eta$ during the increase of $\epsilon$ as 
\begin{align}
  \eta 
  =& -6 - \frac{2}{H\dot \phi} \frac{\partial V}{\partial \phi} + \frac{1}{M_\Pl^2} \frac{\dot \phi^2}{H^2} \label{eq:eta_reform} \\  
  \simeq &  -6 - \frac{4}{3} \frac{m^2}{H^2} \left( -1 + \sqrt{1-\frac{4}{9} \frac{m^2}{H^2}} \right)^{-1}.
  \label{eq:eta_app_constant}
\end{align}
Plugging in $m^2 = \lim_{\phi \rightarrow \phi_1 + 0} V''(\phi)$, we find the value of $\eta$ during the downward step, Eq.~(\ref{eq:eta_app_exp_ease}).
We can also invert this to express $m^2$ as a function of $\eta$ during the transition as
\begin{align}
  m^2 \simeq -\frac{H^2}{4} \left( 6 \eta + \eta^2 \right),
\end{align}
which appears in the main text as  Eq.~(\ref{eq:eta_m_relation}).

\section{Another Type of Potential Step}
\label{app:tanh_step}

In this appendix, we discuss another type of the potential step, described by the hyperbolic tangent function.
Instead of Eq.~(\ref{eq:f_step}), we use the following function as the function $F$:
\begin{align}
	F(\phi;\phi_1,\phi_2,h) = 1 + \frac{h}{2} \left[1 + \tanh\left( \frac{\phi - \phi_1}{\phi_2-\phi_1}  \right) \right].
	\label{eq:f_tanh_step}
\end{align}
The key difference between this case and that of the main text is that the step transition has only a single scale $\phi_2-\phi_1$ which determines the potential derivatives $V^{(n)}$ and evolution of $\epsilon$, $\eta$ across the whole transition.  We shall see next that such a transition cannot simultaneously enhance the power spectrum sufficiently and make the perturbations remain weakly coupled.

\subsection{Downward Step}

\begin{figure}
\centering \includegraphics[width=0.6\columnwidth]{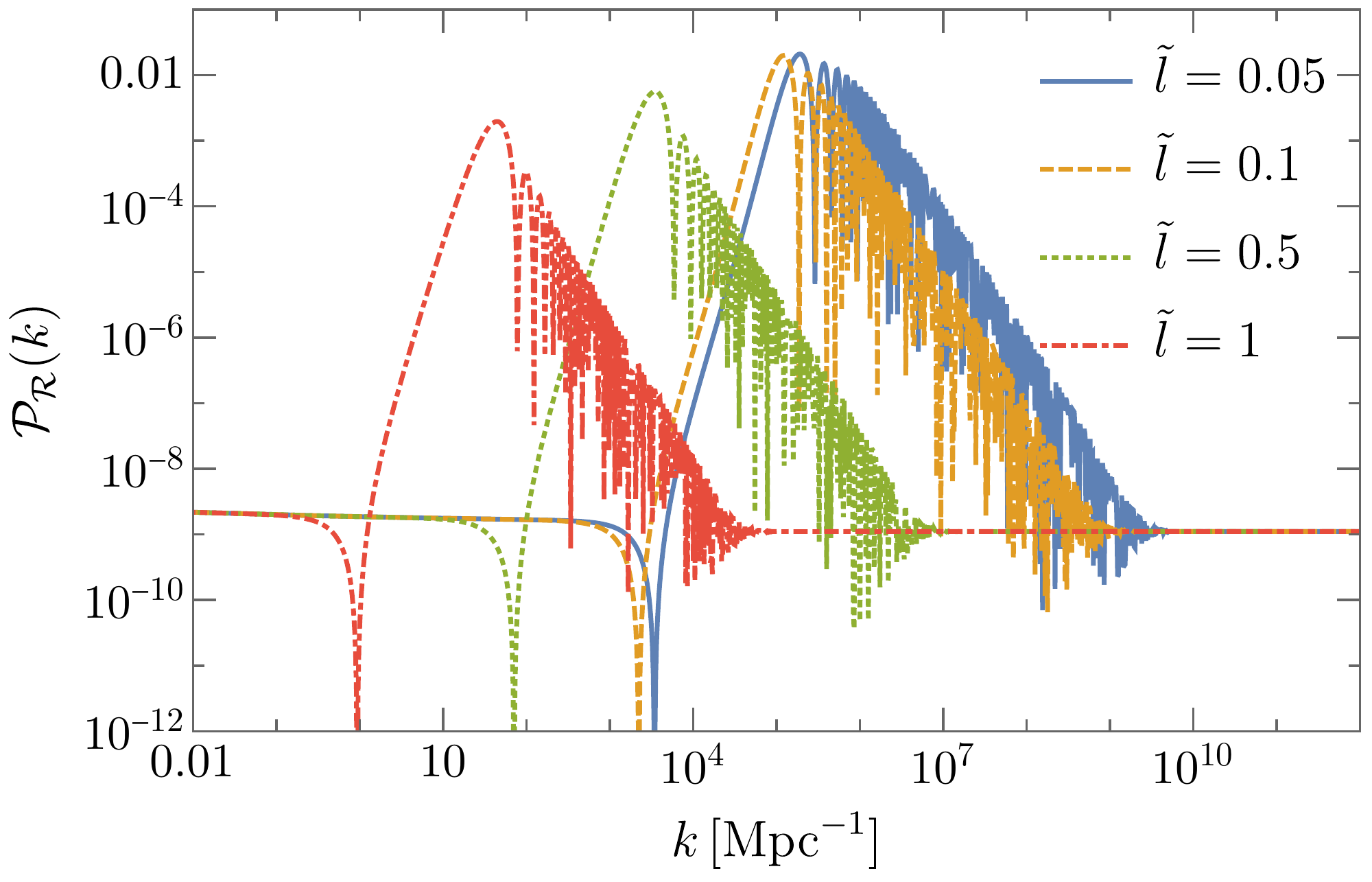}
\caption{ The power spectrum with the downward tanh step given by Eq.~(\ref{eq:f_tanh_step}).
Here, we define $\tilde l \equiv (\phi_2-\phi_1)/(\sqrt{2\epsilon_f}M_\Pl)$ and take the same values for the other parameters as in Fig.~\ref{fig:pot_l01}.
}
\label{fig:ps_tanh}
\end{figure}

Figure~\ref{fig:ps_tanh} shows the power spectrum with the tanh downward step.
We normalize the step width with $\epsilon_f (=\beta^2 \phi_\CMB^2/(2M_\Pl^2))$ because $\epsilon_i$ itself depends on the step width, though $\epsilon_i$ is almost the same as $\epsilon_f$.
The peak scales depend on the step width because the broader step width with fixed $\phi_1$ leads to the earlier step down of the inflaton.
From this figure, we can see that, to realize the $\mathcal O(\epsilon_m/\epsilon_f)$ enhancement with this tanh step, the step width $\phi_2-\phi_1$ must be smaller than $\sqrt{2\epsilon_f}M_\Pl$, otherwise the $\epsilon$ does not change from $\epsilon_i$ to $\epsilon_m$ within less than an e-fold.
If the step width is sharp enough, the peak height of the power spectrum is almost the same as that with the other type of step, whose results are shown in Fig.~\ref{fig:ps_cmb_tra_up}. 
Meanwhile, the small-scale tail of the peak is different from the results in Fig.~\ref{fig:ps_cmb_tra_up}. 
This difference originates from the difference in the evolution of $\eta$.

\begin{figure}  
\centering \includegraphics[width=0.6\columnwidth]{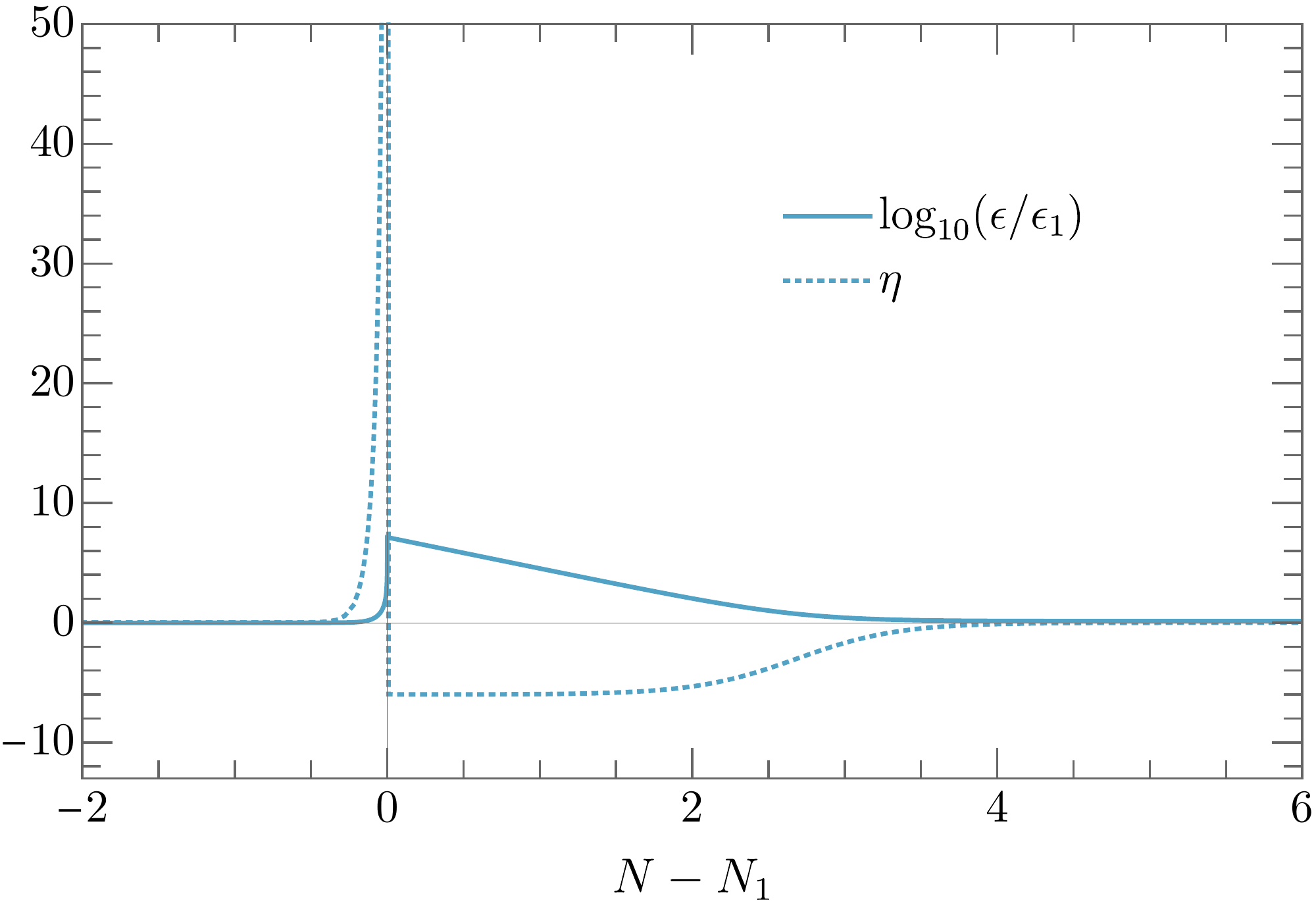}
\caption{ The evolution of $\epsilon$ and $\eta$ for $\tilde l = 0.1$ in Fig.~\ref{fig:ps_tanh}.
}
\label{fig:ep_eta_tanh}
\end{figure}

Figure~\ref{fig:ep_eta_tanh} shows the evolution of $\epsilon$ and $\eta$ that realize the power spectrum in Fig.~\ref{fig:ps_tanh}.
The peak value of $\eta$ is $\eta \sim 2\times 10^4$ (out of Fig.~\ref{fig:ep_eta_tanh}).
This value can be estimated based on the fact that $\eta$ roughly corresponds to the inverse of the e-folds for the $\mathcal O(1)$ change of $\epsilon$ by definition.
Then, we can roughly estimate $\eta \sim \mathcal O(\sqrt{2\epsilon_m}M_\Pl/(\phi_2-\phi_1))$, which is consistent with the numerical result of the peak $\eta$ value. 
Notice that since $\epsilon_m\gg \epsilon_f$ the rapid increase in the inflaton velocity during the step  means that the most rapid change in $\epsilon$ occurs on an e-fold scale $\sim \eta^{-1}$ which is much shorter than the full transition e-fold scale $\sim \tilde l = (\phi_2-\phi_1)/\sqrt{2\epsilon_f} M_{\rm Pl}$.  We shall see that this short timescale leads to strong coupling.

\begin{figure}  
\centering \includegraphics[width=0.6\columnwidth]{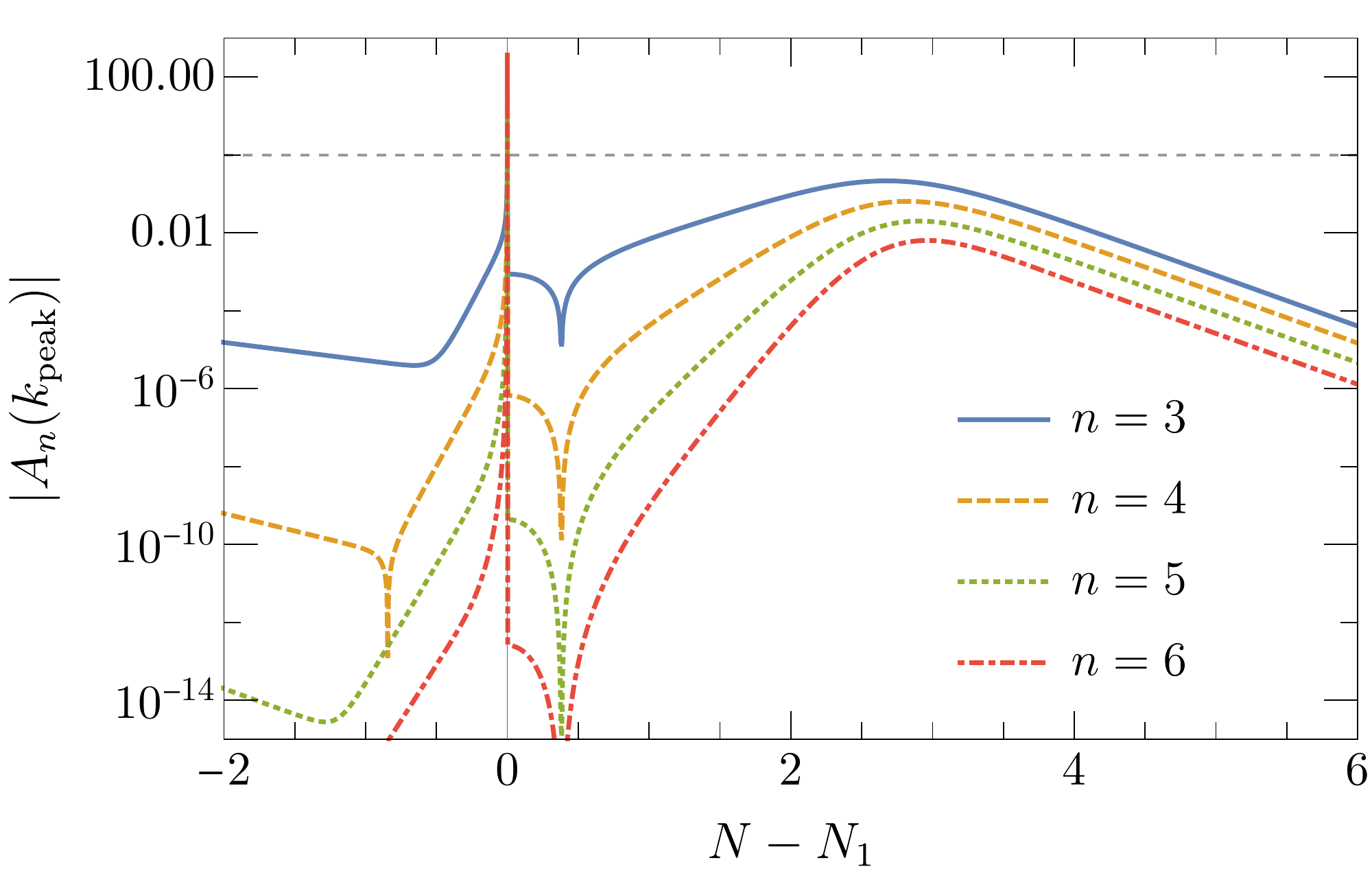}
\caption{ The evolution of $A_n(k_\text{peak})$ for $\tilde l = 0.1$ in Fig.~\ref{fig:ps_tanh}.
}
\label{fig:lnol2_tanh}
\end{figure}

Figure~\ref{fig:lnol2_tanh} shows the evolution of $A_n(k_\text{peak})$, defined in Eq.~(\ref{eq:a_n_def}), for $\tilde l=0.1$ in Fig.~\ref{fig:ps_tanh}.
We can see the sharp spike at $N-N_1 \simeq 0$, which is due to the sharp increase of the $\eta$ in Fig.~\ref{fig:ep_eta_tanh}.
This means that we can no longer use the linearized Mukhanov-Sasaki equation. Instead, we need to perform the lattice simulation to follow the perturbations.
Even if we consider a smaller enhancement of the perturbations and $|A_n(k_\text{peak})|$ is smaller than unity at the transition, the perturbations on the small-scale tail of the power spectrum peak tend to be strongly coupled because the curvature perturbation on the subhorizon scales is proportional to $(-k\tau)$, which leads to $|A_n(k)| > |A_n(k_\text{peak})|$ for $k> k_\text{peak}$ at the transition.
For the above reasons, the perturbation enhancement with a downward tanh step is severely limited if we require that the inflaton fluctuations remain weakly coupled throughout.  The model of the main text avoids this problem by  distinguishing the epoch of particle production from that of the acceleration of the inflaton and setting the appropriate smoothness of the potential for each stage so that neither completes in too small a fraction of an e-fold.

\subsection{Upward Step}

\begin{figure}
\centering \includegraphics[width=0.6\columnwidth]{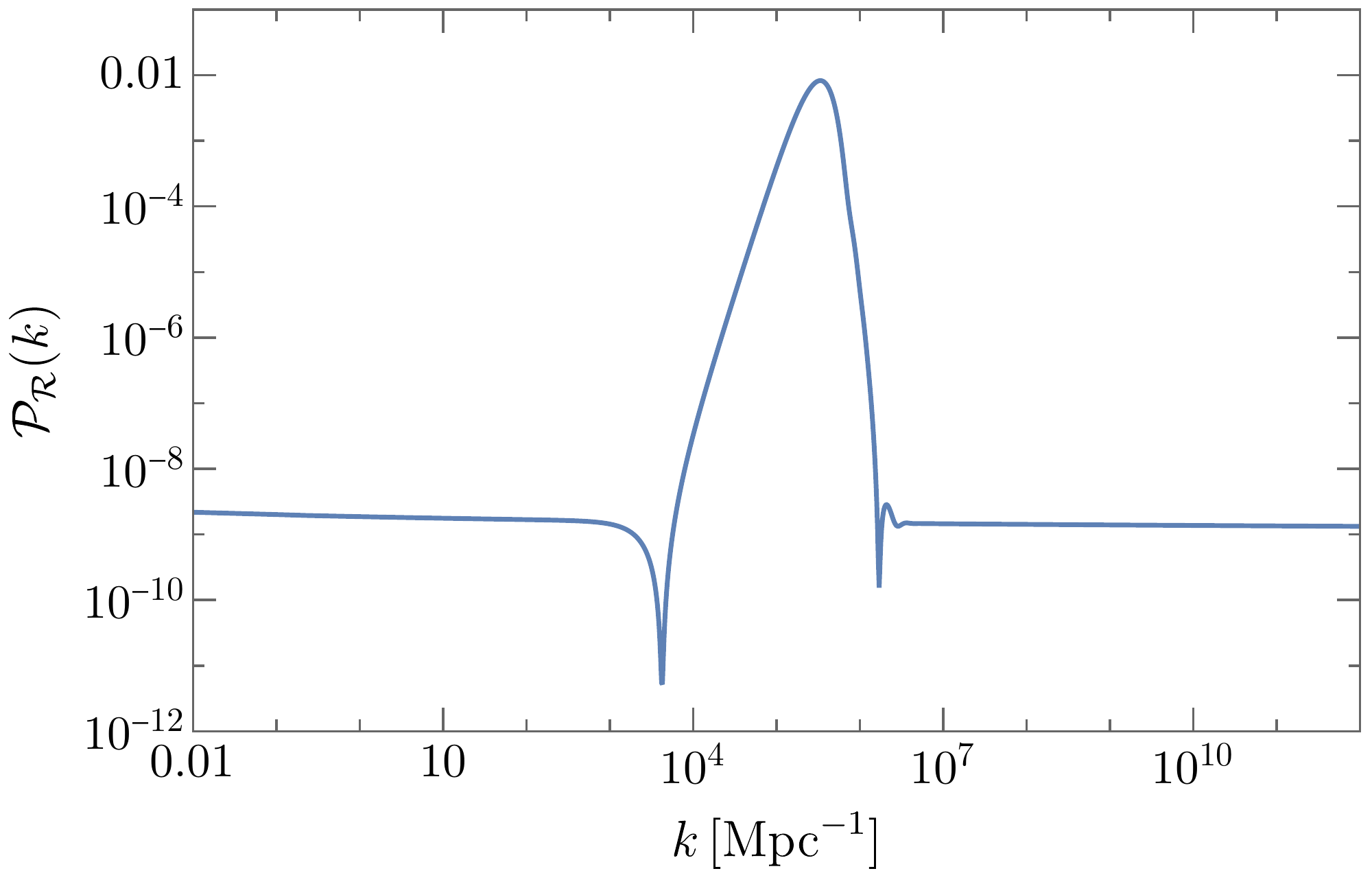}
\caption{ The power spectrum with the upward tanh step, given by Eq.~(\ref{eq:f_tanh_step}).
We take $\tilde l = 0.1$ and the same values for the other parameters as in Fig.~\ref{fig:pot_l01}.
}
\label{fig:ps_tanh_up}
\end{figure}

\begin{figure}  
\centering \includegraphics[width=0.6\columnwidth]{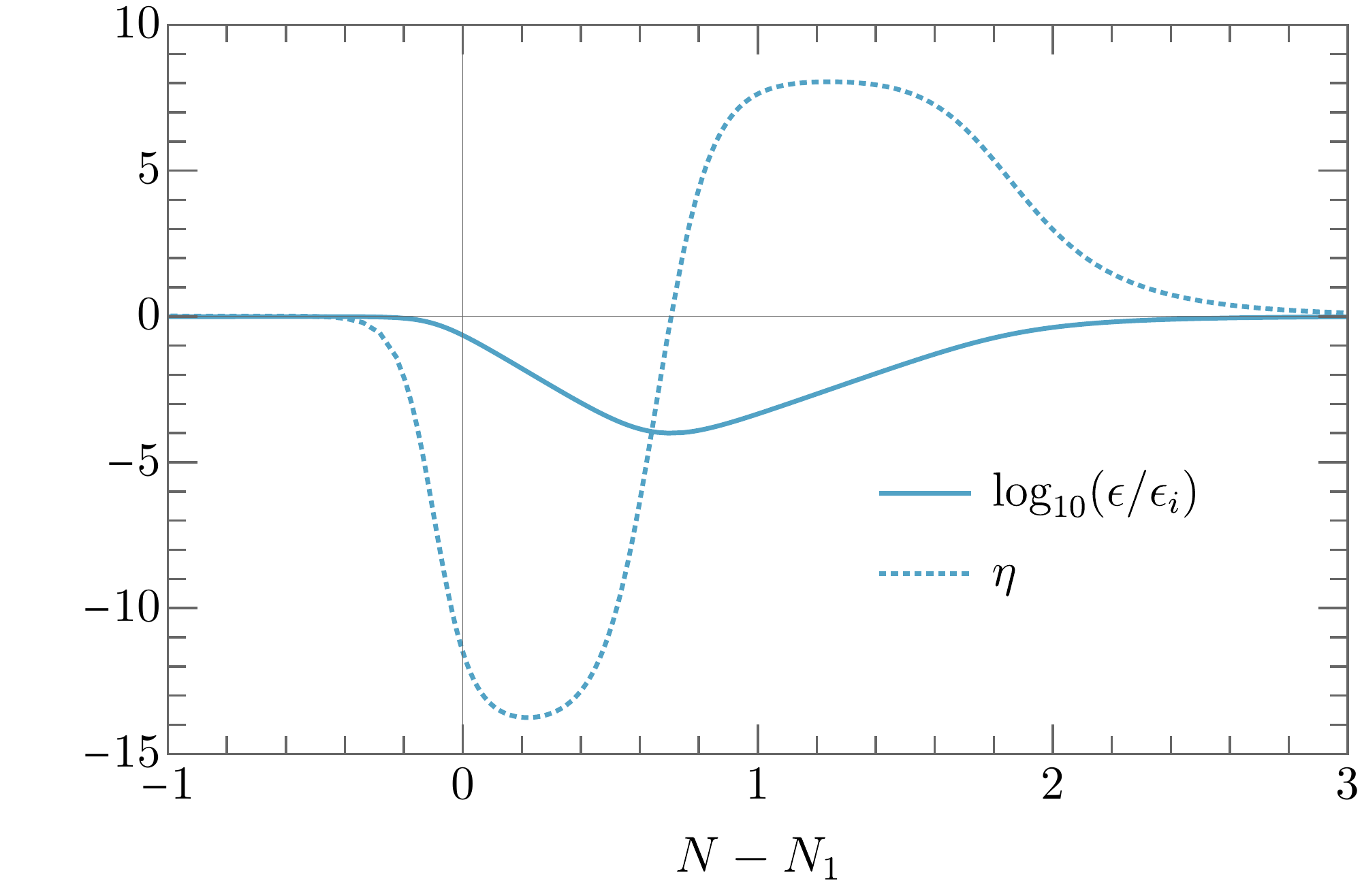}
\caption{ The evolution of $\epsilon$ and $\eta$ for the parameters in Fig.~\ref{fig:ps_tanh_up}.
}
\label{fig:ep_eta_tanh_up}
\end{figure}

\begin{figure}  
\centering \includegraphics[width=0.6\columnwidth]{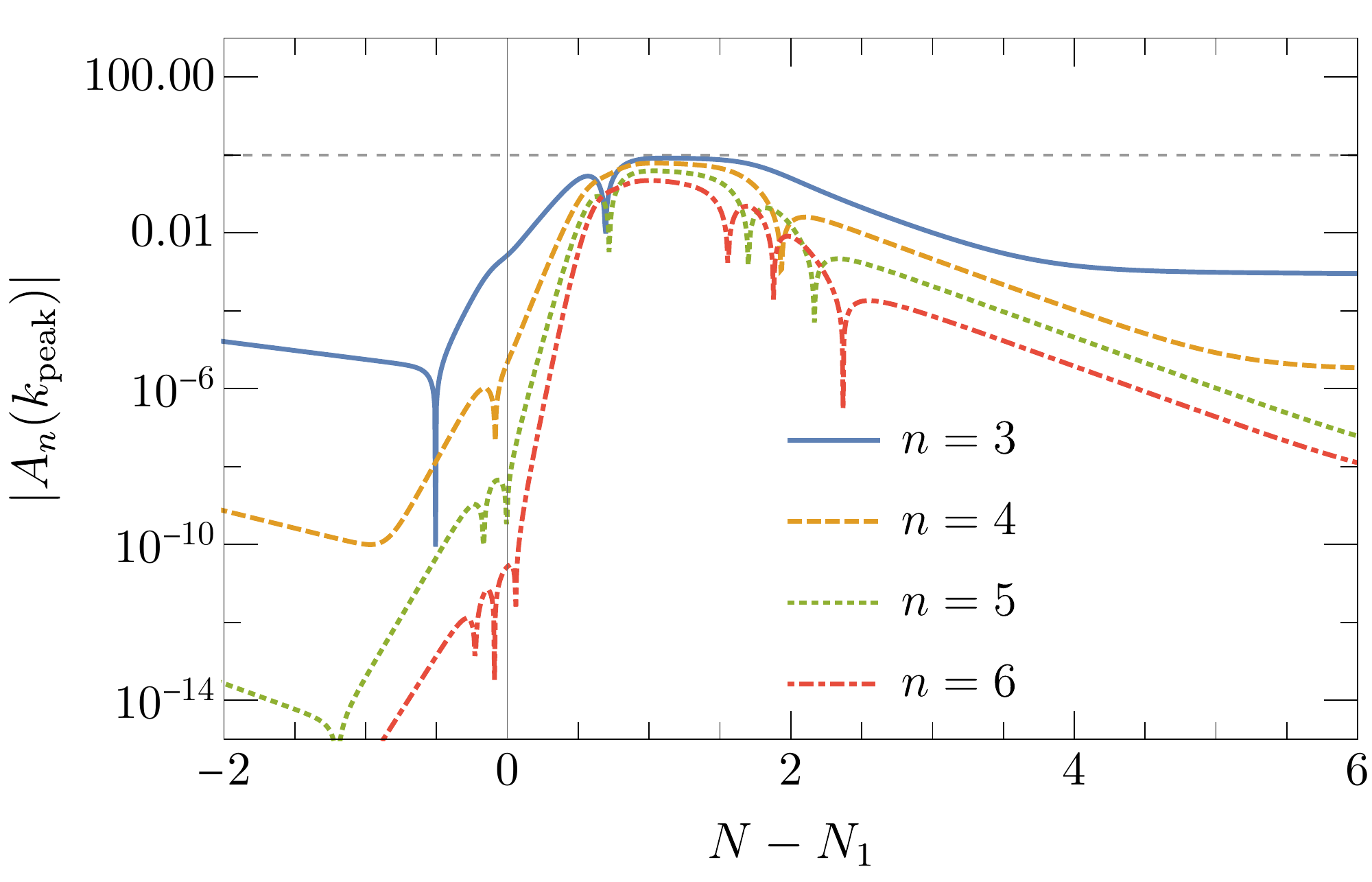}
\caption{ The evolution of $A_n(k_\text{peak})$ for the parameters in Fig.~\ref{fig:ps_tanh_up}.
}
\label{fig:lnol2_tanh_up}
\end{figure}

Figure~\ref{fig:ps_tanh_up} shows the power spectrum in the upward step.
We can see that the spectrum is similar to the results in the main body (Fig.~\ref{fig:ps_cmb_tra_up}).
Also, Figures~\ref{fig:ep_eta_tanh_up} and \ref{fig:lnol2_tanh_up} show the evolution of $\epsilon$ and $\eta$, and $A_n$, respectively.
In particular, Fig.~\ref{fig:lnol2_tanh_up} is very similar to Fig.~\ref{fig:lnol2_up_1em4}. 
From these results, we can see that, unlike in the downward step case, the difference in the step forms does not change the results in the upward step case, in that both cases must nearly violate the strong coupling bound.
This is mainly because the required step widths are almost the same in both forms of the step, as $\phi_2 - \phi_1 \ll \sqrt{2\epsilon_f}M_\Pl (\sim  \sqrt{2\epsilon_i}M_\Pl)$.

\section{Bounce Action for the Inflaton Trapping}
\label{app:bounce_action}

In this appendix, we calculate the bounce action for the tunneling from the local minimum to the true vacuum in our upward step model.
The bounce action determines the exponential factor of the decay rate of the inflaton trapping bubble as $\Gamma \propto \ee^{-S_b}$.
Specifically, what we calculate is the bounce action for the nucleation rate of the true vacuum bubble in the inflaton trapping bubble.
According to Ref.~\cite{Turner:1992tz}, in the case of $\Gamma/H^4 \gtrsim \mathcal O(1)$, the inflaton trapping bubble is expected to be dominated by the true vacuum bubble and will disappear well before the end of the inflation in the other ordinary regions. 
For the inflaton trapping bubble to be a supercritical bubble, $\Gamma/H^4 \ll 1$ is required and this is the motivation for the calculation of the bounce action.

To perform the numerical calculation of the bounce action, we take the potential given by Eq.~(\ref{eq:pot_cmb_to_end_mo_2}).
Here, we assume that the e-folds between the end of the inflation at $\phi_\text{end}$ and the CMB scale at $\phi_\CMB$ is $N_\text{CMB} = 50$ for simplicity and the contribution from $V_\text{end}$ can be neglected in $\phi < \phi_\text{end}$.
In this case, as we will see in the following, the bounce action does not depend on the specific form of the potential around $\phi_\text{end}$.

Based on this concrete potential, we calculate the bounce action, which is given by~\cite{Coleman:1977py,Coleman:1980aw}
\begin{align}
	S_b &= 2\pi^2 \int^\infty_0 \dd r \, r^3 \left[ \frac{1}{2} \left(\frac{\dd \phi_b (r)}{\dd r}\right)^2 + V(\phi_b(r)) - V(\phi_\text{fv}) \right],
	\label{eq:bounce_action}
\end{align}
where $\phi_\text{fv}$ is the value at the false vacuum (the local minimum). 
The equation of motion for the bounce solution is given by~\cite{Coleman:1977py} 
\begin{align}
	\frac{\dd^2 \phi_b}{\dd r^2} + \frac{3}{r} \frac{\dd \phi_b}{\dd r} = \frac{\dd V(\phi_b)}{\dd \phi_b}.
	\label{eq:eom_bounce}
\end{align}
The boundary condition of the bounce solution is given by 
\begin{align}
\frac{\dd\phi_b (0)}{\dd r }= 0, \ \lim_{r\rightarrow \infty} \phi_b(r) = \phi_{\text{fv}}.
\label{eq:boundary_cond}
\end{align}
Once we regard $r$ as the time variable, the bounce solution can be considered as the evolution of $\phi$ in the reverse potential, $-V(\phi)$, and the friction proportional to the inverse of the time.
The solution finally approaches the potential bump in the reverse potential, which corresponds to the false vacuum.

\begin{figure}
\centering \includegraphics[width=0.6\columnwidth]{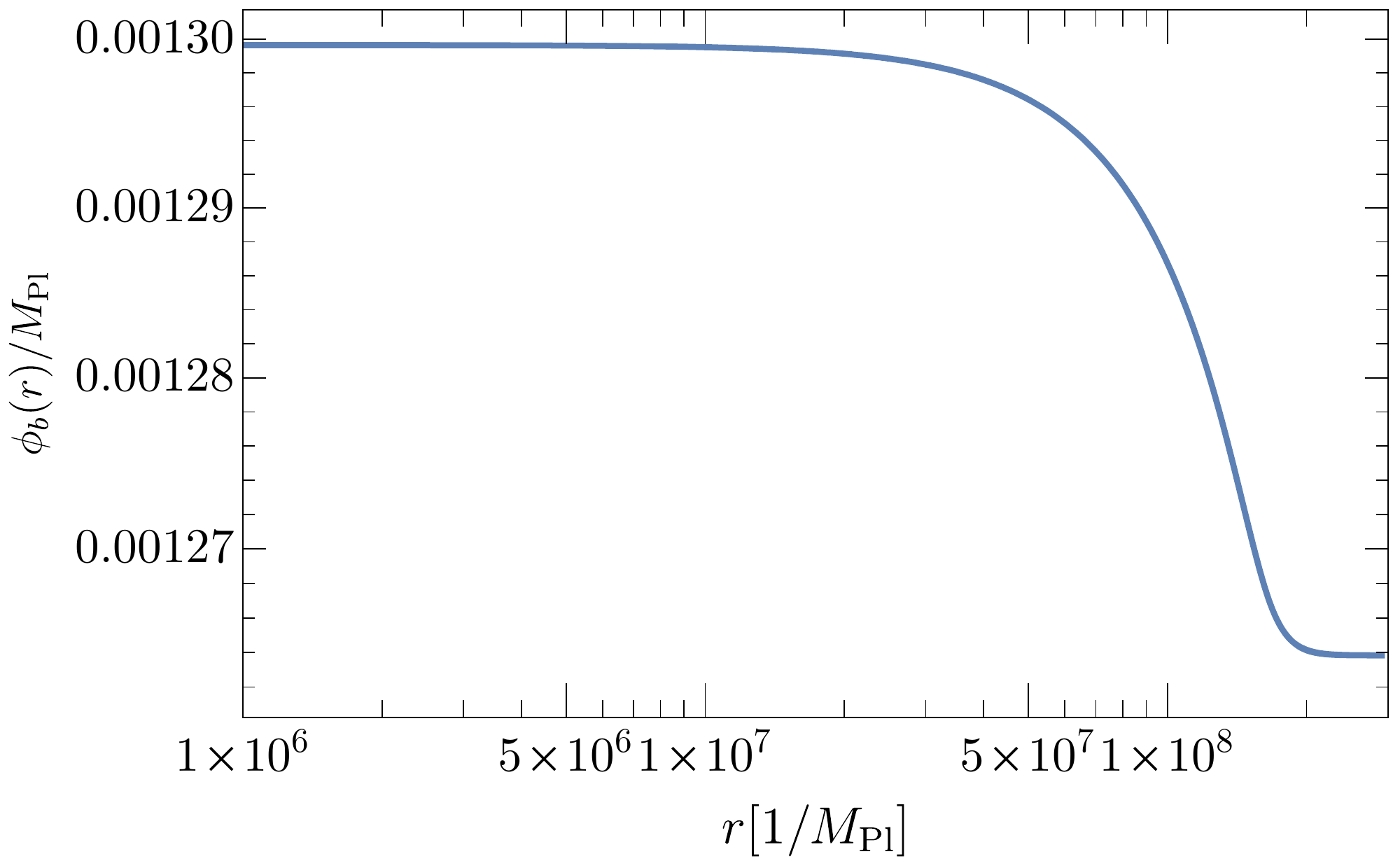}
\caption{ The bounce solution with the parameters for the orange line in Fig.~\ref{fig:ps_smooth_up}.
The equation of motion and the boundary condition for the bounce solution are given by Eqs.~(\ref{eq:eom_bounce}) and (\ref{eq:boundary_cond}), respectively.
}
\label{fig:bounce_sol}
\end{figure}

Figure~\ref{fig:bounce_sol} shows the bounce solution for the potential given by Eq.~(\ref{eq:pot_cmb_to_end_mo_2}) with the parameters for the orange line in Fig.~\ref{fig:ps_smooth_up}, where $\phi_\text{end}/M_\Pl = 2.64\times 10^{-3}$.
From this figure, we can see $\lim_{r\rightarrow 0} \phi_b(r) < \phi_\text{end}$, which means that the bounce solution does not depend on the specific form of $V_\text{end}$.
Substituting this solution into Eq.~(\ref{eq:bounce_action}), we obtain $S_b = 8.7 \times 10^7$, which indicates that the decay rate of the inflaton trapping bubble is significantly suppressed by the exponential factor.

\small
\bibliographystyle{apsrev4-1}
\bibliography{amplification_rise_fall}

\end{document}